\documentclass[journal=chreay,manuscript=review,layout=onecolumn]{achemso}
\usepackage{achemso}
\setkeys{acs}{usetitle = true}
\usepackage{amsmath}

\usepackage[utf8]{inputenc}     

\newcommand{\degree}{$^{\circ}$}
\newcommand{\sgh}{$\sigma$-hole}
\newcommand{\sghs}{$\sigma$-holes}
\newcommand{\etal}{\emph{et al.}}
\newcommand{\ai}{\emph{ab initio}~}

\author{Michal H. Kolář}
\affiliation{Institute of Organic Chemistry and Biochemistry, Academy of 
Sciences of the Czech Republic, Flemingovo nám. 2, 16610 Prague, Czech Republic}
\alsoaffiliation{Institute of Neuroscience and Medicine (INM-9) and Institute
for Advanced Simulations (IAS-5), Forschungszentrum Jülich GmbH, 52428 Jülich,
Federal Republic of Germany}
\email{michal.kolar@uochb.cas.cz}
\phone{+420 220 410 318}

\author{Pavel Hobza}
\affiliation{Institute of Organic Chemistry and Biochemistry, Academy of 
Sciences of the Czech Republic, Flemingovo nám. 2, 16610 Prague, Czech 
Republic}
\alsoaffiliation{Department of Physical Chemistry, Regional Centre of Advanced 
Technologies and Materials, Palacky University, 771 46 Olomouc, Czech Republic}

\title{Computer Modeling of Halogen Bonds and Other $\sigma$-Hole Interactions}

\begin{document}
\tableofcontents


\section*{Abstract}

In the field of noncovalent interactions a new paradigm has recently become popular.
It stems from the analysis of molecular electrostatic potentials and introduces
a label, which has recently attracted enormous attention. The label is \sgh ~and
it was first used in connection with halogens. It initiated a renascence of interest in
halogenated compounds, and later on, when found also on other groups of atoms
(chalcogens, pnicogens, tetrels and aerogens), resulted in a new direction of research of
intermolecular interactions. In the review, we summarize advances from about last
ten years in understanding those interactions related to \sgh. We pay particular
attention to theoretical and computational techniques, which play crucial role
in the field.


\section{Introduction}
\label{sec:introduction}

\begin{figure}[tb]
\includegraphics{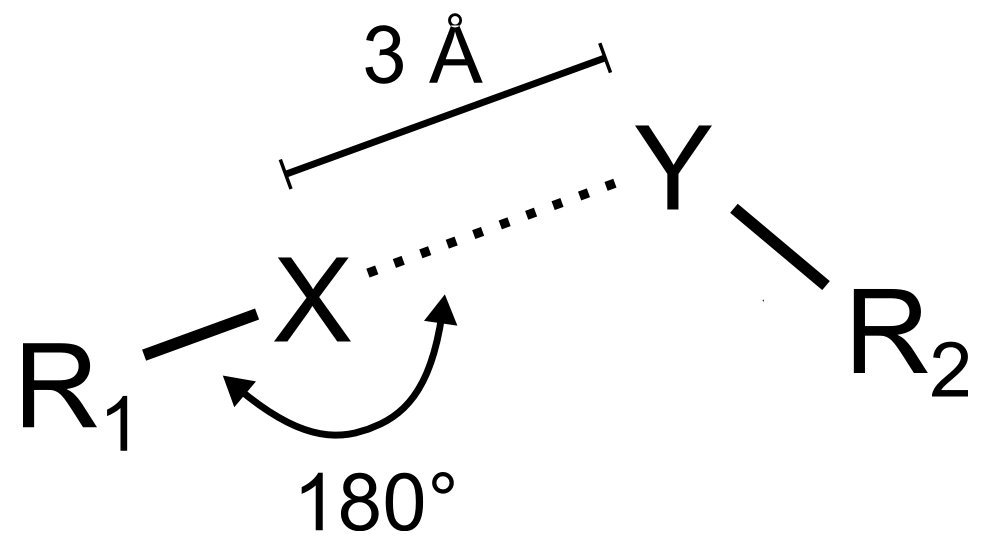}
\caption{Schematic geometry of an idealized halogen bond between a halogen 
atom X (fluorine, chlorine, bromine, or iodine) and an electron-rich atom 
or functional group Y. Typical distance and intermolecular angle are 
depicted.}
\label{fig:xbScheme}
\end{figure}

In 2013, the International Union of Pure and Applied Chemistry (IUPAC) presented 
a recommended definition of the halogen bond (XB). According to Desiraju 
\etal \cite{Desiraju13}, the halogen bond \emph{``occurs when there is evidence of a 
net attractive interaction between an electrophilic region associated with 
a halogen atom in a molecular entity and a nucleophilic region in another,
or the same, molecular entity.''} A representative XB scheme with typical 
geometric features is shown in Figure \ref{fig:xbScheme}. As discussed later, 
its stabilization energy typically reaches several kilocalories per mol.

Halogen bonding is an example of a broader class of noncovalent interactions 
that are characterized by the electrophilic region of one of the two 
interacting partners. These interactions have become known as \sgh ~
bonding. 

The aim of this review is to summarize the last ten years of computational 
studies on XB and other noncovalent interactions involving \sghs. Over
this period, and perhaps even starting a few years beforehand, XB has become
a true celebrity among noncovalent interactions. This is especially evident 
in the field of computational chemistry, which motivated us to sum up the 
most paramount achievements and critically comment on them. A number of 
reviews on XB from a variety of perspectives have been published \cite{Book07,
Metrangolo05, Politzer07, Metrangolo08, Fourmigue09, Politzer10, Legon10,
Erdelyi12, Beale13, Scholfield13, Wilcken13, Gilday15, Scheiner15}. 
The reader might also be interested in a recent review of theoretical 
methods for describing XB \cite{Wolters15}. $\sigma$-hole bonding has not been
reviewed as extensively \cite{Murray09, Politzer12, Clark13, Politzer14b}, 
and we attempt to contribute to this topic, while keeping our major focus on XB.

This review is organized as follows. After several introductory remarks, we 
continue with a subject of turbulent discussions—the nature of halogen 
bonding (Section \ref{sec:nature}). Here, we review the concept of 
\sgh ~with special care to highlight this revolutionary way of 
explaining XB and explain its fundamental importance. Afterwards, we discuss
a few relevant quantum chemical approaches. Section \ref{sec:methods} is 
dedicated to a thorough description of the computational arsenal used to 
tackle XB, covering the range from highly accurate calculations to extremely 
fast and efficient empirical approaches. Finally, in 
Section \ref{sec:studies} we describe several areas of XB research in 
which both computational techniques and experimental work have played roles.

Several factors have contributed to the recent renaissance of XB. The first
is the rapid growth of successful applications of XB in fields such as 
material design \cite{Metrangolo01, Rissanen08}, drug development 
\cite{Lu09, Xu11, Wilcken13}, and catalysis \cite{Bruckmann08, Kniep13}.
These advances opened the door for additional funding and more detailed
research of XB. Another, perhaps even more prominent, factor is the 
character of XB, which, although puzzling, can be explained in a simple 
manner to high school students. One must bear in mind that other noncovalent
interactions have been included in their textbooks for decades, including 
the hydrophobic effect \cite{Tanford80, Chandler05}, the hydrogen 
bond -- a prototypic noncovalent interaction \cite{Pauling60} -- and van der
Waals interactions \cite{London30, Hobza88}. From the students' perspective
is it is always more appealing to explore something which had not been 
known before.

\begin{figure}[tb]
\includegraphics{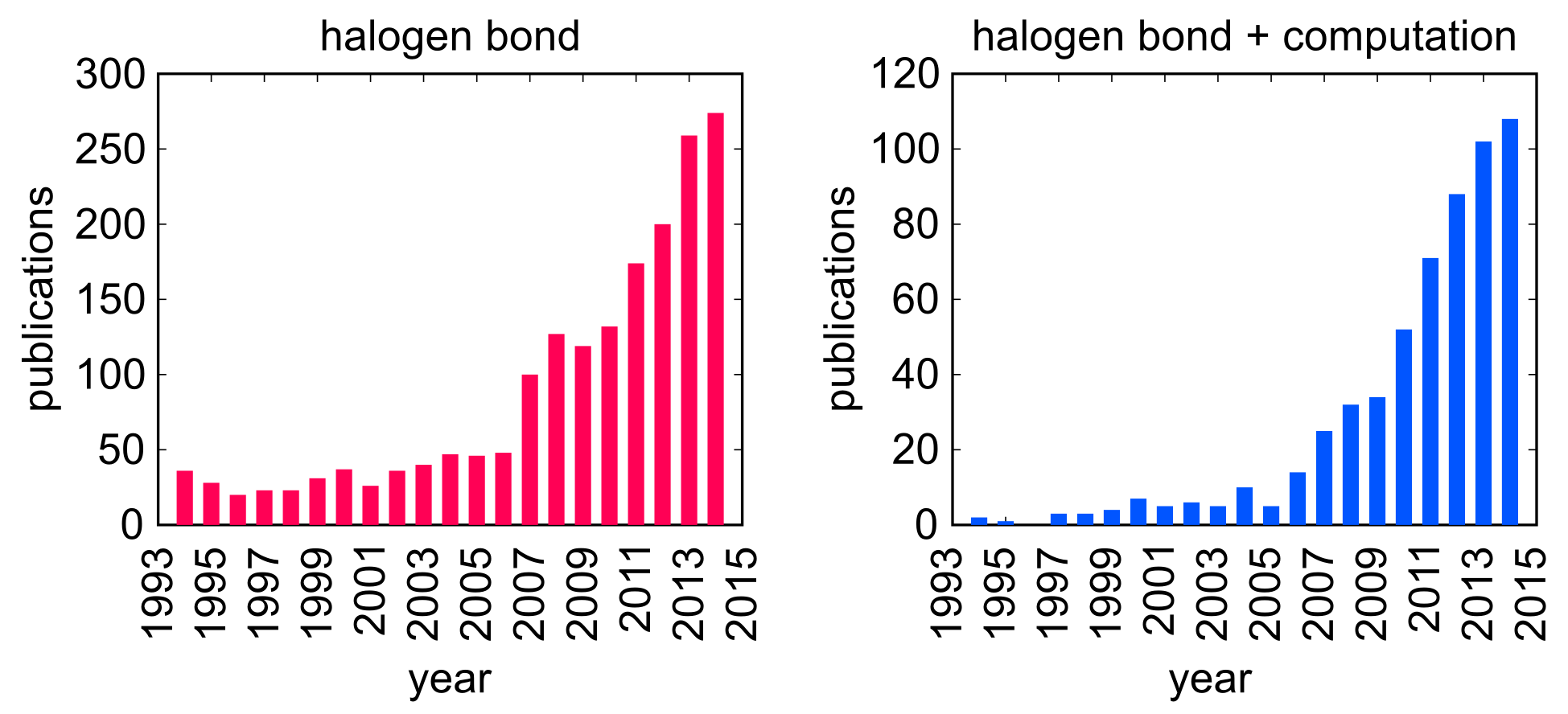}
\caption{Number of publications about XB from the Thomson Reuters Web of 
Science (2015/04). Left: Publications including the keywords ``halogen 
bond'' or ``halogen bonding.'' Right: The same with an additional keyword
from a list evoking a computational or theoretical study (e.g., ``density 
functional theory,'' ``molecular modeling,'' etc.).}
\label{fig:publications}
\end{figure}

The puzzling character of XB arose from the fact that because halogens are 
electronegative atoms, it was not easy to understand why they are attracted 
by electron-rich moieties such as carbonyl oxygens and $\pi$-electron systems. 
To this point, we think that the breakthrough in understanding XB that occurred
about ten years ago was not purely scientific but stemmed more from a marketing
perspective. As discussed in detail below, an elegant way of explaining many 
features of XB -- the so-called \sgh ~-- appeared, or more precisely, 
got a name. A seminal paper by Clark \etal \cite{Clark07} in 2007 triggered
a new epoch of XB research (Figure \ref{fig:publications}).

Finally, from a computational point of view, another factor has played 
an important role -- the ongoing improvement in both software and hardware
\cite{Anton08} used to address questions in chemistry and other natural 
sciences. The level at which one needs to understand a theory to run 
computations has been decreasing. On one hand, in a positive sense, this 
allows people with limited programming and computational science abilities 
to focus on chemistry. On the other hand, it reduces the barrier for doing 
computational science, often negatively affecting the quality and/or impact 
of the output.


\section{Nature of Halogen Bonding}
\label{sec:nature}

The first thing we should address is why it is meaningful to investigate 
the nature of noncovalent interactions. What \emph{kinds} of noncovalent 
interactions nature has? Is it valid to ask? The potential energy part of a 
molecular Hamiltonian contains only the Coulombic term, i.e., the reciprocal
interparticle distance. Consequently, as stated by Feynman \cite{Feynman39}
and argued by Politzer \etal ~in the context of \sgh ~molecular interactions in 
several recent papers \cite{Politzer12, Politzer12b, Politzer14, Politzer15},
the only nature of any interatomic interaction is electrostatics, which 
encompasses many other commonly used contributions, including covalency, 
London dispersion forces, and polarization.

Their arguments are perfectly valid. For instance, from a physical 
perspective, there is no sharp border between covalent and noncovalent 
interactions, and thus there is no physical reason to separate them into 
different groups. On the other hand, it is natural for chemists 
(and not only for them) to distinguish
between the interaction of two carbon atoms in ethane and the interaction of 
carbon atoms in two methane molecules. Therefore the chemist would say that there is 
a chemical bond in between the two carbon atoms, whereas there is nothing 
like chemical bonding in a strict physical view. Likewise, in the gas phase 
the interaction of two methanes is driven by other factors than in water solution.
There would be hardly any doubts that the concept of chemical bond or hydrophobic 
effect \emph{is} useful, however as Clark pointed out, \emph{``We sometimes 
lose track of the fact that many common concepts (e.g., hybridization, molecular
orbitals, and resonance structures) are indeed simply models designed to 
rationalize what we observe.''} \cite{Clark13} It is natural to use schemes and 
simplifying patterns in our daily life observations as well as in chemistry. 
This way of complexity reduction is our main argument for why we think, and
the experience of others seems to justify, that it is advantageous to adopt 
certain interaction energy decompositions to describe noncovalent complexes 
and classify them according to their nature.

In the course of time, a number of schemes have emerged to describe the physical
reality of intermolecular binding (reviewed recently by 
Phipps \etal \cite{Phipps15}). We can understand them as attempts to tell the same
story using variety of words and diction perhaps targeting various listeners. 
Unfortunately, and this must be stressed, the decomposition of interaction
energy \emph{is} ambiguous; there seems to be
no ``more correct'' energy decomposition alike there is no ``more correct'' way of telling
a story. One can only choose the ``more correct'' approach to a listener, or scientific
audience. Consequently when investigating into the nature of molecular binding, 
it is strongly encouraged to stick to particular decomposition scheme, express all 
definitions necessary to understand all energy terms, and keep
in mind that it can be difficult if not impossible to compare results among various
schemes.

The following sections present a few approaches to study the nature of XB and
\sgh ~interactions.

\subsection{$\sigma$-Hole}
\label{ssec:sigmaHole}

\subsubsection{Electrostatic Potential}
\label{sssec:esp}

We begin with the concept that in our opinion has affected XB research most 
significantly -- the concept of \sgh. The name \emph{``\sgh ~bonding''} came to 
apply to a class of noncovalent interactions, when it became clear that \sgh ~is
not related only to halogens. It might be useful to recall a quantity from 
classical electrostatics -- the electrostatic potential (ESP). In atomic units
for a set of atomic nuclei and electrons, the electrostatic potential $V(\textbf{r})$
at spatial position $\textbf{r}$ is defined by equation \ref{eq:esp}.

\begin{equation}
V(\textbf{r}) = \sum_A \frac{Z_A}{\left | \textbf{r}_A - \textbf{r}
\right |} - \int \frac{\rho(\textbf{r}_e)}
{\left | \textbf{r}_e - \textbf{r} \right |} \textrm{d}\textbf{r}_e
\label{eq:esp}
\end{equation}

where $Z_A$ is the charge of nucleus A located at position $\textbf{r}_A$ and $\rho$
is the electron density at position $\textbf{r}_e$. The former positive term accounts
for the ESP generated by the atomic nuclei, while the latter negative term 
stands for the ESP of the electron cloud.

For a fixed electron density, the potential energy $E(\textbf{r}_Q)$ of a probe 
charge $Q$ located at spatial position $\textbf{r}_Q$ is then

\begin{equation}
E(\textbf{r}_Q) = Q V(\textbf{r}_Q)
\label{eq:potentialEnergy}
\end{equation}

It must be noted that, unlike the interaction energy or interaction energy 
contributions derived from any decomposition scheme, the ESP is a physical 
observable. The ESP can be calculated from electronic density either measured
experimentally or calculated from a theoretical model. In current science, 
however, the distance between experiment and theory is diminishing. Without
exaggeration, one can view experimental results as results of models with 
some (varying amount of) experimental input. Examples of ESP measurements 
are provided in the literature \cite{Stewart79, Koritsanszky01}.

For molecules, it is common practice to adopt the Born-Oppenheimer 
approximation \cite{Born27} to calculate the ESP at positions around fixed 
atomic nuclei. It is customary to choose the positions, but it appears 
advantageous to define a molecular surface and project the ESP onto it.
Bader \etal ~proposed a surface with an electron density of 0.001 au 
(i.e., e/bohr$^3$) \cite{Bader87}, which encompasses approximately 96\% 
of the electronic charge of a molecule. This density value became standard
in calculations of molecular electrostatic potential (MEP), although other
values, such as 0.0015 au and 0.002 au, have remained in use. MEPs have proven
useful in understanding molecular properties and their interaction preferences
\cite{Cox81, Pullman81, Politzer13}, and an important finding is that the 
conclusions are often insensitive to the precise choice of electron density
surface.

\subsubsection{Discovery of $\sigma$-Hole}
\label{sssec:discoverySigmaHole}

\begin{figure}[tb]
\includegraphics{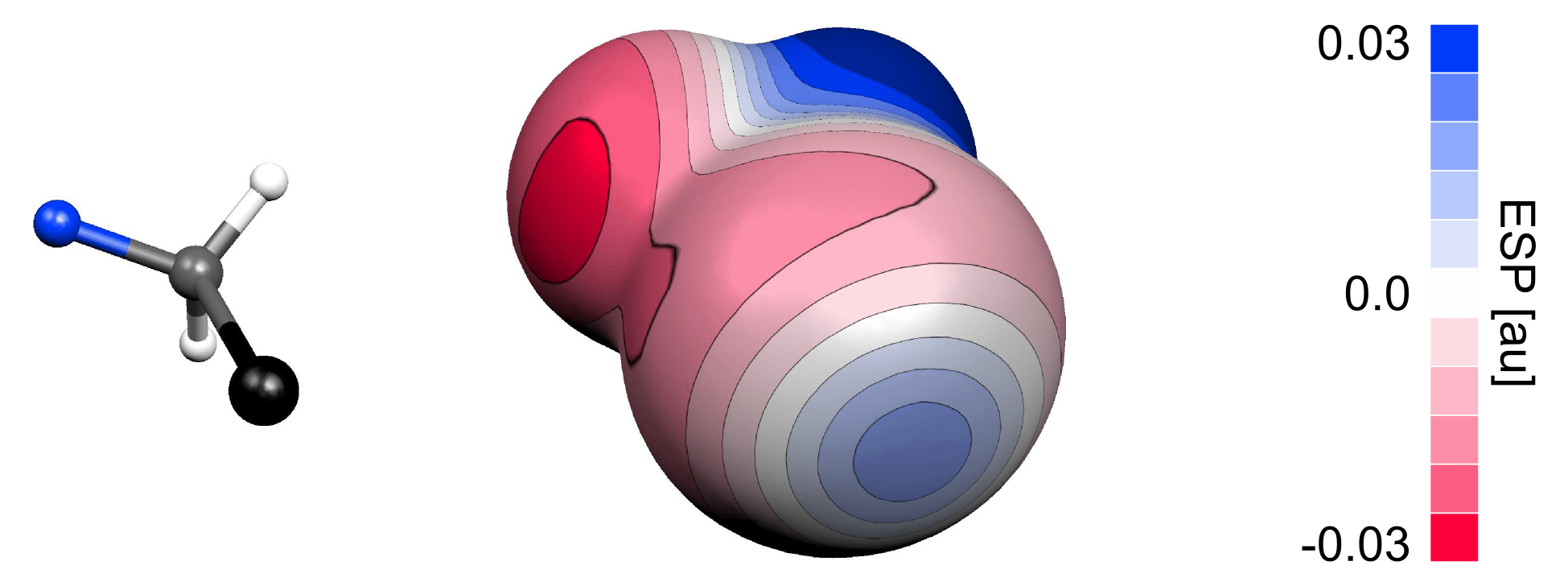}
\caption{The ball-stick model of bromofluoromethane (left). Hydrogen is shown 
in white, carbon in gray, fluorine in blue, and bromine in black. On the right,
the corresponding projection of a molecular electrostatic potential (ESP) 
in atomic units on a surface of 0.001 au electron density is shown. The blue
disc on the surface in the forefront represents the \sgh.}
\label{fig:sigmaHole}
\end{figure}

Analysis of MEP of halogenated methanes in 1992 revealed for the first 
time \cite{Brinck92} that the ESP around halogen atoms is not isotropic but
exhibits regions with both positive and negative values. At that time, the 
positive region was found in the elongation of the C--X covalent bond (where 
X = Cl, Br, or I), whereas the negative region created a concentric belt 
around the C--X bond (Figure \ref{fig:sigmaHole}). 
Based on MEPs calculated at the Hartree-Fock
(HF) level \cite{Roothaan51}, it was possible to interpret interactions in 
crystal structures of halomethanes \cite{Ramasubbu86}. In fact, close 
contacts of halogen atoms with electronegative moieties had been observed much
earlier \cite{Hassel54, Hassel58, Hassel59} but a satisfactory explanation had
been missing.

Such a region of positive ESP was named \sgh ~\cite{Clark07}, because the 
region appears in the elongation of the $\sigma$ bond of the halogen. 
Interestingly enough, the name was introduced 15 years after the first 
reference to halogen ESP anisotropy. A later review on \sghs
\cite{Clark13}  remarked on how fascinating it has been \emph{``$\ldots$to wonder 
why their (XBs') contribution remained unrecognized for so long.''}
It is even more fascinating when one realizes that the discovery of 
anisotropic electron density on halogens dates back to 1960s and 
1970s to the studies of solid chlorine \cite{Collin56, Donohue65, Nyburg79}.

Once introduced, the term \sgh ~was accepted by the XB community very 
quickly.

\subsubsection{$\sigma$-Holes on Halogens and Other Atoms}
\label{sssec:sigmaHolesXb}

Halogens were the first atoms to be shown to carry a \sgh. Because the 
\sgh ~is nothing but a region of ESP with certain properties, it is 
intimately linked to the electron density. In other words, \sgh ~is 
a quite illustrative way of presenting electron density. Literally speaking,
the electrons of a halogenated molecule simply do not like to sit on the extension
of covalent bond to the halogen atom.

\begin{figure}[tb]
\includegraphics{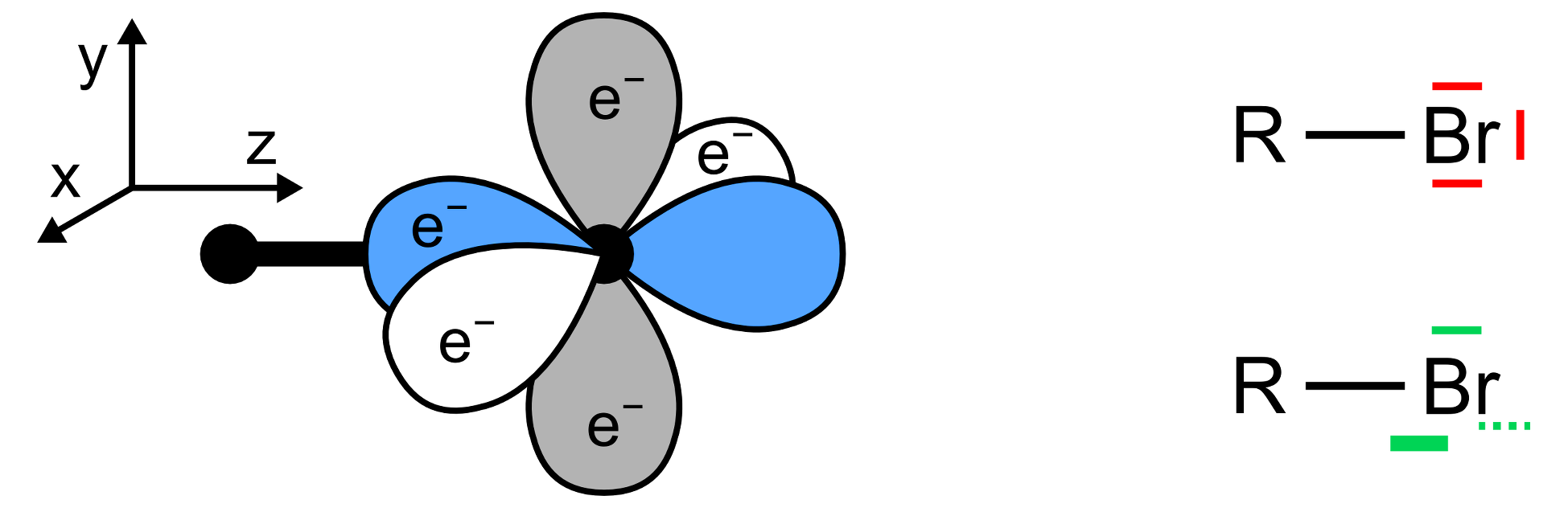}
\caption{The orbital arrangement. Left: the p$_x$ (white), p$_y$ (gray), 
and p$_z$ (blue) valence orbitals on a halogen atom. The electrons (e$^-$)
are localized in p$_x$, p$_y$, and the inner half of the p$_z$ orbital.
The outer half is electron-deficient, which expresses itself as a \sgh.
The remaining two valence electrons are localized in a spherically symmetric 
s-orbital (not drawn). Right: A typical depiction of lone electron pairs
(in red) around bromine (upper structure). It evokes the misleading impression
that the elongation of the R--Br bond is electron-rich. A more correct view, 
which would require three dimensions (lower structure), suggests a region of
electron depletion in the elongation of the R--Br bond. The thick electron pair
is in front of the plane of the paper, while the dotted one is behind (in green).}
\label{fig:orbitals}
\end{figure}

The concept of atomic orbitals provides an illustrative framework to 
understand \sghs: the valence shell of, for example, a chlorine atom
in an organic molecule has a configuration of 3s$^2$ 3p$_x^2$ 3p$_y^2$ 3p$_z^1$,
where the 
z-axis coincides with the direction of the C--Cl bond. Note that hybridization
for halogens and especially for heavier halogens is not significant.
The electron in the p$_z$ orbital is mostly localized in the bond region, which 
results in a lack of electron density in the outer (non-involved) lobe. The 
electron pairs in the p$_x$ and p$_y$ orbitals create a negative region with 
a contribution of the s orbital, as depicted in Figure \ref{fig:orbitals}. 
It is intriguing to see the correspondence with a widely used depiction of 
halogen lone-electron pairs (Figure \ref{fig:orbitals}), which leads to 
incorrect predictions and causes the puzzling character of XBs.

One can ask if such ESP anisotropies are present on other atoms than halogens,
and the answer is yes, definitely. It appears reasonable that the covalent 
bond between any two atoms affects the electron distributions of each of 
them. Thus, for the same reasons as halogens, also chalcogens, pnicogens, tetrels,
and even aerogens (noble gases) may contain \sghs ~in 
the elongations of their covalent bonds with neighboring 
atoms \cite{Murray09, Murray10}. The authors of papers on these topics 
advocate the use of a general name for these binding patterns: \sgh
~bonding. Mostly, however, the denominations halogen (group 17 elements -- F,
Cl, Br, I), chalcogen (group 16 elements -- O, S, Se, 
Te) \cite{Wang11b, Brezgunova13}, pnicogen (group 15 elements -- N, P, 
As, Sb) \cite{DelBene11, Zahn11, Scheiner12}, tetrel (group 14 elements -- C,
Si, Ge, Sn) \cite{Bauza13d, Grabowski14}, and aerogen (group 18 elements -- Ne,
Ar, Kr, Xe) bonding \cite{Bauza15} are used. There is little to no evidence 
that the heaviest atoms of each of the groups (At, Po, Bi, Pb, and Rn, 
respectively) participate in similar kinds of intermolecular interactions. 
Halogen, chalcogen, and pnicogen bonds have one common feature: their 
\sghs ~are easily accessible by electron donors, and the respective 
bonds are thus important structure-making factors. On the other hand, the 
\sgh ~of tetrels is located in the middle of three sp$^3$-hybridized bonds, 
which makes its accessibility rather low. Consequently, halogen, chalcogen, 
and pnicogen bonds may find broad applications in chemistry, biology, and 
material sciences, while potential applications of tetrel bonds are limited.

The \sghs ~of group 17 elements and of other group G (G=16, 15, and 14)
elements differ, which is connected with the systematically monovalent 
character of halogens, as opposed to the other elements, which can be di-,
tri-, or generally polyvalent. Consequently, the halogens possess only one 
\sgh, while other elements can have two, three, or even more 
\sghs.  The single \sgh ~of halogens is located 
opposite to the R--X covalent bond. This location determines 
the structure of halogen bonded complexes; the halogen bond tends to be 
linear, and this linearity is its characteristic property.

\begin{figure}[tb]
\includegraphics{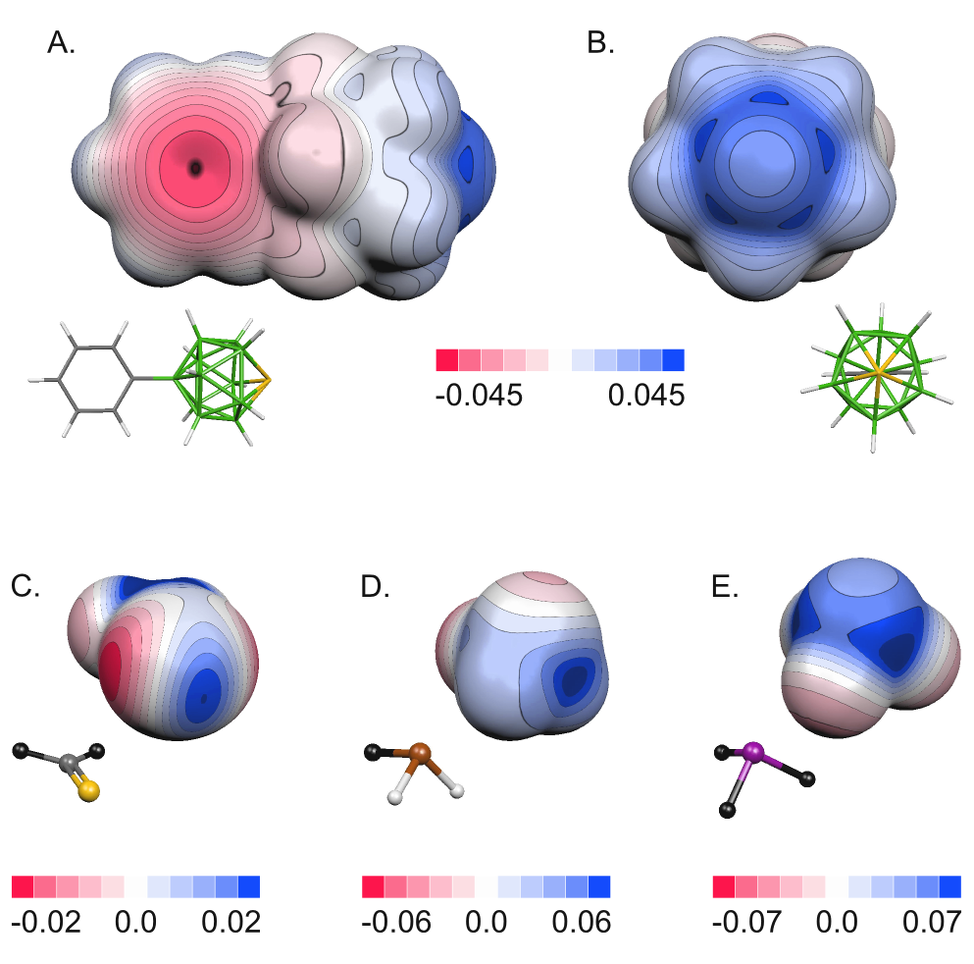}
\caption{The electrostatic potential (ESP) projected on the surface of 0.001 au 
electron density. Side (A) and top (B) views of a thioborane molecule, 
carbonothioyl difluoride (CSF$_2$, C), fluorophosphane (PH$_2$F, D), 
and arsenic trifluoride (AsF$_3$, E). The molecular ball-stick models are 
provided with the following color code: H-white, B-green, C-gray, F-black,
P-brown, S-yellow, As-purple. Note that the molecules are not proportional.}
\label{fig:espSulphur}
\end{figure}

The situation with polyvalent elements is different, which can be demonstrated
with group 16 elements. Figure \ref{fig:espSulphur} shows electrostatic 
potential at two sulfur-containing compounds, F$_2$C=S and 
12-Ph-closo-1-SB$_{11}$H$_{20}$ \cite{Fanfrlik14}. The sulfur atom on the 
first molecule is divalent, but the carbon–sulfur bond is a double bond. 
Consequently, as with the halogens, there is only one \sgh ~located 
on the surface of the sulfur atom. The \sgh ~is clearly visible 
in Figure \ref{fig:espSulphur}, with a magnitude of 0.02 au. It should be 
mentioned that the analog H$_2$C=S carries a negative \sgh ~with a magnitude 
of --0.0049 au (both at the HF/cc-pVDZ level). The chalcogen bonds 
in complexes of the F$_2$C=S would thus be nearly linear, similar 
to halogen bonds.

The situation is dramatically different in thioborane 
(Figure \ref{fig:espSulphur}), in which the sulfur is bound to five 
boron atoms and is positively charged. Interestingly, the ESP of such 
a molecule is highly anisotropic and shows a completely positive region 
on the upper part of the boron cage. The fact that pentavalent sulfur is 
positive is not surprising; what is surprising, however, is that the sulfur
atom carries five more positive \sghs ~on its side. The magnitude 
of these \sghs ~is surprisingly high (0.043 au); it is even higher 
than in the majority of halogenated systems and is comparable to 
the magnitude of the \sgh ~on bromine in pentafluorobromobenzene.
When the phenyl group in this thioborane is replaced with a chlorine atom,
the magnitude of \sgh ~increases (0.049 au), which suggests that
the properties of \sghs ~in group 16 elements can be tuned 
similarly as those of group 17 elements. 

\begin{figure}[tb]
\includegraphics{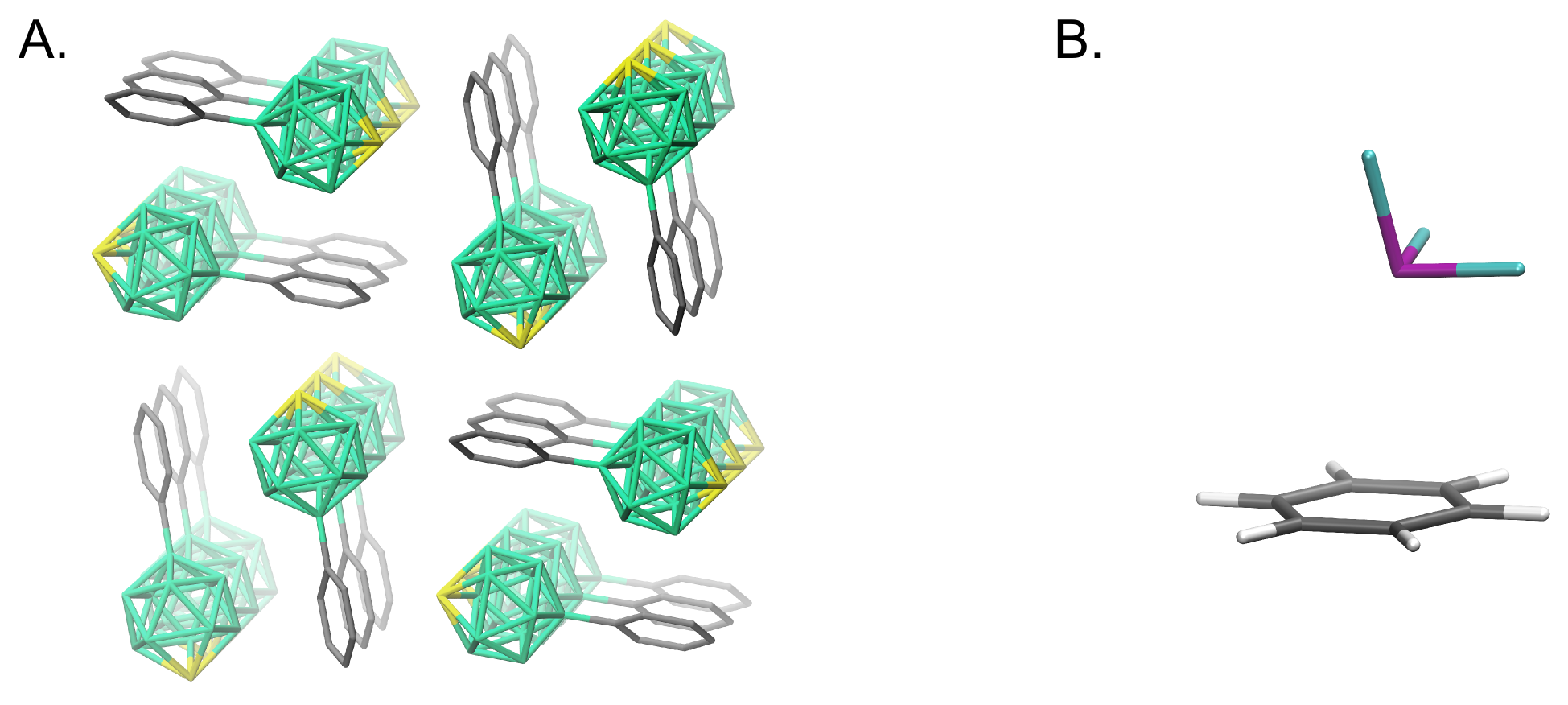}
\caption{Crystal structures of a heterocarborane exhibiting chalcogen 
S$\ldots\pi$ bond (A) (hydrogens omitted for clarity) and a 
benzene$\ldots$AsF$_3$ complex with an As$\ldots\pi$ pnicogen bond (B).
The following color code is used: H-white, B-green, C-gray, F-blue, 
S-yellow, As-purple.}
\label{fig:crystals}
\end{figure}

$\sigma$-holes represent a structure-determining force, and the nearly linear 
arrangement of XB is a consequence. The \sghs ~on sulfur in thioboranes
are, however, not localized on top of sulfur but on its belt, which ranges 
from 120\degree ~to 150\degree ~from the B$_{12}$-S axis. This reflects the 
covalent bonding pattern of the sulfur. Thus, the orientation of the chalcogen
bonded thioboranes in not
linear but bent. The B$_{12}$-S axes in two 12-Ph-closo-1-SB$_{11}$H$_{20}$
molecules in the respective crystal structure are not perpendicular 
(Figure \ref{fig:crystals}), as they should be if the \sgh ~were 
localized on top of the sulfur atom, but have an angle of 155\degree,
which is in full agreement with the prediction of non-linearity of the
chalcogen bond in these thioboranes.

Figure \ref{fig:espSulphur} further shows the ESP of several systems 
containing group 15 elements \cite{Lo15}. The situation is similar as with 
chalcogens discussed above. The \sghs ~of trivalent pnicogens are 
localized on the sides of the atom, whereas the top of the pnicogen is less 
positive. Localization of more positive \sghs ~on the belt of pnicogens
will determine the structure of the respective complexes. As an example, 
Figure \ref{fig:crystals} shows the structure of a AsF$_3\ldots$benzene
complex with an As$\ldots\pi$ pnicogen bond. Assuming the \sgh ~is 
localized on top of the As atom, the complex should have C$_{3v}$ symmetry.
A non-constrained geometry optimization resulted, however, in a complex 
with lower symmetry (C$_1$), which is a consequence of the fact that positive
\sghs ~at As in the AsF$_3$ system are localized on the belt around 
the atom. They preferentially interact with negative $\pi$-electron clouds of
the benzene moiety. This gas-phase-optimized structure fully agrees with 
the experimental X-ray one. When enforcing the C$_{3v}$ structure
in a similar SbF$_3\ldots$hexamethylbenzene complex, 
the respective stabilization energy decreases by 1.1~kcal/mol because 
the $\pi$-electron density of the hexamethylbenzene interacts with the top
of the Sb atom, which is less positive than the respective \sghs ~
localized at the belt of the Sb atom.

The magnitude of the \sghs ~on these pnicogens is very high, even 
higher than that on chalcogens, which makes the pnicogen bonds very stable.
Diiodine possesses the highest magnitude among halogens investigated in this
review (0.049~au); those of AsCl$_3$ and AsF$_3$ equal 0.061 au and 
0.080 au, respectively.  Even higher magnitudes have been found for several
phosphorus molecules: PH$_3$ (0.025 au), PH$_2$F (0.066 au), and P(CN)$_3$
(0.093 au).

\subsubsection{Characteristics of a $\sigma$-Hole}
\label{sssec:characteristics}

Although quite late in view of the early works on \sghs, a unified 
nomenclature of \sgh ~properties was proposed to help in discussions 
of halogen bonding and \sghs. Using a few descriptors, we aimed to 
understand the characteristics of \sghs ~as three-dimensional spatial
objects \cite{Kolar14a, Kolar14b, Kolar14c}. Because of their relative 
simplicity, characteristics have been defined for halogen \sghs ~only,
but the extension to other atoms seems straightforward. A subtle complication,
however, may be seen in the more complicated topology of the chalcogen, 
pnicogen, and tetrel \sghs.

\begin{figure}[tb]
\includegraphics{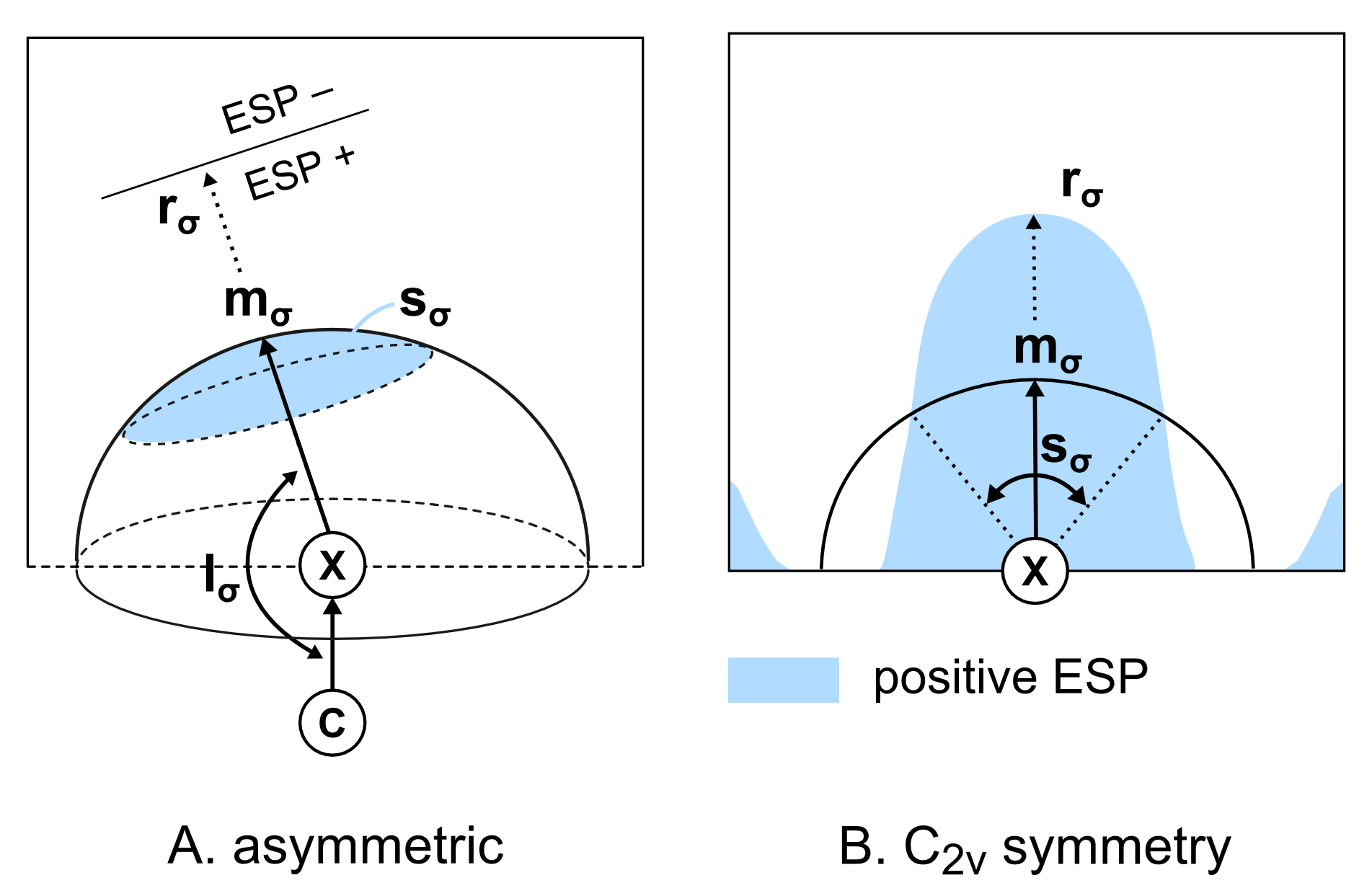}
\caption{A scheme of \sgh ~descriptors. The magnitude (m$_\sigma$),
size (s$_\sigma$), linearity (l$_\sigma$), and range (r$_\sigma$) are 
defined in three dimensions for an asymmetric molecule (left). On the 
right, a two-dimensional case with C$_{2v}$ symmetry has s$_\sigma$ defined
as an angle. The size and range are meaningful only for positive \sghs.}
\label{fig:characteristics}
\end{figure}

The first property of interest is the maximum ESP ($V_{max}$) located on 
top of the halogen atom. This property was originally proposed for hydrogens
about 25 years ago \cite{Murray91}, while its connection with halogens appeared 
only one year after \cite{Brinck92}. 
$V_{max}$ is a value of ESP at the intersection of the 
molecular surface defined by an electron density of 0.001 au \cite{Bader87}
and the elongation of the R--X bond, where X stands for a halogen atom and R
is usually carbon. The value of $V_{max}$ represents the \emph{magnitude of \sgh}
(m$_\sigma$). However, a number of more-or-less confusing
denominations of $V_{max}$ have been also been used. Once the point with 
$V_{max}$--P($V_{max}$) -- has been localized on the molecular surface, the angle 
R--X--P($V_{max}$) can be determined. This represents the \emph{linearity of \sgh}
~(l$_\sigma$, which corresponds to angle $\phi$ in our earlier 
work \cite{Kolar14b}). Although 
the region of positive ESP projected on the molecular surface may have 
a complicated shape, the \emph{size of \sgh} ~(s$_\sigma$) can be defined as an area of 
the halogen MEP with the positive ESP having an approximately circular 
boundary (Figure \ref{fig:characteristics}) \cite{Kolar14b}. Finally, 
the \emph{range of \sgh ~(r$_\sigma$)} is defined as the distance between
the halogen atom and a point at which ESP changes sign from positive 
to negative. The range is determined in the direction X--P($V_{max}$),
and together with the size, it is
meaningful only for positive \sghs. For a molecule with C$_{2v}$ symmetry,
a two-dimensional plot (Figure \ref{fig:characteristics}, right) provides 
a simpler descriptor. Instead of defining the size of \sgh ~as an area, 
one can define it as an angle. While \sgh ~magnitudes have been studied 
intensively, the other parameters have attracted considerably smaller attention.
The parameters for selected halogenated molecules are shown 
in Table \ref{tab:characteristics}.

\begin{table}[tb]
\footnotesize
\begin{tabular}{l r r r r}
\hline
\hline
&magni.  [au]   &  size [\AA$^2$] & range [\AA] & linearity [deg] \\
\hline
Cl$_2$         &  0.0398  & 11.4  & $>$25  &  180.0 \\
Br$_2$         &  0.0450  & 13.7  & $>$25  &  180.0 \\
I$_2$          &  0.0486  & 17.0  & $>$25  &  180.0 \\
H$_2$          &  0.0174  & 8.2   & $>$25  &  180.0 \\
Cl--Ph         &  0.0073  & 1.8   & 2.23   &  180.0 \\
Br--Ph         &  0.0152  & 4.0   & 2.78   &  180.0 \\
I--Ph          &  0.0269  & 7.6   & 4.02   &  180.0 \\
H--Ph          &  0.0230  & 7.5   & $>$25  &  180.0 \\
ClCH$_3$       & --0.0002  & ---  & ---    &  180.0 \\
BrCH$_3$       &  0.0089  & 2.2   & 2.39   &  180.0 \\
ICH$_3$        &  0.0227  & 6.0   & 3.41   &  180.0 \\
CH$_4$         &  0.0141  & 7.4   & $>$25  &  180.0 \\
ClFCH$_2$      &  0.0108  & 3.2   & 2.52   &  178.3 \\
BrFCH$_2$      &  0.0181  & 5.5   & 3.25   &  179.1 \\
IFCH$_2$       &  0.0298  & 9.7   & 5.53   &  179.4 \\
FCH$_3$        &  0.0290  & 8.4   & $>$25  &  168.3 \\
\hline
\hline
\end{tabular}
\caption{The magnitude, size, linearity and range of \sghs ~of selected 
halogenated molecules (at the CAM-B3LYP/def2-QZVP level). For hydrogenated 
analogues, the properties were calculated for the maximum ESP. Ph stands 
for phenyl.}
\label{tab:characteristics}
\end{table}

From the orbital interpretation of \sghs ~(Figure \ref{fig:orbitals}), 
it arises that the lack of electron density in the outer lobe of 
the p$_z$ orbital may depend on both the atomic number of the halogen and 
its chemical environment. At first, it was not clear if the lack of electron
density is manifested by a region of positive ESP or a region of negative ESP
that is, however, less negative than its surroundings. The first evidence 
of ESP anisotropy on halogens has been reported for both of these 
phenomena \cite{Brinck92}.

It has been accepted that \sghs ~are not always positive, although 
in early work the term \sgh ~was reserved for positive ESPs 
only \cite{Clark07}. Brinck \etal \cite{Brinck92} calculated the $V_{max}$ 
on chlorine in CH$_3$Cl to be --0.0033 au (at the HF/6-31G* level). In this 
case, the \sgh ~is, however, less negative than the rest of the MEP on
the sides of the chlorine. Negative \sghs ~are often present on 
fluorinated or chlorinated compounds. On the other hand, a computational 
analysis of more than 2,500 commercially available halogenated drug-like 
organic molecules revealed \cite{Kolar14b} that only a small fraction (less
than 1\%) contains negative \sghs; the others are positive (at 
the B3LYP/def2-QZVP level). A positive \sgh ~is, however, not a necessary
prerequisite for a molecule to participate in XB as a halogen donor. 
It was shown, for example, that a XB complex of CH$_3$Cl with formaldehyde 
has a stabilization energy of 1.17~kcal/mol despite a negative 
\sgh ~\cite{Rezac12a}. This provides evidence that the electrostatic 
interaction between the \sgh ~and an electron acceptor is not the only 
stabilizing contribution. An important part of halogen bond stabilization 
originates in systematically attractive dispersion energy.

There are two main factors that affect the characteristics of \sghs:
i) the nature of the halogen, defined by its atomic number, and ii) its 
chemical environment. Table \ref{tab:characteristics} portrays both effects.

\begin{figure}[tb]
\includegraphics{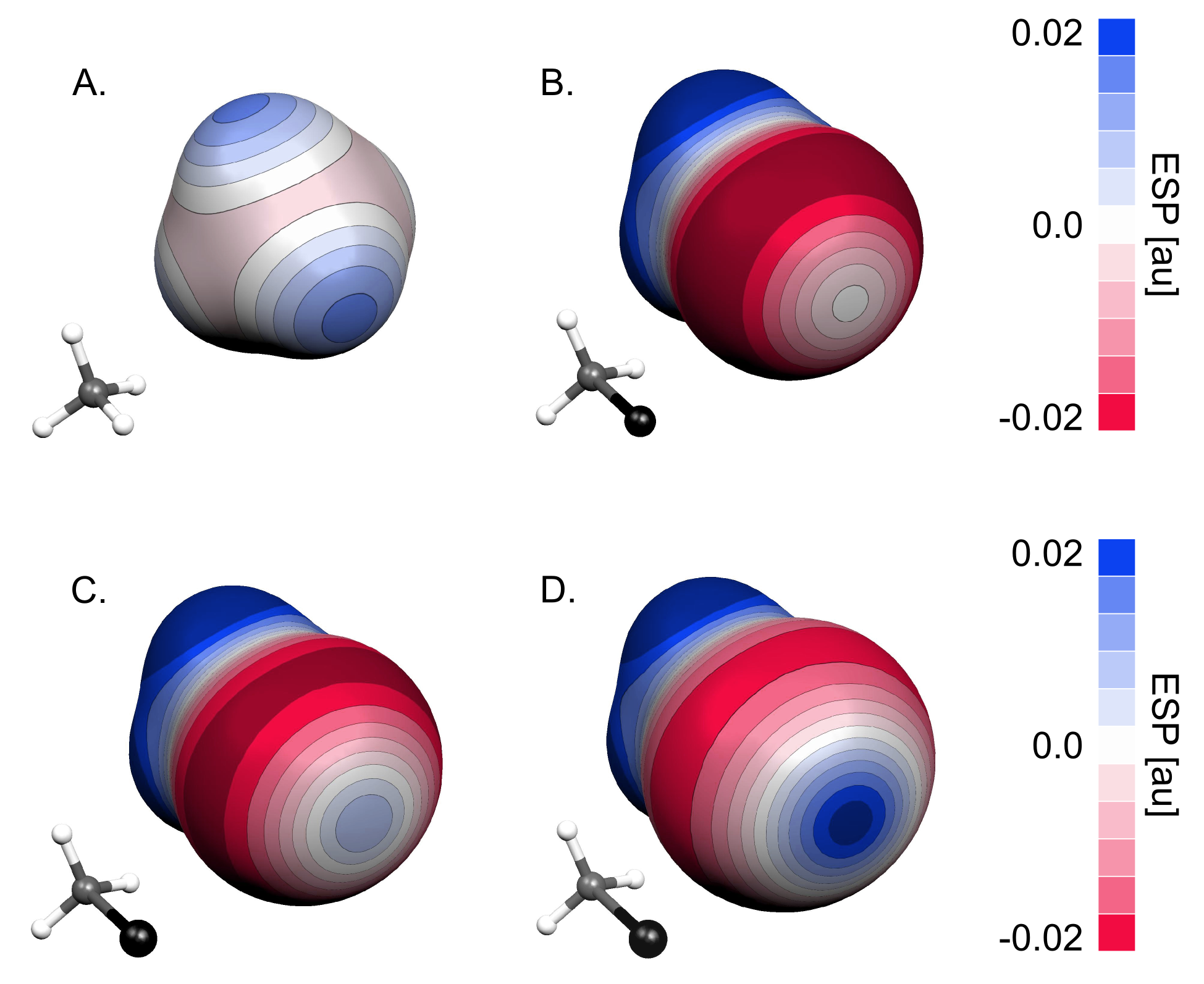}
\caption{The electrostatic potential (ESP) projected on a surface of 0.001 
au electron density of methane (A), chloromethane (B), bromomethane (C), and 
iodomethane (D). The ball-stick models of the molecules are provided with 
hydrogen in white, carbon in gray, and halogen in black. Note the same ESP 
scale across the surfaces with positive blues and negative reds.}
\label{fig:methaneSigmaHole}
\end{figure}

The magnitude of \sghs ~increases with increasing atomic number of 
the halogen in the order F $<$ Cl $<$ Br $<$ I \cite{Clark07, Politzer10, Kolar14b}.
Figure \ref{fig:methaneSigmaHole} shows the differences in \sghs ~for methane
and halogenated methanes. The reason for these differences is the increasing 
polarizability and decreasing electronegativity when going from lighter 
to heavier halogens. 

A fluorine atom has the lowest polarizability and the highest 
electronegativity, which earlier led to the incorrect hypothesis that 
fluorine does not possess a positive \sgh ~and thus is unable to create
halogen bonds \cite{Clark07}. However, calculations later proved 
the opposite \cite{Lu07, Metrangolo11a}, and several experiments eventually 
provided evidence of halogen bonds involving fluorine \cite{Legon99}
(for review, see Mentrangolo \etal \cite{Metrangolo11b}). According to 
a recent report \cite{Eskandari15}, there exist differences in the properties 
of the electron density of fluorine and other halogens. The authors distinguish
between a fluorine bond and a traditional halogen bond of Cl, Br, and I. At the
same time, however, they admit that some fluorine interactions (e.g., 
the F$_2\ldots$NH$_3$ interaction) are very close to the traditional halogen 
bond, which reflects somewhat negatively on their classification.

The positive \sgh ~of fluorine has been localized on molecules including
F$_2$, ClF, BrF$_3$, and FCN. Apparently, the molecular fragments bound 
to fluorine have a common feature -- they are all strongly electron-withdrawing,
which seems to be a prerequisite for fluorine to possess a positive \sgh.
It is also the reason that fluorine in organic molecules (when bound to 
a carbon atom) typically has negative \sghs ~and is unlikely to 
participate in XB.

It was soon recognized that \sgh ~characteristics are sensitive to the 
chemical environment. Neighboring electron-withdrawing chemical groups 
(e.g., --F, --CN) usually make the \sgh ~more positive and bigger 
(compare ClCH$_3$ and ClFCH$_2$ in Table 1), whereas electron-donating 
groups (e.g., --CH$_3$) make the \sgh ~less positive and 
smaller \cite{Riley08, Riley11, Kolar14a}. This is especially pronounced 
on halogens bound to an aromatic ring. Furthermore, Riley \etal ~emphasized
that the \sgh ~on iodobenzene has the same magnitude as the one on 
bromo-2,6-difluorobenzene \cite{Riley11}. According to them, in an application
in which halogen bonding of iodobenzene might not be feasible (e.g., for steric
reasons), bromo-2,6-difluorobenzene could be used instead. Similarly, it has 
been shown that the effect of two methyl groups in meta-positions can be 
compensated for by a fluorine in a para-position with respect to a halogen 
on a benzene ring \cite{Kolar14a}. This means that the \sgh ~magnitudes 
on halobenzenes are very close to the magnitudes on 
3,5-dimethyl-4-fluorohalobenzenes. Interestingly, the effect of aromatic-ring
substitution on the magnitude of a halogen \sgh ~is comparable to 
the effect on the positive ESP of a hydrogen atom in non-halogenated aromatic
analogs \cite{Kolar14a}. 

The \sgh ~magnitude and size correlate well in both 
symmetric \cite{Kolar14a} and asymmetric organic molecules \cite{Kolar14c}.
An interaction energy decomposition affirmed that halogens with higher 
\sgh ~magnitude have lower local polarizabilities \cite{Riley13b}. 
An intriguing feature of halogen \sghs ~is that they are often 
surrounded by a region of negative ESP, which assigns them both electrophilic
and nucleophilic character. The Protein Data Bank (PDB) contains a number 
of instances in which a halogen participates in a halogen bond and 
a hydrogen bond at the same time, and the two interactions form an angle 
of approximately 90\degree \cite{Voth09}.

By tuning the size of \sgh ~with neighboring chemical substitutions, 
it is possible to modify the ratio of the two characters. For instance, when
going from bromobenzene to para-cyano-bromobenzene, the size of \sgh ~
doubles \cite{Kolar14a}, and it can be deduced that the electrophilic character
of the bromine increases at the expense of the nucleophilic one. This contrasts
with the same modification of a hydrogen analogue, i.e., benzene vs. cyanobenzene.
While the effect on the magnitudes is comparable for halo- and hydrogen-analogues,
the effect on the sizes is dramatically different due to the missing negative 
belt on the hydrogen.

It must be stressed that the presence of a \sgh ~is not induced by any 
interacting partner; it is a sole molecular property. However, 
the characteristics of \sghs ~are indeed sensitive to their electric 
environment. Hennemann \etal ~demonstrated a notable polarization effect on 
\sghs ~of water \cite{Hennemann12}. The polarizability in the direction 
of R--H bonds \cite{Fiedler11}, and similarly of R--X bonds, seems to be rather
simple, which turns into their strong response to an external electric field 
caused either by a test point charge \cite{Hennemann12} or by other molecules. 
For instance, the polarization of the \sgh ~may explain the XB of CH$_3$Cl,
which carries a negative \sgh ~in an isolated state, but a positive one
induced by an interacting partner \cite{Clark13, Clark15}.

\subsubsection{Calculations of $\sigma$-Holes}
\label{sssec:sigmaHoleCalculations}

The ESP is normally calculated from the electron density derived at the 
\ai or density functional theory (DFT) level. Previous studies reflect 
a large ambiguity in method choice, and the basis on which one should choose 
the proper one remains unclear. Here, we provide a comparison of computational
methods and basis sets that focuses on \sgh ~properties, partially 
because we were unable to find a similar comparison in the literature and 
because we feel this comparison complements our discussion of \sghs ~
in this review.

Table \ref{tab:methods} summarizes the magnitudes of \sghs ~of selected
molecules calculated by several QM and DFT methods of variable complexity 
using the def2-QZVP basis set \cite{Weigend05}. The choice of methods was 
more-or-less arbitrary, and we admit that there certainly exist other DFT 
functionals or QM methods that would give comparable or even 
better results. A modest 
criterion here was the popularity and availability of the particular method.
The column order follows the computational demands, with Hartree-Fock being 
the fastest and MP2 the slowest. For selected molecules, even more demanding
quadratic configuration interaction with single and double excitations 
(QCISD) \cite{Pople87} values were also calculated, and these might be 
considered as the reference values. The calculations were carried out in 
the Gaussian09 program package, revision D.01 \cite{Frisch09}, with the 
help of the EMSL Basis Set Exchange database \cite{Schuchardt07}.

\begin{table}[tb]
\footnotesize
\begin{tabular}{l c c c c c c}
\hline
\hline
           & HF            & PBE1PBE       & B3LYP         & CAM-B3LYP     & MP2           & QCISD \\
\hline
Cl$_2$     & 0.0433 (1.00) & 0.0403 (1.00) & 0.0395 (1.00) & 0.0398 (1.00) & 0.0401 (1.00) & 0.0395 (1.00) \\
Br$_2$     & 0.0478 (1.10) & 0.0456 (1.13) & 0.0448 (1.13) & 0.0450 (1.13) & 0.0446 (1.11) & 0.0445 (1.13) \\
I$_2$      & 0.0521 (1.20) & 0.0486 (1.21) & 0.0475 (1.20) & 0.0486 (1.22) & 0.0468 (1.17) & 0.0482 (1.22) \\
H$_2$      & 0.0190 (0.44) & 0.0177 (0.44) & 0.0171 (0.43) & 0.0174 (0.44) & 0.0182 (0.45) & 0.0178 (0.45) \\
Cl--Ph     & 0.0079 (1.00) & 0.0078 (1.00) & 0.0074 (1.00) & 0.0073 (1.00) & 0.0078 (1.00) & --- \\
Br--Ph     & 0.0158 (2.00) & 0.0155 (1.99) & 0.0152 (2.05) & 0.0152 (2.08) & 0.0143 (1.83) & --- \\
I--Ph      & 0.0290 (3.67) & 0.0267 (3.42) & 0.0265 (3.58) & 0.0269 (3.68) & 0.0236 (3.03) & --- \\
H--Ph      & 0.0253 (3.20) & 0.0238 (3.05) & 0.0222 (3.00) & 0.0230 (3.15) & 0.0234 (3.00) & --- \\
ClFCH$_2$  & 0.0114 (1.00) & 0.0105 (1.00) & 0.0106 (1.00) & 0.0108 (1.00) & 0.0105 (1.00) & 0.0111 (1.00) \\
BrFCH$_2$  & 0.0181 (1.59) & 0.0175 (1.67) & 0.0177 (1.67) & 0.0181 (1.68) & 0.0167 (1.59) & 0.0179 (1.61) \\
IFCH$_2$   & 0.0317 (2.78) & 0.0281 (2.68) & 0.0288 (2.72) & 0.0298 (2.76) & 0.0263 (2.50) & 0.0290 (2.61) \\
FCH$_3$    & 0.0312 (2.74) & 0.0285 (2.71) & 0.0276 (2.60) & 0.0290 (2.69) & 0.0291 (2.77) & 0.0292 (2.63) \\
\hline
\hline
\end{tabular}
\caption{The magnitudes of \sgh ~(in atomic units) calculated at various
\ai non-empirical QM and DFT levels. The def2-QZVP basis set with 
the effective core potentials on iodine was used systematically. For 
hydrogenated analogues, the magnitude was calculated as the maximum ESP. 
The relative magnitudes normalized with respect to the chlorinated 
analogues are provided in parentheses. Ph stands for phenyl.}
\label{tab:methods}
\end{table}

The variations in the magnitudes are very small. HF gives slightly more 
positive ESP and the second-order Møller-Plesset perturbation theory 
(MP2) \cite{Moller34} more negative values than the QCISD. All three DFT 
functionals are very close to the reference QCISD data. It is worth stressing
that the MP2 values are noticeably less accurate than the triplet of popular 
DFT functionals. Importantly, Table \ref{tab:methods} shows in parentheses 
the relative values normalized with respect to the 
chlorinated analogues. The trend 
that the \sgh ~magnitude increases with increasing atomic number of the 
halogen is well-described, even with the least demanding method. However, 
when compared to QCISD, the MP2 variations of ESP within the halogens are 
contracted. The DFT functionals, as well as HF, perform rather well for 
the series of diatomics, whereas the fluoro-halogen-methanes are best 
described by the DFT. The differences between the functionals are minor. 
Overall, in terms of MEPs, none of the presented methods with the def2-QZVP
basis set \cite{Weigend05} would yield misleading results. This is contrary
to XB interaction energies and geometries as discussed below 
in Section \ref{sssec:interactionEnergies}. 

\begin{figure}[tb]
\includegraphics{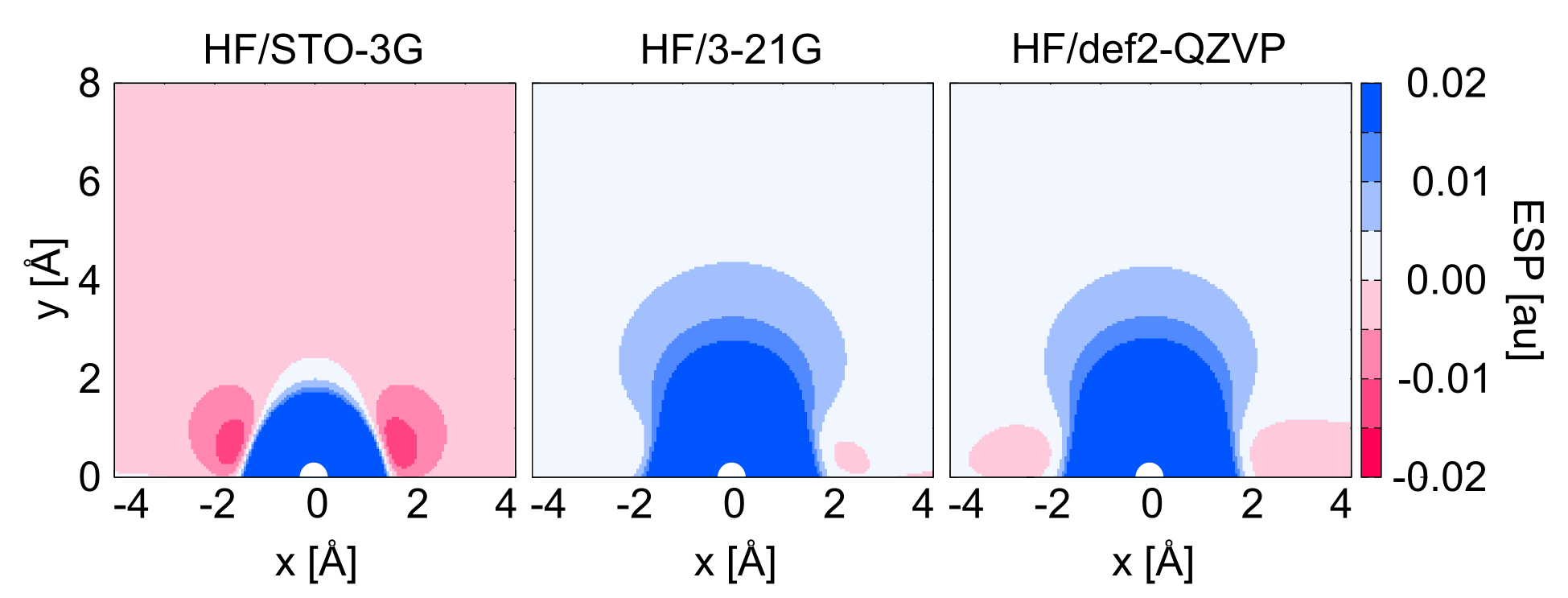}
\caption{The projection of the electrostatic potential (ESP) in atomic 
units of fluorobromomethane to the plane defined by atoms F, C, and Br. The 
position of the bromine is represented by a white semicircle; the C--Br bond 
is parallel to the y-axis and goes to negative values. Three basis sets 
were used with the Hartree-Fock
(HF) method. Note that the scale goes from negative reds, through neutral 
white, to positive blues.}
\label{fig:noSigmaHole}
\end{figure}

It is broadly believed that no \sgh ~can be obtained with a minimal 
basis set. Figure \ref{fig:noSigmaHole} shows that this 
assumption is not entirely correct. The 
ESP maps of CF$_3$Br were calculated at the HF level with the minimal STO-3G, 
small double-zeta 3-21G, and huge def2-QZVP basis sets. The ESP was projected
onto the plane defined by the atoms F, C, and Br. The bromine atom was 
located at [0, 0], and the C--Br bond is parallel to the y-axis. 
The anisotropy of the ESP is somehow captured by the minimal basis set, but 
the absolute values are significantly off. While there is a visible difference
between the ESPs calculated with the minimal (39 basis functions) and 
double-zeta (59 basis functions) basis sets, the difference between 
the double-zeta and quadruple-zeta (303 basis functions) basis sets 
is surprisingly modest. Consequently, the minimal basis set seems to yield 
a \sgh ~that is not very positive, or more likely negative. The use of 
a basis set of double-zeta or better quality is therefore strongly encouraged
to describe \sghs.

The basis set dependence of the \sgh ~magnitudes is given 
in Table \ref{tab:basisSets}.
The CAM-B3LYP functional was used \cite{Yanai04} with a set of systematically 
improving basis sets of def2- series \cite{Weigend05, Rappoport10}. The 
relative values normalized with respect to the chlorine analogues are provided
in Table \ref{tab:basisSets} in parentheses. Again, the magnitudes behave 
as expected; they are least positive for chlorinated compounds and most 
positive for the iodinated ones. Here, however, the relative 
magnitudes are notably basis-set dependent.
A smaller basis set yields larger variations in the magnitudes. 
The triple-zeta basis set with diffuse functions gives practically
identical relative magnitudes as the one of quadruple-zeta quality.

\begin{table}[tb]
\footnotesize
\begin{tabular}{l c c c c}
\hline
\hline
                 & SVP           & SVPD          & TZVP          & TZVPD \\
\hline
Cl$_2$           & 0.0336 (1.00) & 0.0404 (1.00) & 0.0409 (1.00) & 0.0407 (1.00) \\
Br$_2$           & 0.0437 (1.30) & 0.0466 (1.15) & 0.0478 (1.17) & 0.0456 (1.12) \\
I$_2$            & 0.0526 (1.57) & 0.0542 (1.34) & 0.0527 (1.29) & 0.0502 (1.23) \\
H$_2$            & 0.0164 (0.49) & 0.0148 (0.37) & 0.0157 (0.38) & 0.0167 (0.41) \\
Cl--Ph           & 0.0039 (1.00) & 0.0065 (1.00) & 0.0065 (1.00) & 0.0076 (1.00) \\
Br--Ph           & 0.0153 (3.92) & 0.0157 (2.42) & 0.0147 (2.26) & 0.0158 (2.08) \\
I--Ph            & 0.0306 (7.85) & 0.0291 (4.48) & 0.0274 (4.22) & 0.0274 (3.61) \\
H--Ph            & 0.0185 (4.74) & 0.0210 (3.23) & 0.0214 (3.29) & 0.0218 (2.87) \\
ClFCH$_2$        & 0.0045 (1.00) & 0.0127 (1.00) & 0.0096 (1.00) & 0.0118 (1.00) \\
BrFCH$_2$        & 0.0156 (3.47) & 0.0208 (1.64) & 0.0173 (1.80) & 0.0189 (1.60) \\
IFCH$_2$         & 0.0313 (6.96) & 0.0329 (2.59) & 0.0299 (3.11) & 0.0302 (2.56) \\
FCH$_3$          & 0.0256 (5.69) & 0.0275 (2.17) & 0.0274 (2.85) & 0.0278 (2.36) \\
\hline
\hline
\end{tabular}
\caption{The magnitudes of \sgh ~(in atomic units) calculated at 
the CAM-B3LYP level using a systematically improving basis set from the 
def2- series \cite{Weigend05, Rappoport10}. For hydrogenated analogues, 
the magnitude was calculated as the maximum ESP. The relative magnitudes
normalized with respect to the chlorinated analogues are provided 
in parentheses. Ph stands for phenyl.}
\label{tab:basisSets}
\end{table}

\subsubsection{Interaction Energies and $\sigma$-Holes}
\label{sssec:interactionEnergies}

The importance of the \sgh ~for XB lies in the fact that numerous
XB features can be deduced from the \sgh ~properties. One of the 
XB features of paramount importance is its stabilization energy, 
the energy that is released upon formation of a halogen bond or 
is needed to break it. The higher the stabilization energy, the stronger
the XB is said to be. Valerio \etal ~observed a correlation between 
the number of fluorines in the vicinity of the halogen atom and the 
stabilization energy of the halogen-bonded complex \cite{Valerio00}.
At that time, however, such behavior was not related to the ESP.

A number of studies on \sghs ~identified a relationship between
the magnitude of \sgh ~on the halogen donor and the stabilization
energy of the respective 
XB \cite{Lu07b, Riley08, Shields10, Riley11, Politzer13b}. It was 
claimed that the more positive the \sgh, the stronger interaction
it creates. It was later shown that \sgh ~donors other than 
halogens can be included into the regression analysis without 
disturbing the correlation, given a single electron donor \cite{Politzer12}.

The strength of any kind of noncovalent interaction depends not only on one
of the partners (e.g., the halogen donor in the case of XB) but also on 
the other. The correlation between \sgh ~magnitude and stabilization
energies is considerably worse if a variety of electron donors is included.
Politzer \etal ~have pointed out that a good correlation is obtained between
the stabilization energy and the product of the maximum ESP on the halogen 
and the minimum ESP on the electron donor \cite{Politzer13b}.

\begin{figure}[tb]
\includegraphics{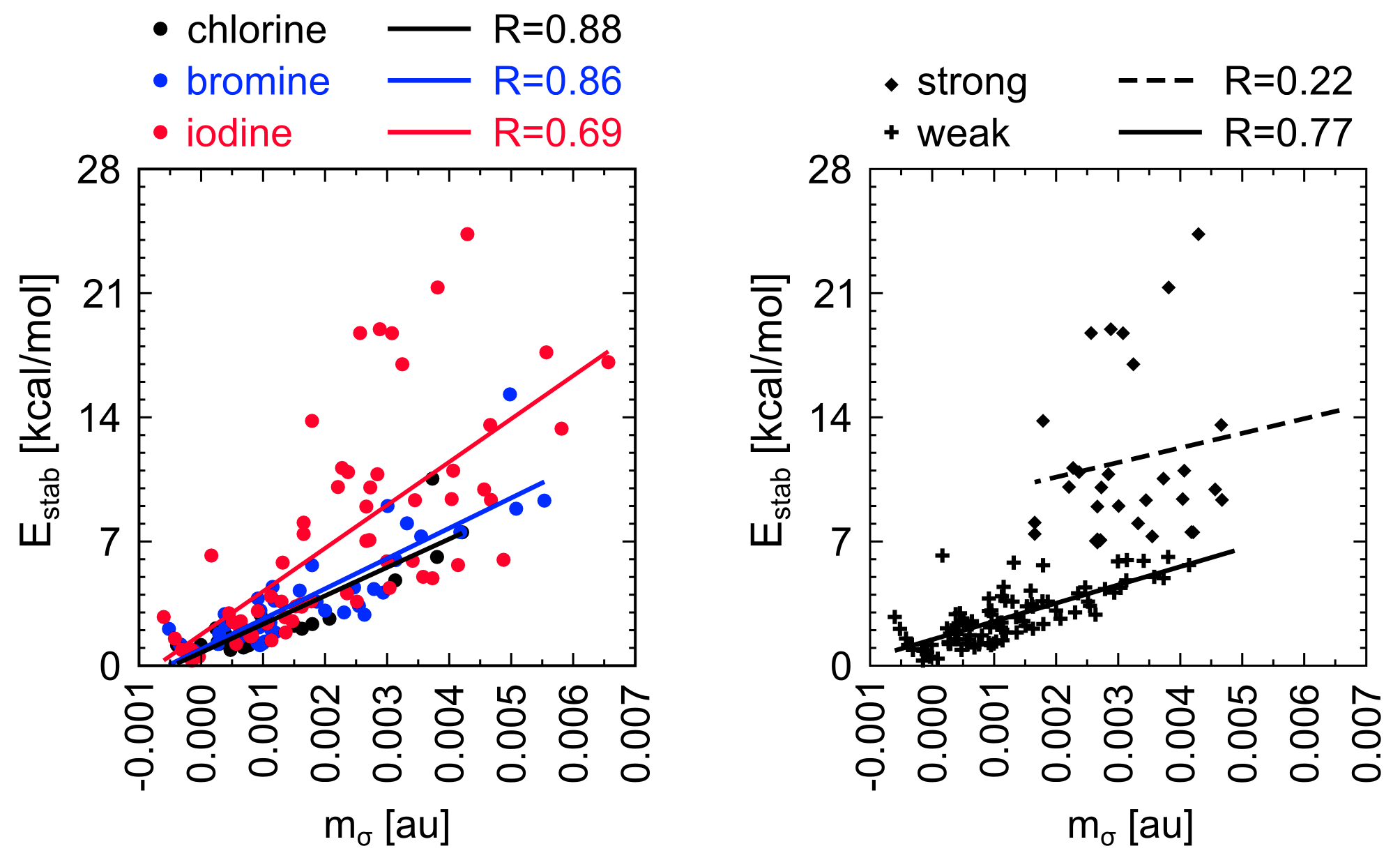}
\caption{The dependence of the stabilization energy of XB complexes 
on the product of maximal ESP on the halogen donor and minimal ESP on 
the electron donor. The linear regression fits are provided with the 
correlation coefficients. The stabilization energy of 7~kcal/mol was 
used for an arbitrary division into classes of strong and weak XB complexes.}
\label{fig:correlationTopics}
\end{figure}

Figure \ref{fig:correlationTopics} summarizes the XB complexes investigated 
in our previous work \cite{Kolar14c}. The stabilization energy calculated 
at the best available level of theory (mostly CCSD(T)/CBS. For details,
see the original work.) is plotted against the product of maximal ESP 
on the halogen donor and minimal ESP on the electron donor, both calculated
at the PBE0/aug-cc-pVDZ level on the 0.001 au electron density isosurface. 
The quality of a regression fit decreases with the atomic number of 
the halogen involved. The correlation coefficient (R) is 0.88 for 
chlorinated compounds, 0.86 for brominated, and 0.69 for iodinated. 
The arbitrary division of the set according to the strength of the XB 
revealed that the stabilization energy of weaker complexes tends 
to correlate better with the product than that of stronger complexes. 
Two factors may contribute to such behavior. First, the quality of the 
stabilization energy, which was calculated at various levels due to 
the variable size of the complexes, and second, the charge-transfer 
character of the strongest complexes, mostly involving iodine atoms, 
which diverges from the presumed electrostatic origin of XB.

\subsection{Interaction Energy Decompositions}
\label{ssec:decompositions}

Today, determination of accurate stabilization energies for \sgh ~
bonded complexes is a routine procedure, and techniques like CCSD(T) at 
the complete basis set (CBS) limit provide highly accurate stabilization 
energies for complexes with up to about 50 atoms. The total stabilization 
energy does not tell us, however, the leading stabilization energy term. 
In other words, it does not tell us which energy component (if any) is 
dominant. This information may be important not only for understanding 
the nature of this bonding but also for optimizing/maximizing the strength 
of the respective \sgh ~bonding. Such classification of noncovalent 
interactions according to their nature may be practical, for example,
in the area of biological sciences.

The stabilization of an R--X$\ldots$Y halogen bond, where X is Cl, Br, or I;
R is an arbitrary chemical group (mostly carbon atom); and Y is an electron donor
(O, N, S, $\pi$-electrons, etc.), is explained by the existence of a positive
\sgh ~on top of the halogen atom.

Following the reliable CCSD(T)/CBS calculations, the stabilization energy 
of the most stable complex with a halogen bond from Table \ref{tab:saptXb}
(iodobenzene$\ldots$trimethylamine) amounts to 5.8~kcal/mol, and this 
stabilization energy is comparable to stabilization of strong hydrogen 
bonds. Much larger stabilization energies were, however, found (at the same 
theoretical level) for complexes of small halogen donors, including 17.1 
and 15.3~kcal/mol for FI$\ldots$NH$_3$ and FBr$\ldots$NH$_3$, 
respectively \cite{Kozuch13}. Similarly large stabilization energies (8.0 
and 15.0~kcal/mol) were found (again at the same theoretical level) for 
crystals of large organic molecules with diiodine, 
1,3-dithiole-2-thione-4-carboxyclic acid$\ldots$I$_2$ and DABCO$\ldots$I$_2$
complexes \cite{Deepa14}. Where do these large stabilization energies come 
from? Is the nature of stabilization in these complexes the same as 
in others?

The attraction in halogen bonded complexes was originally assigned to 
electrostatic attraction between the positive \sgh ~and a negative 
electron donor, and the recent IUPAC definition of a halogen 
bond \cite{Desiraju13} is in accord with this view. However, we recently 
highlighted the important role of dispersion 
interaction \cite{Riley08b, Riley13}, which is the attraction caused by
the interaction of temporary electric multipole with an induced electric 
multipole.

This is understandable in the light of the fact that in any halogen bond,
two atoms with high polarizability (halogen and electron donor) are 
located close together
(closer than the sum of their van de Waals radii). This somewhat contradicts
the previously mentioned definition of a halogen bond, which states,
\emph{``The forces involved in the formation of the halogen bond are 
primarily electrostatic, but polarization, charge transfer, and dispersion
contributions all play an important role.''} To explain this contradiction,
we investigated in detail the nature of different halogen bonds, from 
the weakest ones with stabilization energy around a few~kcal/mol to 
the strongest ones with stabilization energy higher than tens of~kcal/mol.
We deduced the nature of the halogen bond on the basis of dominant energy 
terms; this means that the total stabilization energy in halogen bonded 
complexes should be decomposed to single energy components.

Several computational methods have been suggested to decompose the total 
stabilization energy. Symmetry-adapted perturbation theory (SAPT) provides
a well-defined decomposition of total interaction energies into various components.
With respect to halogen bonding, a number of other methods have been used 
to specify the nature of the bonding. After SAPT, we will discuss a few 
others, among which the quantum theory of atoms in molecules (QTAIM) is perhaps 
the most popular.

\subsubsection{Symmetry-Adapted Perturbation Theory}
\label{sssec:sapt}

Several energy decomposition schemes have been developed. We prefer 
SAPT \cite{Jeziorski94}, which provides a well-defined procedure for decomposing 
the total interaction energies into components of the first and second 
perturbation order. The first- and second-order perturbations
composed of systematically repulsive 
exchange-repulsion energy, systematically attractive induction and 
dispersion terms and polarization/electrostatic energy. The latter can be 
attractive or repulsive, depending on the orientation of subsystems.

The SAPT decomposition does not recognize the electron donor -- electron 
acceptor energy (referred to as charge-transfer), which is significant only when a good 
electron donor interacts with a good electron acceptor. This type 
of interaction occurs in halogen bonded complexes indeed, especially if the heaviest
halogen (iodine) takes part. In the SAPT expansion, the charge-transfer energy
is mostly covered by the induction energy. It reflects the fact that charge-transfer
is only a model for physical event (electric polarization) as reported by Stone and 
Misquita \cite{Stone09}. Thus it is not easy to separate 
the classical induction energy (multipole -- induced multipole) from 
the charge-transfer energy. The only way to describe this systematically 
attractive energy term at least qualitatively is to use an alternative decomposition
scheme, such as Natural Bond Orbital (NBO) analysis \cite{Reed88}. Within this 
scheme, the E$^{(2)}$ charge-transfer energy depends on the energy difference 
between localized orbitals of electron donor and acceptor as well as 
on the overlap of these orbitals (thus taking the geometry of the complex
into account). The problem is that this energy is of different scale and 
cannot be directly compared with the SAPT energy terms. For a 
relative comparison, such as for comparison of complexes with a common electron
donor and a series of halogen donors, the E$^{(2)}$ energy provides 
a reliable relative charge-transfer estimate.

The use of the original SAPT method is limited to rather
small complexes, and this was the reason for introduction of the DFT-SAPT 
method \cite{Hesselmann02a, Hesselmann02b, Misquita02, Jansen01, 
Hesselmann03, Hesselmann05, Hesselmann06, Podeszwa06, Williams01}, which 
allows consideration of complexes with up to about 40 atoms. 
The intramolecular treatment is conducted using the DFT, and therefore 
suffers from inaccurate energies of the virtual orbitals. This drawback
should be corrected by performing a gradient-controlled shift procedure
(see Ref. \cite{Jansen01}).

The total interaction energy ($\Delta E$) is decomposed (cf, eq. \ref{eq:sapt})
into first-order polarization/electrostatic ($E_{Pol}$) and 
exchange-repulsion ($E_{ER}$) terms, and second-order induction ($E_I$), 
dispersion ($E_D$), exchange-induction ($E_{Ex-I}$), and 
exchange-dispersion ($E_{Ex-D}$) terms. Finally, the $\delta HF$ term, which 
accounts for higher than second-order terms covered by the Hartree-Fock 
approach, is added.

\begin{equation}
\Delta E =   E_{Pol}  + E_{ER} + E_{I} + E_{D} + E_{Ex-I} 
+ E_{Ex-D} + \delta HF
\label{eq:sapt}
\end{equation}

The exchange-induction and exchange-dispersion energies are often presented 
as sums with the parent induction and dispersion energies, and finally, 
the $\delta HF$ term is summed up with induction energy. From a brief 
inspection of eq. \ref{eq:sapt}, we find that the charge transfer energy,
which is the energy stabilizing complexes of electron donor (small 
ionization potential) and electron acceptor (small electron affinity),
is not explicitly accounted for. It is implicitly covered by 
the second-order induction energy and higher-order energies covered in 
the $\delta HF$ term. The total induction energy thus contains not only
the classical induction energy term (permanent multipole - induced 
multipole) but also the charge transfer (electron donor - electron
acceptor) energy.

An important advantage of the original SAPT method was the fact that 
the method was basis set superposition error (BSSE)-free. However, due
to inclusion of the $\delta HF$ term calculated from the HF interaction
energy (where the BSSE was considered), the resulting DFT-SAPT 
interaction energy is not BSSE-free.

\begin{table}[tb]
\footnotesize
\begin{tabular}{l c c c c c c c}
\hline
\hline
Complex            &  X-bond      &  $\Delta E$ & $E_{Pol}$ & $E_{ER}$ &  $E_I$ & $E_D$  & X$\ldots$Y/$\Delta$r \\
\hline
H$_3$CCl$\ldots$OCH$_2$           &  Cl$\ldots$O    &   --1.07   &   --1.11  &  2.36  &   --0.38  &   --1.94  &  3.30/0.03 \\
Cl$_2\ldots$Cl$_2$                &  Cl$\ldots$Cl   &   --1.20   &   --1.27  &  3.02  &   --0.63  &   --2.32  &  3.30/0.20 \\
(CH$_2$BrOH)$_2$                  &  Br$\ldots$Br   &   --1.44   &   --0.98  &  1.94  &   --0.31  &   --2.09  &  3.80/0.10 \\
(CH$_2$BrOH)$_2$                  &  Br$\ldots$O    &   --1.56   &   --2.48  &  4.30  &   --0.50  &   --2.88  &  3.10/0.27 \\
H$_3$CBr$\ldots$OCH$_2$           &  Br$\ldots$O    &   --1.79   &   --2.19  &  3.40  &   --0.50  &   --2.49  &  3.17/0.2 \\
C$_6$Cl$_6\ldots$C$_6$Cl$_6{}^a$  &  Cl$\ldots$Cl   &   --2.10   &   --1.20  &  3.57  &   --0.10  &   --4.37  &  --- \\
Br$_2\ldots$Br$_2$                &  Br$\ldots$Br   &   --2.47   &   --2.53  &  4.27  &   --0.76  &   --3.45  &  3.40/0.30 \\
C$_6$Br$_6\ldots$C$_6$Br$_6{}^a$  &  Br$\ldots$Br   &   --2.90   &   --2.30  &  5.80  &   --0.30  &   --6.10  &  --- \\
H$_3$CI$\ldots$OCH$_2$            &  I$\ldots$O     &   --2.63   &   --3.53  &  4.79  &   --0.94  &   --2.96  &  3.21/0.29 \\
I$_2\ldots$I$_2$                  &  I$\ldots$I     &   --3.44   &   --3.61  &  5.58  &   --1.09  &   --4.32  &  3.70/0.26 \\
Br$_2\ldots$C$_6$H$_6$            &  Br$\ldots\pi$  &   --4.23   &   --5.01  &  9.83  &   --2.87  &   --5.85  &  --- \\
F$_3$CI$\ldots$OCH$_2$            &  I$\ldots$O     &   --4.35   &   --6.43  &  7.38  &   --1.83  &   --3.47  &  3.01/0.49 \\
C$_6$(CH$_3$)$_6\ldots$Br$_2$     &  Br$\ldots\pi$  &   --6.06   &   --5.43  &  9.60  &   --2.07  &   --7.79  &  --- \\
C$_6$H$_5$I$\ldots$NC$_3$H$_9$    &  I$\ldots$N     &   --7.02   &  --12.76  & 15.65  &   --3.42  &   --6.50  &  2.97/0.56 \\
HCN$\ldots$BrF                    &  Br$\ldots$N    &   --8.24   &  --14.76  & 17.22  &   --5.41  &   --5.30  &  2.52/0.88 \\
H$_3$N$\ldots$ClF                 &  Cl$\ldots$N    &   --9.59   &  --28.31  & 42.61  &  --16.49  &   --7.40  &  2.34/0.96 \\
Br$_2\ldots$NC$_5$H$_5$           &  Br$\ldots$N    &  --10.89   &  --19.97  & 24.15  &   --7.60  &   --7.47  &  2.57/0.83 \\
H$_3$N$\ldots$IC$_4$H$_2$NO$_2{}^a$ &  I$\ldots$N   &  --14.25   &  --25.37  & 26.16  &   --8.30  &   --6.74  &  2.69/0.84 \\
I$_2\ldots$NC$_6$H$_{15}{}^a$     &  I$\ldots$N     &  --25.20   &  --46.63  & 57.85  &  --22.21  &  --16.34  &  2.55/0.95 \\
\hline
\hline
\end{tabular}
\caption{Decomposition of DFT-SAPT/aug-cc-pVTZ interaction energies ($\Delta E$) 
of complexes with halogen bonds into electrostatic 
($E_{Pol}$), exchange-repulsion ($E_{ER}$), dispersion ($E_D$),
and induction ($E_{I}$) energy; energies in~kcal/mol. Ph stands for phenyl.
$^a$ aug-cc-pVDZ basis set}
\label{tab:saptXb}
\end{table}

Table \ref{tab:saptXb} shows the DFT-SAPT interaction energies and their 
components for various electroneutral XB complexes where the type of XB 
is depicted. Note that dihalogen bonds of the X$\ldots$X type are included
in addition to standard halogen bond of the X$\ldots$Y type. Throughout 
the study (if not otherwise mentioned), the aug-cc-pVTZ basis set was 
used. For a few of the largest complexes, the smaller aug-cc-pVDZ basis 
set was applied. The energy components evaluated with these two basis sets
are very similar, with the exception of the dispersion term, which is 
underestimated by 10--15\% with the smaller basis set. Complexes are 
ordered by increasing stabilization; the stabilization energy of 
the strongest complexes is substantial, more than 25~kcal/mol. 
A stabilization energy of 7~kcal/mol was considered (arbitrarily) as 
the border between weak and medium/strong halogen bonded complexes. 
The same limit was considered in our previous study \cite{Kolar14c}, 
in which we investigated 129 XB complexes: 38 with stabilization energies 
larger than 7~kcal/mol and 91 with stabilization energies between 0.3 and 
7~kcal/mol. In the following paragraphs, the composition of total interaction
energy and complex properties will be discussed separately for the two 
classes of XB complexes.

We start with the X$\ldots$Y intermolecular distances, which for all XB 
complexes (Table \ref{tab:saptXb}) are shorter than the sum of the respective
van der Waals radii (vdW distance), which amounts to: 3.27 (Cl$\ldots$O), 
3.37 (Br$\ldots$O), 3.50 (I$\ldots$O), 3.30 (Cl$\ldots$N), 
3.40 (Br$\ldots$N), 3.53 (I$\ldots$N), 3.55 (Cl$\ldots$S), 
3.65 (Br$\ldots$S), 3.78 (I$\ldots$S), 2.70 (F$\ldots$F), 
3.50 (Cl$\ldots$Cl), 3.70 (Br$\ldots$Br), and 3.96 (I$\ldots$I) with all 
values in \AA. The shortest distance of 2.34 \AA ~was found for 
H$_3$N$\ldots$ClF complexes, and the contraction of the vdW distance was 
substantial, approximately 1 \AA. For the six strongest complexes, 
the intermolecular distance was shorter than 3 \AA, and the respective 
contraction was larger than 0.5 \AA. For five of these six complexes, 
the contraction was even larger than 0.8 \AA. Large contraction of 
the interatomic distance is, as expected, connected with a very large 
repulsion from the exchange-repulsion term.

By investigating the nature of stabilization in the group of medium/strong 
complexes, we found that the polarization/electrostatic energy is mostly 
dominant. Very large (attractive) polarization energies are connected with 
very large magnitudes of the \sghs ~of the respective halogen donors. 
However, because the value of polarization energy and magnitude of 
the \sgh ~do not correlate, it is clear that other attractive 
electrostatic components (e.g., the dipole -- dipole electrostatic term) 
also contribute. The other attractive energy terms, induction and dispersion,
are large but contribute less than half of the polarization term. 
Surprisingly, in five of six cases, the induction energy was the second 
largest stabilization term. In these cases, the stabilization clearly stems 
from the charge-transfer interaction and not from the standard induction term
of the multipole -- induced multipole type. Diiodine and other dihalogens 
such as dibromine and ClF are very good electron acceptors, which can be 
seen from their lowest unoccupied molecular orbital (LUMO) energies: 
--0.144, --0.140, and --0.127 au, respectively. Some organic compounds 
containing halogens, like IC$_4$H$_2$O$_2$N, also are very good electron
acceptors and possess very low LUMO energy (--0.106 au).

The second group of XB complexes consists of 13 complexes with stabilization
energy between 1.07 and 6.06~kcal/mol (Table \ref{tab:saptXb}). In these 
cases, the XB is systematically longer than those in the first group, and 
is larger than 3 \AA. The contractions of XB distance with regard to the 
van der Waals radii are lower and do not exceed 0.5 \AA; within the weakest
complexes (below 2~kcal/mol), it is less than 0.2 \AA. The dominant 
stabilization term is, surprisingly, the dispersion energy. Only in two of 
13 cases is the polarization term the largest in magnitude. In these two 
cases, the halogen donor molecule contains iodine (iodomethane and 
trifluoroiodomethane), and the magnitude of the \sgh ~is rather high 
(0.022 and 0.049 au). In the remaining 11 complexes, the dominant dispersion
energy is followed by the polarization and induction terms. The induction 
energy is typically lower than 1~kcal/mol.

The contribution of the SAPT-like dispersion energy to 
the stability of halogen-bonded
complexes is surprising, and it is much larger than that of similar 
hydrogen-bonded (HB) complexes. This can be easily explained by the fact that 
in XB two heavy atoms (the halogen and the electron donor) with high 
polarizabilities are in close contact (see intermolecular distances 
in Table \ref{tab:saptXb}), whereas in the case of HB, only the light 
hydrogen and the electron donor are in close contact. In the XB complexes,
the dispersion energy of the contact atom pair contributes as much as 40\%
of the total dispersion energy between halogen donor and electron 
acceptor \cite{Kolar14c}. This significant contribution is related to 
the rather short X$\ldots$electron donor distance. Two mutually consistent
views may explain
this low interatomic distance: i) the attractive interaction between 
the positive \sgh ~on the halogen and the negative electron donor
and ii) the lower exchange-repulsion in the direction of the XB between
the two subsystems, which is due to polar flattening \cite{Kerdawy13}.

\begin{figure}[tb]
\includegraphics{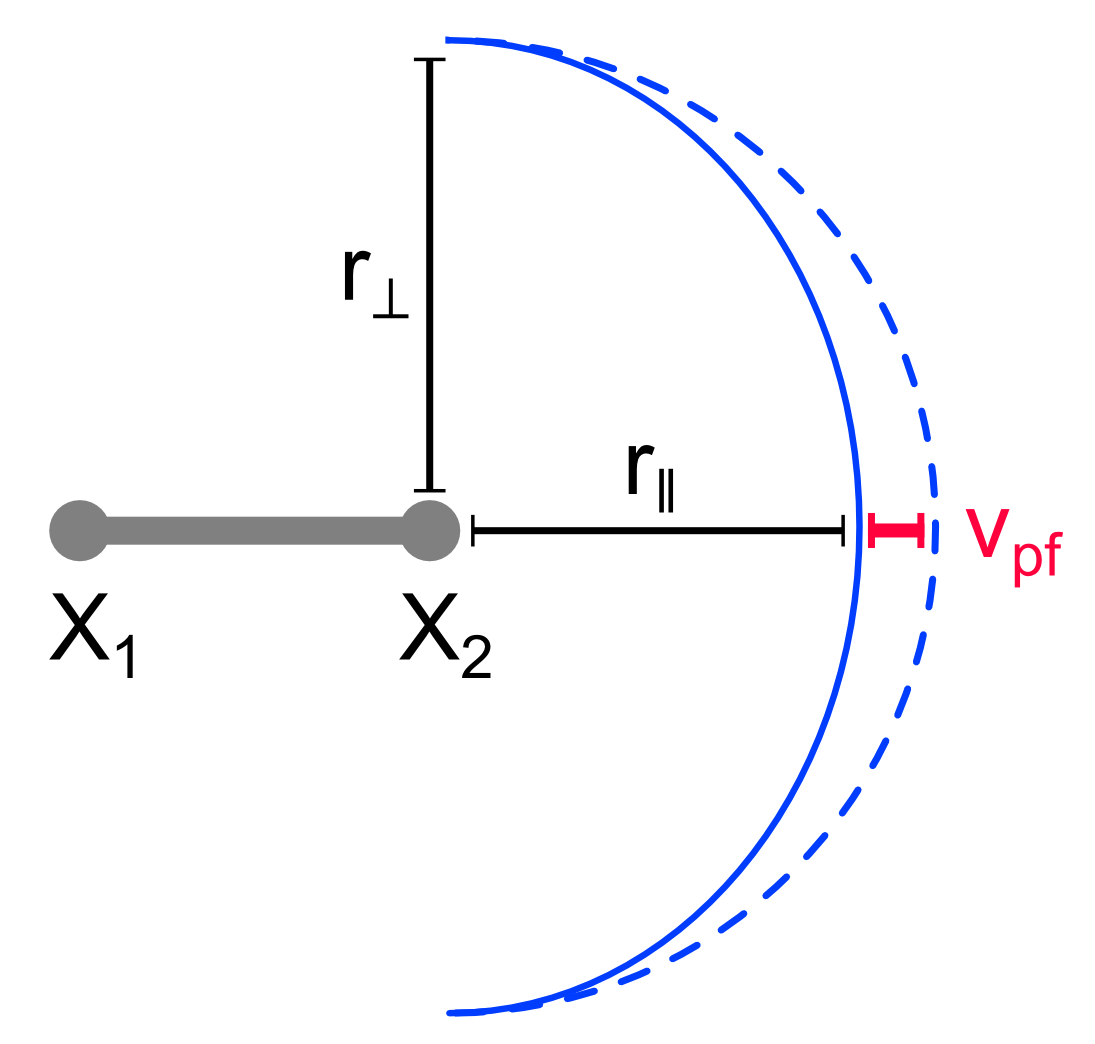}
\caption{A scheme of polar flattening of an X$_1$--X$_2$ dihalogen 
molecule. The electron isodensity surface is not spherically symmetric
(dashed blue line), but flattened in the elongation of the covalent bond 
(solid blue line). The value of such flattening ($v_{pf} = r_{\bot} - 
r_{\parallel}$)
is dependent on the electron density. Figure adapted with permission from 
Ref. \cite{Sedlak15}. Copyright 2015 American Chemical Society.}
\label{fig:polarFlattening}
\end{figure}

The first point is intuitive -- the absence of a positive 
\sgh, the Coulomb repulsion would separate both subsystems and 
the intermolecular distance would thus be larger; due to the r$^{-6}$
dependence, the dispersion energy would decrease even more significantly.
The second point is slightly extraordinary and originates in a non-spherical
electron density around the halogen atom covalently bound to another atom. 
This so-called polar flattening (Figure \ref{fig:polarFlattening}) 
is a consequence of the fact that a halogen atom is covalently bound 
to another atom \cite{Nyburg79}. We should note that the effect of polar 
flattening is not limited to a halogen atom covalently bound to another 
atom, but exists for any atom that is covalently bound to another one. 
This effect is due to a flattening of the spherical electron density of 
an isolated atom in the direction of a covalent bond, and it is most 
pronounced in the case of monovalent atoms.

It can be argued, however, that the polar flattening and \sgh ~are two 
ways of expressing the same -- the electronic situation around a molecule --
in the former case at the level of electron density, whereas at the latter
case at the level of electrostatic potential. Then one must accept that
there is no causality between polar flattening and \sgh.

Recently, we investigated \cite{Sedlak15} the role of polar flattening 
in several model halogen-bonded complexes and concluded that it is not 
negligible and reaches 0.1 - 0.2 \AA ~and 10 - 15\%  for absolute and relative
values of intermolecular distance, respectively. We compared the real situations
with a hypothetical ones, where the halogens were spherically symmetric.
These geometry changes would induce stabilization energy changes that would be
relatively small due to compensation of attractive and repulsive 
energy terms. The shortening of the intermolecular distance caused 
by polar flattening would be responsible for a relative increase 
in stabilization energy up to 15\%.

In summary, the characteristic properties of XB complexes are due to 
the concerted action of polarization/electrostatic, dispersion, and 
induction energy terms in the sense of SAPT decomposition.
While for the strongest XB complexes ($>$ 7~kcal/mol),
the dominant polarization energy is followed by induction (here 
charge-transfer is significant) and dispersion terms, for weak and moderate 
XB complexes (1 -- 7~kcal/mol), the dispersion energy is dominant. Very 
similar conclusions were drawn from the whole set of 129 XB complexes 
investigated previously \cite{Kolar14c}. Thus, we conclude that in our opinion
the IUPAC definition of a halogen bond (\emph{``The forces involved in 
the formation of the halogen bond are primarily electrostatic, but 
polarization, charge transfer, and dispersion contributions all play 
an important role.''}) \cite{Desiraju13} is not sufficiently accurate and does not 
satisfactorily and fully describe the unique phenomenon of 
XB.

In the previous part, we investigated the nature of halogen bonds in 
different complexes with halogen and dihalogen bonds. Next, we extend 
our efforts to complexes with chalcogen and pnicogen bonds. Bleiholder 
\etal ~performed an early SAPT analysis of chalcogen-bonded 
complexes \cite{Bleiholder06} before the terms \emph{``chalcogen bond''}
and \emph{``\sgh ~bond''} entered into use. On a set of complexes 
containing homoatomic contacts of O, S, Se, and Te, they found that in 
terms of SAPT components, the induction and dispersion energies are 
the leading terms in most cases. They noticed that the competition between
chalcogen bonding, hydrogen bonding, and other interactions is less equable 
with increasing atomic number of the chalcogen, i.e., Te-complexes prefer 
chalcogen bonds over hydrogen bonds more strongly than O- or S-complexes.

A SAPT comparison of hydrogen, pnicogen, chalcogen, and halogen bonds in 
four complexes containing NH$_3$ revealed that the polarization and induction 
terms are dominant \cite{Scheiner13}. The role of dispersion was claimed to
be rather low, but not negligible. It must be noted that the stabilization 
energies of these complexes are between 6 and 11.6~kcal/mol (at 
the MP2/aug-cc-pVDZ level), so the importance of polarization and induction 
terms is in line with the DFT-SAPT decompositions of XB complexes.

Here, we continue with an analysis similar to the XB one based on DFT-SAPT 
calculations. Table \ref{tab:saptOther} summarizes the DFT-SAPT energies 
for selected complexes with a chalcogen or pnicogen bond. 
The chalcogen-bonded thioborane dimer shows a very large stabilization 
energy, in which the dispersion contribution is clearly dominant. However,
the electrostatic term is also substantial, and stabilization of this 
structure is largely due to these two attraction terms. The stacking 
structure of the dimer is less stable due to a less attractive electrostatic
term; note that the dispersion energy is almost as important as it is in the
first structure. Similar conclusions confirming a dominant role of dispersion
and, to lesser extent, electrostatic energy were found for complexes of 
neutral closo-heteroboranes possessing chalcogen or pnicogen 
bonds \cite{Pecina15}. All remaining complexes in Table \ref{tab:saptOther} 
exhibit pnicogen bonds, and among them, the benzene$\ldots$SbX$_3$ complexes 
are the most stable. Their exceptional stabilization is due to the concerted
action of electrostatic and dispersion energy. For all other complexes 
with a pnicogen bond, the dispersion energy is clearly dominant.

\begin{table}[tb]
\footnotesize
\begin{tabular}{l l c c c c c}
\hline
\hline
Complex     & type of bond    &   $\Delta E$ & $E_{Pol}$ & $E_{ER}$ & $E_D$ & $E_I$ \\   
\hline
Ph-SB$_{11}$ dimer       & B-S$\ldots\pi$  &  --8.8  &  --5.8  & 10.8  &  --12.0 &  --2.0 \\
Ph-SB$_{11}$ dimer       & Stacking        &  --6.6  &  --2.4  & 9.4   &  --14.8 &  --0.8 \\
C$_6$H$_6\ldots$AsF$_3$  & As$\ldots\pi$   &  --2.0  &  --1.6  & 4.6   &  --4.2  &  --0.8 \\
C$_6$H$_6\ldots$AsCl$_3$ & As$\ldots\pi$   &  --2.9  &  --2.2  & 6.4   &  --6.2  &  --0.9 \\
C$_6$H$_6\ldots$SbF$_3$  & Sb$\ldots\pi$   &  --6.9  &  --5.8  & 5.4   &  --4.4  &  --2.0 \\
C$_6$H$_6\ldots$SbCl$_3$ & Sb$\ldots\pi$   &  --6.9  &  --6.1  & 7.1   &  --5.9  &  --2.0 \\
C$_6$F$_6\ldots$AsF$_3$  & As$\ldots\pi$   &  --2.0  &  --1.6  & 4.6   &  --4.2  &  --0.8 \\
C$_6$F$_6\ldots$AsCl$_3$ & As$\ldots\pi$   &  --2.9  &  --2.2  & 6.4   &  --6.2  &  --0.9 \\
C$_6$F$_6\ldots$SbF$_3$  & Sb$\ldots\pi$   &  --3.1  &  --3.2  & 4.6   &  --3.9  &  --0.6 \\
C$_6$F$_6\ldots$SbCl$_3$ & Sb$\ldots\pi$   &  --3.3  &  --3.2  & 6.0   &  --5.3  &  --0.8 \\
\hline
\hline
\end{tabular}
\caption{Decomposition of DFT-SAPT/aug-cc-pVDZ interaction energies ($\Delta E$) 
of complexes with chalcogen and pnicogen bonds into electrostatic 
($E_{Pol}$), exchange-repulsion ($E_{ER}$), dispersion ($E_D$),
and induction ($E_{I}$) energy; energies in~kcal/mol. Ph stands for phenyl.}
\label{tab:saptOther}
\end{table}

Replacing benzene with another electron donor (e.g., NH$_3$) results 
in even stronger pnicogen bonds. The situation with other pnicogen bonds
might be different. Recently, Setiawan \etal \cite{Setiawan14} published 
an extended study on pnicogen bonds in complexes involving the group Va 
elements N, P, and As. They investigated a set of 36 homo- (E$\ldots$E) 
and heterodimers (E$\ldots$E’), where E stands for a group 15 element, at 
the $\omega$B97X-D/aug-cc-pVTZ level. They showed that a subset of 10 dimers
was mainly stabilized by pnicogen covalent bonding characterized by 
significant charge transfer (CT) from a donor lone pair orbital to the 
acceptor $\sigma^\ast$ orbital (through-space anomeric effect), while all 
the other dimers were stabilized by electrostatic interactions. These 
conclusions were, however, not based on SAPT calculations.

For the pnicogen set, we performed DFT-SAPT calculations similar to those
shown in Tables \ref{tab:saptXb} and \ref{tab:saptOther}. For the 10 dimers 
shown to be stabilized by pnicogen covalent bonding, 
the polarization/electrostatic energy was systematically dominant; 
the induction energy was important for all these complexes, but in all cases
it was comparable to the dispersion energy. The polarization/electrostatic 
energy was dominant in 65\% and dispersion energy in 35\% of the remaining 
complexes. Evidently, for pnicogen complexes, the polarization/electrostatic
energy is mostly dominant, followed by dispersion and induction energies. 
The important role of polarization/electrostatic energy is related to the 
high magnitudes of the \sgh ~for these systems. The H$_3$N$\ldots$P(CN)$_3$
complex had the highest DFT-SAPT stabilization energy (8.5~kcal/mol) of 
all complexes investigated. Its polarization/electrostatic energy 
(--19.9~kcal/mol) was followed by induction and dispersion energies 
(--7.2 and --6.2~kcal/mol, respectively). As mentioned above, the magnitude
of the \sgh ~for P(CN)$_3$ is very high (0.093 au), likely the highest
among all complexes mentioned in this review.

For the above-mentioned small model complexes with pnicogen bonds, 
the polarization/electrostatic energy is dominant. To elucidate the role 
of polarization/electrostatic and dispersion energies, further investigation
of a more extended set of larger complexes with pnicogen bonds is needed. 
From the Cambridge Structural Database, we selected 95 structures 
exhibiting close contacts between pnicogens (As and Sb) and electron donors 
($\pi$-electrons, N, O, S, Se, Te, Cl, I). The stabilization energies of all
of these complexes were determined at the DFT-D3/BLYP/def2-TZVPP level. 
A comparison of the dispersion energy with the total stabilization energy 
can shed light on the relative role of the two contributions.

The first results were encouraging. For 10 As$\ldots\pi$ and 30 Sb$\ldots\pi$ 
complexes, the dispersion energy is systematically larger than 
the electrostatic one. The respective stabilization energies fall within 
a range between 4.9 and 16~kcal/mol. For 9 As$\ldots$X and 46 Sb$\ldots$X
complexes, the dispersion energy is larger in 3 and 17 cases, and 
the respective stabilization energies were within the 3.7 -- 32.8~kcal/mol 
range. Altogether, from 95 complexes with a pnicogen bond, the dispersion 
energy was dominant in 60 cases, while the electrostatic energy was dominant
only in 35 cases. Further systematic studies based on DFT-SAPT calculations 
are required for final estimation of the role of polarization/electrostatic 
and dispersion energies, but the preliminary results show that the dispersion 
energy seems to be dominant in the majority of complexes with pnicogen bonds.

In conclusion, the characteristic properties of halogen, chalcogen, and 
pnicogen bonding are due to the concerted action of polarization/electrostatic
and dispersion energy terms in the SAPT sense. In the case of halogen bonded 
complexes with a heavier halogen, mainly iodine, the induction/charge-transfer
energy is also significant. The structure of \sgh ~complexes is due to specific 
polarization/electrostatic energy while the stability is 
due to non-specific dispersion energy, an important portion of which originates
in the interaction of the atom that carries the \sgh ~and the electron 
donor.

\subsubsection{Other Schemes}
\label{sssec:otherSchemes}

NBO \cite{Reed88} analysis is one way to estimate charge transfer contribution. As 
mentioned above, the values tend to be too high, so this should serve as 
a qualitative approach. For example, it was shown that the charge transfer 
in complexes in which fluorine stands for an electron acceptor is rather low
except for the F$_2\ldots$NH$_3$ complex \cite{Lu07}. This contrasts with 
complexes of bromobenzene with several electron donors, in which the charge 
transfer was found to be significant \cite{Lu07b}. The stabilization of 
various X$_2$ complexes has been studied using 
NBO \cite{Munusamy11, Sedlak14}. X$_2$ molecules are intrinsically 
different from N$_2$ (dinitrogen), which has consequences for their 
binding preferences. 

NBO population analysis was used in the seminal paper of Clark 
\etal \cite{Clark07}, who demonstrated on CF$_3$X (X = F, Cl, Br or I)
that the level of hybridization of the valence-shell s and p orbitals is 
very low (about 10\%) for heavier halogens, but notably higher for 
fluorine. Also, the contribution of the valence p orbital to the bonding 
$\sigma_{C-X}$ orbital is significant, which justifies the intuitive view of 
the valence-shell configuration of s$^2$ p$_x^2$p$_y^2$p$_z^1$ (in which
the z-axis coincides with the C--X bond).  

In the context of noncovalent binding, the quantum theory of atoms and 
molecules \cite{Bader85, Bader87} has become particularly popular
despite some controversial conclusions based on it \cite{Matta03, Poater06}.
QTAIM stems from topological analysis of the electron density $\rho$, its 
gradient $\nabla \rho$, and Laplacian $\Delta \rho$ (sometimes denoted
$\nabla^2 \rho$). The noteworthy feature 
of the electron density is that, in contrast to the 4N-dimensional wave 
function, it is only three-dimensional.

Generally, bond critical points (BCPs) are identified for halogen bonds. 
The electron density at BCPs represents a meaningful measure of the XB 
strength \cite{Lu07b}. Halogen bonding is an interaction between a (3, -1)
type critical point (CP) -- a hole on the electron acceptor -- and a (3, -3) 
CP -- a lump on the electron donor \cite{Eskandari10}. In terms of QTAIM, 
a number of similarities have been found between halogen, dihalogen, and 
hydrogen bonding \cite{Grabowski12}. QTAIM was also used to justify the 
existence of XB in fluorinated compounds \cite{Eskandari15}.

Let us finalize this section by listing several studies of halogen and 
\sgh ~bonding using less common quantum chemical approaches: PIXEL 
scheme \cite{Gavezzotti08}, intermolecular perturbation theory \cite{Allen97},
Kohn-Sham molecular orbital analysis \cite{Palusiak10}, Fukui 
function \cite{Wolters12, Kozuch13}, noncovalent interaction 
plot \cite{Pinter13, Kozuch13}, energy decomposition analysis (EDA), and 
Voronoi deformation density \cite{Wolters12}. All of these bring certain 
insights into halogen bonding, but mutual comparison is not possible due 
to natural differences and nomenclature of the schemes. Some of the methods
are reviewed in more detail in Ref. \cite{Wolters15}.


\section{Computational Methods}
\label{sec:methods}

\subsection{Benchmark Calculations and Performance of Various Computational 
Methods}
\label{ssec:benchmark}

Benchmark stabilization energies of halogen-bonded complexes can be obtained 
similarly to those of other noncovalently bound molecular clusters by 
exploiting the coupled cluster (CC) theory. The CC theory, introduced 
to quantum chemistry by the Czech physicist Jiří Čížek more than 
a half-century ago \cite{Cizek66, Bartlett07}, relies on the exponential 
formulation of the wave operator and its expansion into cluster excitation 
operators. One of the most important features of the CC theory is the fact 
that it is systematically improvable upon inclusion of a higher excitation 
operator \cite{Paldus72}. Accurate energies are, however, obtained only if 
a sufficiently large AO basis set is adopted. Then, the following hierarchy
of accuracy is (mostly) valid:

CCSD $<$ CCSD(T) $<$ CCSDT $<$ CCSDT(Q) $<$ CCSDTQ $< \ldots <$ FCI,

where S, D, T, and Q stand for single, double, triple, and quadruple electron
excitation, respectively. Placing a letter in parentheses indicates that 
a non-iterative perturbative treatment takes place, while in other cases 
fully iterative treatment is adopted. FCI stands for the full configuration
interaction, and the respective energy represents the most accurate value 
that can be obtained within the basis set used \cite{Szabo96}. 
The interaction energy obtained at the FCI level thus represents 
the benchmark value that other methods can only approach. Triple excitations
play an important role in covalent and noncovalent interactions, and thus 
the CCSD(T) scheme represents a widely applicable method. The CCSD(T) method,
characterized by iterative inclusion of the single and double excitations 
with non-iterative, perturbative inclusion of triple (4th order) and
single (5th order) excitations,
is especially successful for ground state energies and the calculation of
properties for systems with single-reference character. 
As demonstrated \cite{Bartlett07}, CCSD(T) is the \emph{``golden standard''}
of CC-approximations due to its outstanding accuracy-to-computational-costs 
ratio.

As mentioned above, accurate results are only obtained if an extended AO 
basis set or a complete basis set (CBS) limit is applied. The CBS limit is 
now commonly determined by an extrapolation technique, and Helgaker 2- 
or 3-point extrapolation \cite{Halkier98} is frequently adopted. The 2-point
extrapolation form contains one parameter obtained from previous calculations.
The 3-point one is parameter-free, and this technique was used 
in determination of stabilization energies in the A24 dataset \cite{Rezac13},
which contains 24 small noncovalent complexes of different types 
(hydrogen-bonded complexes, dispersion-bound complexes, etc.).

Interaction energies in this dataset were determined directly as 
the difference between total CCSD(T) energy of the complex and the sum 
of subsystem energies. Three systematically improved AO basis sets were used:
aug-cc-pVTZ, aug-cc-pVQZ, and aug-cc-pV5Z. Augmented basis sets were used, 
and the smallest basis set was of triple-zeta quality. The function 
counterpoise procedure of Boys and Bernardi \cite{Boys70} was used throughout
the study. The obtained energies represent a highly accurate benchmark that 
serves for testing and/or parameterization of lower-level methods and techniques
that allow treatment of more extended noncovalently bound complexes. The use 
of the CCSD(T) with pentuple-zeta basis set is limited to complexes containing
no more than five heavy atoms. For larger complexes, the calculation procedure
should be modified, and instead of direct determination of interaction energy 
from total CCSD(T) energies, a composite 
scheme \cite{Koch98, Hobza02, Sinnokrot02} can be used successfully.

The applicability of the composite scheme (eq. \ref{eq:ccsdt}) is based on 
the different basis set size dependence of the energy components of total 
CCSD(T) energy, i.e., the HF, MP2, and $\Delta$CCSD(T) energies. HF energy,
which can be calculated relatively easily, depends least on the basis set 
size, and an extended basis set (e.g., aug-cc-pVQZ) can be used. MP2 energy
is extrapolated to the CBS limit, and an aug-cc-pVDZ$\rightarrow$aug-cc-pVTZ or, 
preferentially, an aug-cc-pVTZ$\rightarrow$aug-cc-pVQZ scheme can be 
utilized. Finally, 
the computationally most difficult term, the $\Delta$CCSD(T) one, calculated 
as the difference between CCSD(T) and MP2 energies, is determined in the smaller
basis set. The lesser dependence of the $\Delta$CCSD(T) term is due to 
the similar dependence of MP2 and CCSD(T) energies on the size of the basis 
set. The size of the basis set used to compute the $\Delta$CCSD(T) term has 
a large effect on the quality of the final energies, and usually 
the aug-cc-pVDZ or, preferentially, the aug-cc-pVTZ basis set are 
utilized (Equation \ref{eq:ccsdt}).

\begin{equation}
\footnotesize
\begin{aligned}
\Delta E(CCSD(T)/CBS) & \approx \Delta E(HF/aug-cc-pVQZ) + \Delta E(MP2/CBS) +\\
& + \left [ \Delta E(CCSD(T)/aug-cc-pVDZ) - 
\Delta E(MP2/aug-cc-pVDZ) \right ]
\end{aligned}
\label{eq:ccsdt}
\end{equation}

The present scheme allows treatment of much larger complexes (containing up 
to $\sim$40 atoms) than the previous one and still provides highly 
accurate values. The average unsigned error for these two approaches for
the A24 dataset \cite{Rezac13} is only 1.04\%. This excellent accuracy is,
however, the result of error compensation, and other combinations of basis 
sets provide larger errors.

The performance of computational methods has been tested on several data 
sets of halogenated complexes. The importance of going beyond MP2 for 
halogen-bonded complexes was recognized rather early in the work of 
Karpfen \cite{Karpfen00}, who calculated geometries and stabilization 
energies of complexes of dihalogens with NH$_3$ at the CCSD(T) level. MP2
cannot provide benchmark stabilization energies; thus, its use for assessing
the reliability of other \ai and DFT methods cannot be recommended. 
More precisely, the quality of MP2 has been tested, and for some applications
it may provide sufficiently accurate results. These are, however, not of 
a benchmark quality. Consequently, we limit our attention to studies in 
which the CCSD(T), preferentially with a basis set extrapolated to the CBS
limit, was used.

The X40 data set from our laboratory \cite{Rezac12a} focuses on XB and 
non-XB complexes of halogenated molecules and thus acts as a non-biased
set for general purposes. In this regard, the X40 data set is analogous 
to a more general S66 data set \cite{Rezac11a}. The X40 data set has mainly
been used to characterize the reliability of \ai non-empirical QM 
methods, but also for parameterization of lower-level computational methods.
On the other hand, the XB18 and XB51 data sets developed by Kozuch and 
Martin \cite{Kozuch13} contain XB complexes only. In addition to using 
\ai methods, they heavily inspected the reliability of density 
functional schemes, thus nicely complementing our X40. Furthermore, 
the XB18 set has been used to assess XB geometries.

Other \sgh ~interactions have not been investigated as extensively, 
apart from an assessment of DFT functionals by Bauzá \etal \cite{Bauza13b}.

\subsubsection{X40 Data Set}
\label{sssec:x40}

\begin{figure}[tb]
\includegraphics{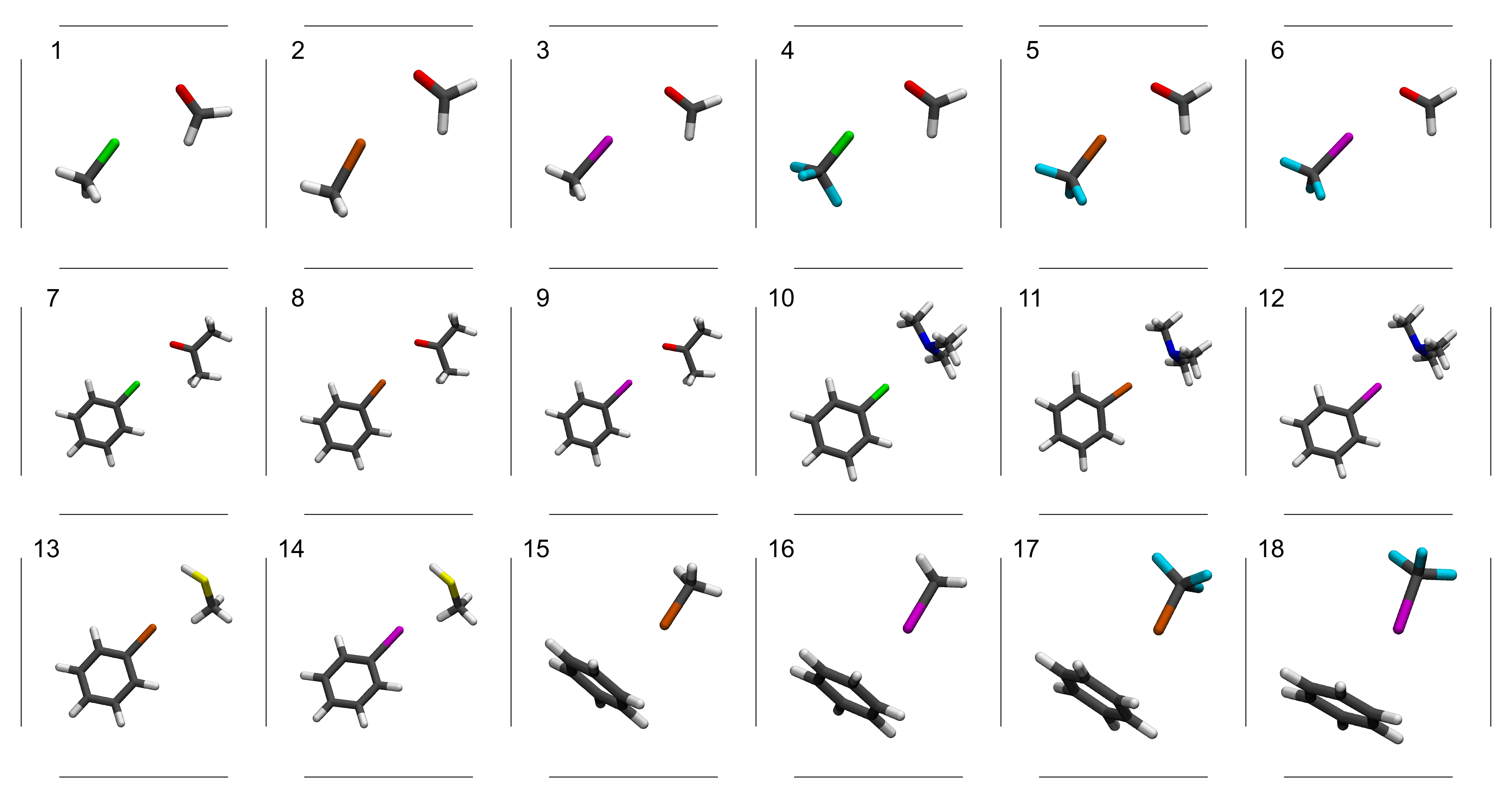}
\caption{18 halogen-bonded complexes from the X40 
set \cite{Rezac12a}. Color code: H-red, N-navy, C-gray, O-red, F-cyan, 
S-yellow, Cl-green, Br-orange, I-purple.}
\label{fig:x40}
\end{figure}

Benchmark 10-point dissociation curves were determined for the representative
set of model halogen-bonded complexes shown in Figure \ref{fig:x40}. They 
represent a subset of a broader series of halogen-containing complexes called 
X40 \cite{Rezac12a}. The complex geometries were optimized at 
the counterpoise-corrected MP2/cc-pVTZ level with the effective core 
potentials on Br and I atoms. As previously stated \cite{Rezac12a}, 
this \emph{``basis set was found to provide the most favorable error 
compensation when used with MP2, and the resulting geometries are in good
agreement with the CCSD(T)/CBS reference.''} \cite{Rezac12c} From 
the optimized geometries, the dissociation curves were constructed 
by multiplying the intermolecular distance by factors between 0.8 and 2.0,
while keeping the mutual orientation of the monomers fixed.

The stabilization energies were determined at the HF and $\Delta$CCSD(T) 
levels with the aug-cc-pVQZ and aug-cc-pVDZ basis sets, respectively, and
the MP2 energy was extrapolated using the aug-cc-pVTZ$\rightarrow$aug-cc-pVQZ scheme.
To test the accuracy of this setup, we calculated interaction energies 
also at the higher theoretical level. Specifically, the HF and MP2 energies 
were calculated at the same level, but the critical $\Delta$CCSD(T) term was 
determined with the aug-cc-pVTZ basis set (for hydrogens, only the cc-pVTZ 
basis set was used). The improvement in accuracy over the previous level 
was negligible, with a root-mean-square difference of 0.06~kcal/mol.

The model complexes (1 -- 18) contain various combinations of halogen 
donors (Cl, Br, I) and acceptor atoms (O, N, S) (Figure \ref{fig:x40}). 
The benchmark interaction energies are given in Table \ref{tab:xb40Complexes}.
The effect of the chemical environment on the strength of a halogen bond is 
demonstrated in the series of complexes with oxygen acceptors (formaldehyde 
and acetone). The weakest interaction is observed in halomethanes, while 
the same halogen bond in aromatic halides (halobenzenes) is stronger. 
In trifluorohalomethane complexes, the withdrawal of electrons from the atom
involved in the halogen bond results in an increase in the magnitude of 
the \sgh ~and a reversal of the dipole moment, which make the interaction
much stronger. The halogen bonds involving nitrogen in complexes 
of halobenzenes with trimethylamine are stronger than the respective 
complexes with oxygen, whereas these with sulfur (e.g., methanethiol) have 
a stability between those of the oxygen and nitrogen analogues. 
The iodobenzene$\ldots$trimethylamine complex is the strongest, with 
stabilization energy approaching 6~kcal/mol. Heavier halogens (Br and I)
interact with the aromatic systems via both dispersion and electrostatic 
interaction between the \sgh ~and the $\pi$ electrons. These interactions 
are moderately strong in complexes of bromo- and iodomethane and benzene 
(15, 16). The electrostatic interaction can be strengthened by enlarging 
the magnitude of the \sgh, for example, by fluorination. This is 
the case for trifluorobromomethane and trifluorochloromethane interacting
with benzene (complexes 17 and 18). 

\begin{table}[tb]
\footnotesize
\begin{tabular}{r l c c}
\hline
\hline
No. & complex                                    & XB pair    &   $\Delta E$ [kcal/mol] \\
\hline
 1  & chloromethane$\ldots$formaldehyde          & Cl$\ldots$O       &     1.17 \\
 2  & bromomethane$\ldots$formaldehyde           & Br$\ldots$O       &     1.72 \\
 3  & iodomethane$\ldots$formaldehyde            &  I$\ldots$O       &     2.38 \\
 4  & trifluorochloromethane$\ldots$formaldehyde & Cl$\ldots$O       &     2.25 \\
 5  & trifluorobromomethane$\ldots$formaldehyde  & Br$\ldots$O       &     3.10 \\
 6  & trifluoroiodomethane$\ldots$formaldehyde   &  I$\ldots$O       &     4.08 \\
 7  & chlorobenzene$\ldots$acetone               & Cl$\ldots$O       &     1.49 \\
 8  & bromobenzene$\ldots$acetone                & Br$\ldots$O       &     2.43 \\
 9  & iodobenzene$\ldots$acetone                 &  I$\ldots$O       &     3.46 \\
 10 & chlorobenzene$\ldots$trimethylamine        & Cl$\ldots$O       &     2.11 \\
 11 & bromobenzene$\ldots$trimethylamine         & Br$\ldots$N       &     3.78 \\
 12 & iodobenzene$\ldots$trimethylamine          &  I$\ldots$N       &     5.81 \\
 13 & bromobenzene$\ldots$methanethiol           & Br$\ldots$S       &     2.32 \\
 14 & iodobenzene$\ldots$methanethiol            &  I$\ldots$S       &     3.08 \\
 15 & bromomethane$\ldots$benzene                & Br$\ldots\pi$     &     1.81 \\
 16 & iodomethane$\ldots$benzene                 &  I$\ldots\pi$     &     2.48 \\
 17 & trifluorobromomethane$\ldots$benzene       & Br$\ldots\pi$     &     3.11 \\
 18 & trifluoroiodomethane$\ldots$benzene        &  I$\ldots\pi$     &     3.91 \\
\hline
\hline
\end{tabular}
\caption{The CCSD(T)/CBS benchmark stabilization energies (in~kcal/mol) 
for halogen-bonded complexes. The interacting pair of atoms is provided.}
\label{tab:xb40Complexes}
\end{table}

The reference dataset is presented to provide accurate values of model 
halogen bonded complexes. The other, even more important goal is to use 
this reference to assess the accuracy of selected post-HF methods for 
investigation of XB complexes. Table \ref{tab:xb40Methods} lists 
root-mean-square errors (RMSEs) and mean unsigned errors (MUEs) for 
the 18 halogen-bonded complexes shown in Table \ref{tab:xb40Complexes} and 
the 22 remaining halogen-containing complexes of the X40 dataset. For 
the sake of comparison with other types of noncovalent interactions, 
RMSEs are also listed for our largest dataset (S66) containing 8-point 
dissociation curves of 66 noncovalent complexes \cite{Rezac11a}. Note that 
the RMSE and MUE were calculated with respect to the stabilization energies
corresponding to the equilibrium interatomic distances.

\begin{table}[tb]
\begin{tabular}{l c c c c}
\hline
\hline
method               &  RMSE   &  RMSE (\%) & MUE  & S66 RMSE \\
\hline
MP2/aug-cc-pVDZ      &  0.55   &  15  & 0.43  &  0.79 \\
MP2/CBS              &  0.70   &  19  & 0.43  &  0.69 \\
SCS-MP2/CBS          &  0.40   &  11  & 0.35  &  0.87 \\
SCS-MI-MP2/CBS       &  0.39   &  10  & 0.26  &  0.38 \\
MP3/CBS              &  0.45   &  12  & 0.32  &  0.62 \\
MP2.5/CBS            &  0.18   &  5   & 0.12  &  0.16 \\
CCSD/CBS             &  0.48   &  13  & 0.40  &  0.70 \\
SCS-CCSD/CBS         &  0.19   &  5   & 0.12  &  0.25 \\
SCS-MI-CCSD/CBS      &  0.06   &  2   & 0.05  &  0.08 \\
BLYP-D3/def2-QZVP    &  0.39   &  10  & 0.29  &  0.25  \\
\hline
\hline
\end{tabular}
\caption{: The root-mean-square error (RMSE; in~kcal/mol and \% of average 
interaction energy) and mean unsigned error (MUE; in~kcal/mol) for 
the 18 halogen-bonded complexes shown in Table \ref{tab:xb40Complexes} and 
the 22 remaining halogen-containing complexes of the X40 dataset. The RMSE 
in the S66 dataset is listed for comparison.}
\label{tab:xb40Methods}
\end{table}

Proper treatment of the ESP anisotropy seems to be a key prerequisite 
for reliable XB calculations. This is, however, not a difficult requirement,
and any HF, post-HF, or DFT method should suffice when a larger-than-minimal
basis set is used. The semiempirical QM (SQM) methods are the exception among
wave-function theories, because they are unable to correctly yield positive 
\sghs ~on halogens. Consequently, the SQM methods fail in describing 
halogen-bonded complexes. Note that this is also the case of molecular 
mechanics, which by definition does not describe \sghs.

There exist a number of less computationally intensive techniques that 
address \sgh ~and yield reasonably accurate results for \sgh ~
complexes. In the following paragraphs, we will briefly comment on the most
important methods. MP2 has been used for many years with widely varying 
results that are strongly dependent on the basis 
set \cite{Kolar07, Riley07, Cybulski07}. With a small basis set, 
the method may give reasonable results, unfortunately due to unpredictable
compensation of errors. When used with large basis sets (or the CBS limit),
MP2 typically produces accurate binding energies for hydrogen-bonded 
complexes and dispersion interactions involving aliphatic species, but 
overestimates binding energies for dispersion interactions involving 
aromatic groups \cite{Riley12}. Several attempts appeared to improve 
the accuracy of MP2. Among these, the spin component scaled variants 
(SCS-MP2) \cite{Grimme03} significantly improved results for noncovalent 
complexes. The SCS-MI-MP2 method of Distasio \etal \cite{Distasio07}, 
which is parameterized specifically for noncovalent interactions, yields
binding energies that are significantly more accurate for a variety 
of interaction types. The SCS technique was also used with a more advanced
method covering a higher portion of correlation energy, the coupled 
clusters with single and double excitations (CCSD). 
The SCS-CCSD \cite{Takatani08} and SCS-MI-CCSD \cite{Pitonak10} methods 
were parameterized and successfully tested. The latter method in particular
provides highly accurate results for diverse noncovalent complexes 
including halogen bonded ones.

Another type of post-MP2 correction makes use of the fact that, while MP2 
tends to overestimate binding energies, MP3 tends to underestimate them. 
The magnitudes of the over/underestimation predicted by the MP2 and MP3 
methods, respectively, are very similar. This is the basis for 
the MP2.5 \cite{Pitonak09} method, which includes an MP2.5 correction, 
the average of binding energies obtained by MP2 and MP3, [$\Delta$E(MP3)
-- $\Delta$E(MP2)]/2. This method provides very good estimates of 
the stabilization energies of different types of noncovalent complexes, 
which closely match the more expensive SCS-MI-CCSD energies. The calculation
of MP3 energy, which very roughly corresponds to one iteration in 
the iterative CCSD calculation, is a critical point. Typically, 
approximately 20 iterations are needed to converge the respective CCSD 
calculation. Finally, for extended noncovalent complexes, for which wave 
function calculations are impractical, the DFT method augmented by 
the dispersion term provides reasonable results.

Of the methods listed in Table \ref{tab:xb40Methods}, the most 
computationally demanding is the CCSD method, but such demands are not 
reflected by the accuracy. On the other hand, the SCS-CCSD and particularly
the SCS-MI-CCSD method significantly improve upon the pure CCSD. 
The SCS-MI-CCSD method provides RMSE and MUE very close to the estimated 
accuracy of the CCSD(T)/CBS benchmark method. MP3 and MP2 methods are 
lower in computational complexity, but neither performs well. However, 
the MP2.5 results are much better: RMSE decreases to 0.18~kcal/mol, 
which is only 5\% of the average interaction energy. Note that the more 
demanding SCS-CCSD method provides results comparable to those of MP2.5.
Better results are only obtained when the parameterized SCS-MI-CCSD/CBS 
method is adopted. An appreciable feature of the MP2.5/CBS, the average
of MP2/CBS and MP3/CBS, is that it is parameter-free. The MP2, following
expectations, shows large errors that are reduced when the SCS technique
is adopted.

In view of the X40 data set, the DFT augmented with dispersion correction
represents a reliable tool for description of noncovalent interactions. 
This finding, however, has not been reflected by other studies. Among many 
options, BLYP in a large basis set in combination with D3 correction of 
Grimme \cite{Grimme11} was assessed. This method was tested with the S66 
dataset and yielded very promising results (see Table \ref{tab:xb40Methods}),
but the error for present halogen-bonded and halogenated complexes was 
somewhat large, although still better than errors obtained with MP3 or CCSD.
It must be noted, however, 
that as we will discuss later, the favorable performance of 
the BLYP-D3/def2-QZVP method is likely due to significantly supportive 
compensation of errors.

Among different computational methods tested for halogen-bonded and 
halogen-containing complexes, the SCS-MI-CCSD and MP2.5 methods exhibit 
the best performance. The use of the latter method is especially attractive 
in light of its considerably better computational economy; MP2.5 is faster 
than SCS-MI-CCSD by more than one order of magnitude. 

Finally, let us comment on the applicability of all these methods 
to extended complexes. Recently, we investigated the performance of 
different WFT and DFT methods to determine 3-body effects \cite{Rezac15}, 
which becomes topical when moving to extended complexes. Let us stress 
that 3-body non-additive effects originate in the induction/polarization 
and dispersion energies; all other terms (such as electrostatic terms) are 
fully additive. The induction non-additivity is covered in the SCF or DFT 
energies, while the dispersion non-additivity is covered only when performing
higher than 2nd order perturbative calculations. The SCS-MI-CCSD and MP2.5 
methods exhibited excellent performance; all other methods, including all 
DFT methods, provided less satisfactory results, indicating that these 
methods should be used for extended complexes with care.

\subsubsection{XB18 and XB51 Data Sets}
\label{sssec:kozuch}

Kozuch and Martin designed two data sets to test the reliability of 
\ai non-empirical QM and DFT methods in reproducing XB complex 
geometries and interaction energies \cite{Kozuch13}. The first set, 
named XB18, contains 18 complexes of halogen diatomics (HBr, HI, Br$_2$,
I$_2$, FBr, FI, ClBr, ClI, BrI) with small electron donors (OCH$_2$ and
NCH), which are small enough to test geometry optimization. They obtained 
reference geometries by gradient optimization at CCSD(T)/aug-cc-pVQZ 
(with the ECPs on Br and I), while the reference interaction energies 
at the CCSD(T)/CBS limit level were extrapolated from quadruple-zeta and 
pentuple-zeta basis sets according to equation \ref{eq:kozuch}.

\begin{equation}
E_{CBS} = E_L + \frac{E_L - E_{L-1}}{\left ( \frac{L}{L-1} \right )^\alpha - 1},
\label{eq:kozuch}
\end{equation}

where $L$ is the cardinal number of the basis set (here $L=5$), 
$\alpha=5$ for singlet coupled pairs and connected triples, and $\alpha=3$
for SCF and CCSD triplet coupled pairs. Note that there is a typo in 
equation 1 of Ref. \cite{Kozuch13}.

In addition to a few \ai non-empirical QM methods, 42 DFT functionals 
have been examined, covering the whole Jacob’s ladder 
of accuracy \cite{Perdew01}. The calculations were performed with 
the aug-cc-pVQZ basis set. Due to the lack of dynamic correlation, the HF 
method failed to reproduce XB geometries (characterized by the XB length). 
Similar to the results with the X40 data set, MP2 tended to notably shorten 
XB lengths, which could be corrected for by separate scaling of same and 
opposite spin components in the SCS-MP2 or SCS-MI-MP2 methods. Among the DFT 
functionals, M06-2X \cite{Zhao08}, BMK \cite{Boese04}, and 
CAM-B3LYP \cite{Yanai04} yielded accurate XB lengths.

Kozuch and Martin also compiled a broader set of 51 complexes called XB51 
to evaluate the accuracy of interaction energies (Table \ref{tab:kozuch}). 
For some reason, the researchers did not include chlorinated XB donors, and 
selected only two XB acceptors containing oxygen in the XB51 set.
The low number of oxygen
acceptors is a drawback, because oxygen is the most abundant
 XB acceptor in biomolecular XB complexes. On the other hand, they paid
special attention to covering a wide range of stabilization energies. 
The geometries of XB51 complexes were optimized at the $\omega$B97X/aug-cc-pVTZ 
level \cite{Chai08}, while the benchmark stabilization energies were 
calculated at a composite CCSD(T)/CBS(MP2) level (Equation \ref{eq:kozuch2}).

\begin{equation}
E_{CBS(MP2)} = E_{MP2}(CBS) + E_{CCSD(T)}(aVTZ) - E_{MP2}(aVTZ),
\label{eq:kozuch2}
\end{equation}

where the MP2/CBS term was extrapolated from the aug-cc-pVQZ and aug-cc-pVa5Z
basis sets according to equation \ref{eq:kozuch}. The benchmark stabilization 
energies are summarized in Table \ref{tab:kozuch}.

\begin{table}[tb]
\begin{tabular}{l c c c c l c c c}
\hline
\hline
      & \multicolumn{3}{l}{X acceptor} & &  &  \multicolumn{3}{l}{X donor} \\
\hline
X donor & PCH   & NCH   & NH$_3$  & ~ & X acceptor  & MeI   & Br$_2$& FI \\
\hline
PhBr    & 0.85  & 1.15  & 2.02    & ~ & FCCH        & 0.50  & 0.74  & 0.29 \\
MeI     & 0.85  & 1.42  & 2.73    & ~ & PCH         & 0.85  & 1.18  & 2.74 \\
PhI     & 0.92  & 1.87  & 3.33    & ~ & NCH         & 1.42  & 2.87  & 5.97 \\
F$_3$CI & 0.89  & 3.61  & 5.88    & ~ & FMe         & 1.70  & 3.61  & 9.33 \\
Br$_2$  & 1.18  & 3.61  & 7.29    & ~ & OCH$_2$     & 2.39  & 4.41  & 9.94 \\
NBS$^a$ & 1.19  & 4.32  & 8.02    & ~ & NH$_3$      & 2.73  & 5.95  & 13.36 \\
FCl     & 1.16  & 4.81  & 10.54   & ~ & OPH$_3$     & 3.34  & 7.29  & 17.11 \\
NIS$^b$ & 1.53  & 5.91  & 10.99   & ~ & Pyr$^c$     & 3.61  & 9.00  & 17.66 \\
FBr     & 2.07  & 7.53  & 15.30   & ~ & HLi         & 3.62  & 9.07  & 20.34 \\
FI      & 2.74  & 9.33  & 17.11   & ~ & PdHP$_2$Cl  & 5.05  & 23.11 & 33.79 \\
\hline
\hline
\end{tabular}
\caption{The benchmark stabilization energies (in~kcal/mol) of the XB51 data 
set at the CCSD(T)/CBS level. $^a$N-bromosuccinimide, $^b$N-iodosuccinimide,
$^c$pyridine}
\label{tab:kozuch}
\end{table}

The wave function methods mostly performed in accordance with the results 
obtained with the X40 data set. HF underestimated the XB stabilization energies
by about 4~kcal/mol on average; MP2 overestimated the stabilization by values 
between 1.24 and 0.70~kcal/mol for aVDZ and CBS(Q5), respectively. However, 
the RMSE for the MP2 method was very close to 1~kcal/mol regardless of 
the basis set. Spin-component-scaled variants performed better; the RMSE 
was 0.44 and 0.33~kcal/mol for SCS-MP2 and SCS(MI)-MP2, respectively.

\begin{figure}[tb]
\includegraphics{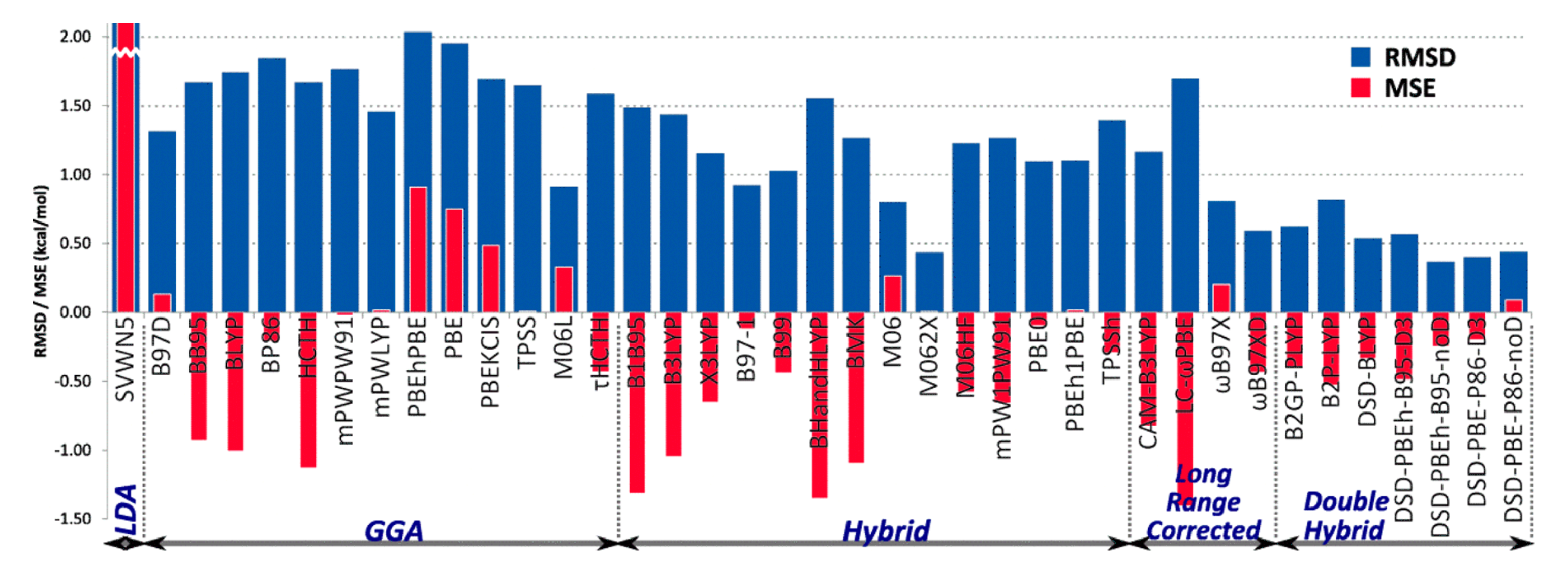}
\caption{The root-mean-square deviation (RMSD) and mean signed error (MSE) 
for XB51 stabilization energies calculated with respect to the CCSD(T)/CBS 
reference. Adapted with permission from Ref. \cite{Kozuch13}.
Copyright 2013 American Chemical Society.}
\label{fig:kozuchMethods}
\end{figure}

Extraordinary attention has been paid to DFT. Figure \ref{fig:kozuchMethods}
shows the root-mean-square deviations (RMSD) and mean signed errors (MSE) 
for the stabilization energies. As a clear leader, M06-2X was identified 
with an RMSE of 0.41~kcal/mol. It was noted that the correct description 
of charge-transfer was more important than the description of dispersion 
energy for a reasonable performance of a DFT functional. The amount of exact
exchange was critical, as justified by M06-2X, with 54\% HF exchange. It was
further shown that the inclusion of a Grimme’s-like dispersion correction 
term \cite{Grimme04, Grimme11} did not generally improve the performance, 
no matter what kind of damping scheme was used. This is surprising because
the BLYP-D3/def2-QZVP method gave reasonable agreement with reference data
from the X40 set. It was pointed out that the delocalization error, which 
is particularly severe in semilocal functionals, leads to overstabilization
of XB when empirical pairwise correction is used. As recently suggested by
Otero-de-la-Roza \etal \cite{Otero14}, the use of exchange-hole dipole 
moment dispersion correction \cite{Becke05} with a functional with low 
delocalization error may solve the problem.

\subsubsection{Benchmarking Other $\sigma$-Hole Interactions}
\label{sssec:other}

The two above-mentioned benchmark sets for halogen bonding provide sufficient
data for method assessment and development. The situation for other \sgh
~interactions is not as clear-cut. A systematic large-enough CCSD(T)/CBS test 
set is missing, so the best available data can be perhaps found in the work 
of Bauzá \etal \cite{Bauza13b}. They compiled a set of halogen-, chalcogen-,
and pnicogen-bonded complexes and used it to test the quality of \ai 
non-empirical QM methods and DFT functionals.

Both the reference geometries and stabilization energies were determined at 
the CCSD(T)/aug-cc-pVTZ level. The difference from CCSD(T)/CBS is expected to 
be very low. The reference stabilization energies and equilibrium distances 
between the pair of closest atoms are summarized in 
Table \ref{tab:bauza} \cite{Bauza13c}.

\begin{table}[tb]
\begin{tabular}{l l l l c l l l}
\hline
\hline
    & \multicolumn{3}{l}{stabilization energy} & ~  & \multicolumn{3}{l}{equilibrium atom-atom distance} \\
\hline
          & Cl$^-$  & Br$^-$  & NH$_3$ &  ~ & Cl$^-$ &  Br$^-$  & NH$_3$ \\
\hline
FCl       & 28.98   & 26.77   & 9.39   &  ~ & 2.317  &  2.452   & 2.320 \\
FBr       & 35.48   & 32.73   & 13.39  &  ~ & 2.440  &  2.588   & 2.355 \\
SF$_2$    & 21.18   & 17.45   & 6.33   &  ~ & 2.459  &  2.459   & 2.556 \\
SeF$_2$   & 31.07   & 26.68   & 10.65  &  ~ & 2.490  &  2.678   & 2.453 \\
F$_2$CS   & 7.29    & 5.99    & 1.48   &  ~ & 2.490  &  3.165   & 3.300 \\
F$_2$CSe  & 10.73   & 8.90    & 2.38   &  ~ & 2.899  &  3.113   & 3.230 \\
F$_3$PS   & 6.37    & 5.10    & 1.25   &  ~ & 3.046  &  3.274   & 3.297 \\
F$_3$PSe  & 12.14   & 9.93    & 2.46   &  ~ & 2.870  &  3.098   & 3.169 \\
F$_3$P    & 15.09   & 11.68   & 4.11   &  ~ & 2.665  &  2.973   & 2.883 \\
F$_3$As   & 25.76   & 20.86   & 7.57   &  ~ & 2.575  &  2.824   & 2.674 \\
\hline
\hline
\end{tabular}
\caption{The reference stabilization energies (in~kcal/mol) and equilibrium 
geometries (in \AA) of a \sgh ~benchmark set derived at 
the BSSE-corrected CCSD(T)/aug-cc-pVTZ level.}
\label{tab:bauza}
\end{table}

\begin{figure}[tb]
\includegraphics{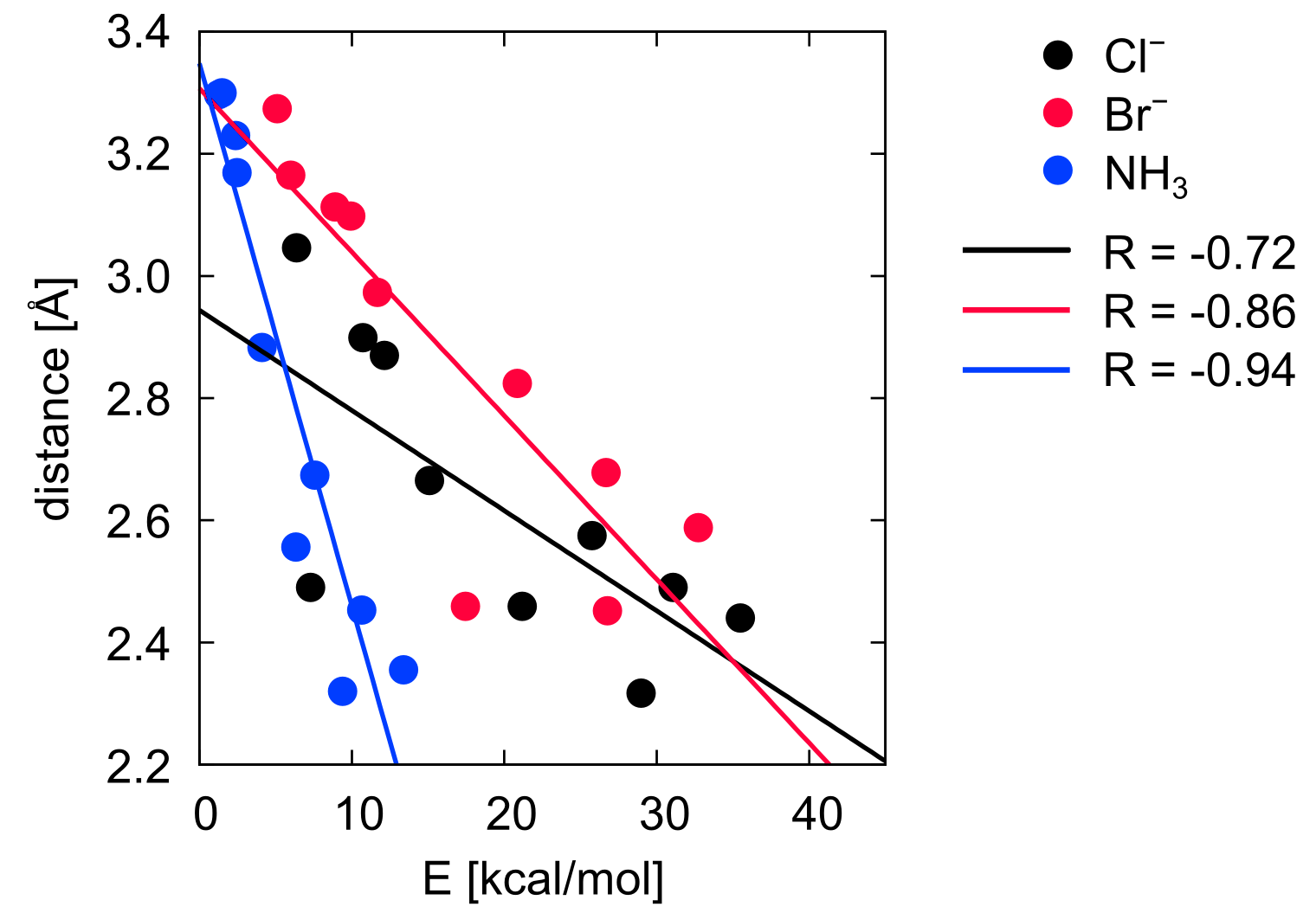}
\caption{Dependence of the stabilization energy on the interatomic distance 
(circles). Linear regression fits are shown as lines together with 
the correlation coefficients R.}
\label{fig:bauzaCorrel}
\end{figure}

Figure \ref{fig:bauzaCorrel} shows the relationship between the stabilization
energies and interatomic distances -- the \sgh-bonding lengths. There 
is no general trend reflected by the overall correlation coefficient 
R of --0.20. When, however data from a single electron donor are considered, 
an inverse correlation (anticorrelation) is apparent. NH$_3$ complexes 
show very good correlation coefficient (R = --0.94),
which is better than those of both charged series (R of --0.72 and --0.86 
for Cl$^-$ and Br$^-$ complexes, respectively).

Apart from the HF method, all the other methods overestimated 
the stabilization energy as shown by mean signed error between 0.74 (RI-MP2)
and 4.96~kcal/mol (BP86-D3). Notably, MP2 and RI-MP2 performed rather 
well (RMSE of 1.18 and 1.23~kcal/mol, respectively). Among the DFT 
functionals, M06-2X was clearly best, with an RMSE of 1.73~kcal/mol, 
followed by B3LYP (RMSE = 2.23~kcal/mol). Similarly to the XB benchmark 
sets, the dispersion-corrected functionals (BLYP-D3, BP86-D3) performed 
poorly (RMSE = 5~kcal/mol), especially when the complex was charged. 
The results of BLYP-D3/aug-cc-pVTZ are in disagreement with the X40 results.
It should be, however, mentioned that the X40 data set contains only neutral
complexes, and, furthermore, that a larger basis set (def2-QZVP) was applied
in that case. Interestingly, the M06-2X functional was rather weak 
in description of pnicogen bonded complexes. However, the test set 
compiled by Bauzá \etal ~is rather imbalanced: it contains 6 halogen-bonded,
18 chalcogen-bonded, and 6 pnicogen-bonded complexes, which could distort 
the final statistical evaluation.

Regarding complex geometries, the authors recommend M06-2X or MP2 at augmented
triple-zeta basis set.

\subsection{Semiempirical Methods}
\label{ssec:sqm}

Semiempirical QM methods (SQM) can be applied to more extended molecular 
systems than QM methods, while still retaining the basic advantage of 
the QM approach, namely the solution of the Schrödinger equation. 
Today's successful SQM methods stem from the neglect of diatomic differential
overlap approximation (NDDO) of the Hartree-Fock 
Hamiltonian \cite{Pople65, Pople67}. In addition to the AM1 \cite{Dewar85}
and PM3 \cite{Stewart89} methods, PM6 has become popular, likely thanks to
the availability of parameters for a large fraction of the periodic 
table \cite{Stewart07}. 

The density-functional tight-binding scheme (DFTB), represented by 
the DFTB2 \cite{Elstner98} and DFTB3 \cite{Gaus11} models, may also 
be considered a semiempirical quantum chemical method. Unlike NDDO-based
methods, DFTB is derived from the Taylor expansion of total energy within
the framework of density functional theory. The parameter sets for halogens
for the second-order expansion (DFTB2) were developed only 
recently \cite{Kubar13} without much focus on noncovalent interactions. 
Halogen parameters for DFTB3, the newer variant based on the Taylor expansion
into the third order, have been provided as well \cite{Kubillus14}. 
The overbinding of halogen bonds in DFTB3 may be corrected for with 
an empirical correction \cite{Kubillus14} that has been parameterized 
and benchmarked using the set of equilibrium and non-equilibrium interaction
energies from the X40 data set \cite{Rezac12a}.

The applicability of SQM methods to noncovalent interactions is generally 
limited, and the main problem concerns description of dispersion-bound 
complexes and, perhaps surprisingly, description of hydrogen-bonded 
complexes \cite{Li14}. SQM methods can be modified to provide a reliable 
description of these interactions, and the easiest way seems to be to 
include some empirical corrections. In our laboratory, we have developed 
several versions of such corrections that were parameterized against accurate
\ai interaction energies. The latest version, called D3H4, provides 
an accurate description of various types of noncovalent interactions,
including hydrogen bonding and dispersion interactions \cite{Rezac12b}, 
and has been utilized in a semiempirical-based computer-aided drug design
scheme \cite{Lepsik13}.

Most of the QM methods describe the \sgh ~satisfactorily, providing 
that a larger-than-minimal AO basis set is used. The SQM methods use 
a specially optimized sub-minimal basis set and therefore cannot properly
describe the anisotropic density on the halogen atom, which contributes 
notably to the distinguishing features of the interaction between halogen 
and electron donor. To our surprise \cite{Rezac11b}, the standard PM6 
method yields interaction energies in various XB complexes close to 
the benchmark values. Later, however, we found that the potential energy
surface of these 
complexes can be completely inaccurate, and the reliable characteristics 
were obtained by accident. The problem is connected to a known failure 
of SQM methods -- underestimated repulsion. Consequently, the nonbonded 
interatomic distances are too short, which leads to greatly overestimated
stabilization energies in noncovalent complexes, particularly in 
XB complexes. In analogy with corrections to HB and dispersion interactions,
we introduced a correction for XB. We developed an empirical correction 
parameterized to small complexes for which accurate QM data were 
available \cite{Rezac11b}, similar to development of HB correction. 
Specifically, we used the MP2/aug-cc-pVDZ dissociation curves for 
halobenzene$\ldots$acetone and halobenzene$\ldots$trimethylamine; this 
method provides interaction energies for halogen bonded complexes 
in close agreement with the benchmark CCSD(T) calculations. The correction
was parameterized for chlorine, bromine and iodine.
The method was tested on a set of complexes in which substitutions on 
an aromatic ring modulate the strength of the halogen bond, and it 
reproduces the reference QM interaction energies with reasonable accuracy.
Furthermore, the method was incorporated into a semiempirical scoring 
function and validated on a series of protein kinase CK2 complexes 
containing polyhalogenated ligands \cite{Dobes11}.

In summary, the latest version of the corrected SQM method, called 
PM6-D3H4X, provides an accurate description of a wide range of noncovalent 
interactions, including dispersion, hydrogen bonding, and XB. An important 
advantage of this technique is that it can be directly applied to other 
types of \sgh ~bonding. Currently, we are working on extension of 
the PM6 method to chalcogen- and pnicogen-bonding. In principle, the D3H4X
corrections could be combined with other SQM methods such as 
OM-x \cite{Weber00}, RM1 \cite{Rocha06}, and SCC-DFTB \cite{Elstner98},
but to the best of our knowledge, such enhancements have been pursued only
for a DFTB variant \cite{Kubillus14}. In our opinion, the use of corrections
with PM6 carries a major advantage—the linear scaling algorithm 
MOZYME \cite{Stewart96}, which can be applied to calculations of several 
thousands of atoms. The PM6-D3H4X version of the original PM6 method is 
now available in the MOPAC program package \cite{Stewart09}.

\subsection{Empirical Force Fields}
\label{ssec:ff}

Motivated by the utility of XB in drug 
design \cite{Lu09, Hernandes10, Wilcken13}, efforts to develop computational
methods even less demanding than SQM have been strengthened. The method 
of choice in computer-aided drug development and bio-oriented computational
research is molecular mechanics (MM) with pairwise effective interaction 
potentials. In a typical MM approach, one treats the molecules as a set 
of structureless particles (atoms) connected with harmonic springs, which 
allows investigation of up to 106 atoms on wide range of time 
scales \cite{Karplus02}.

The molecular mechanical noncovalent interactions are explicitly accounted 
for those atoms that are separated by about three covalent bonds (i.e., 
the springs), depending on the MM scheme. The use of partial atomic charges
representing the electrostatic interactions and Lennard-Jones (LJ) potential
to describe van der Waals interactions is rather general among MM 
variants \cite{Verlet67, Nicolas79} (Equation \ref{eq:ffNonbonded}).

\begin{equation}
E_{NB} = E_{elec} + E_{LJ} = \frac{1}{4 \pi \varepsilon} \frac{Q_1Q_2}{r}
+ \epsilon \left [ \left ( \frac{\sigma}{r} \right ) ^{12}
- \left( \frac{\sigma}{r} \right ) ^6 \right ],  
\label{eq:ffNonbonded}
\end{equation}

where the nonbonded interaction energy $E_{NB}$ is the sum of 
the electrostatic $E_{elec}$ and LJ term $E_{LJ}$, and $r$ stands for 
the interparticle (interatomic) distance, $Q_1$ and $Q_2$ are the partial
atomic charges, $\varepsilon$ is electric 
permittivity, and $\epsilon$ and $\sigma$ are two interatomic pair 
parameters of the LJ potential.

Notably, the atomic charges do not correspond to any physical observable, 
and there is a large ambiguity in their 
derivation \cite{Bessler90, Wiberg93, Sigfridsson98}. On the other hand, 
the concept of partial charges is essential for contemporary molecular 
modeling. Regarding biomolecular simulations and drug design, a few 
techniques, such as Restricted Electrostatic Potential 
(RESP) \cite{Bayly93} and AM1-BCC \cite{Jakalian00, Jakalian02}, have become
widely used in charge derivation. A common drawback of describing 
electrostatics by atomic partial charges is their inability to produce 
locally anisotropic electrostatic potentials. From equation \ref{eq:ffNonbonded},
it is also apparent that the LJ potential energy contribution 
is spherically symmetric.

\begin{figure}[tb]
\includegraphics{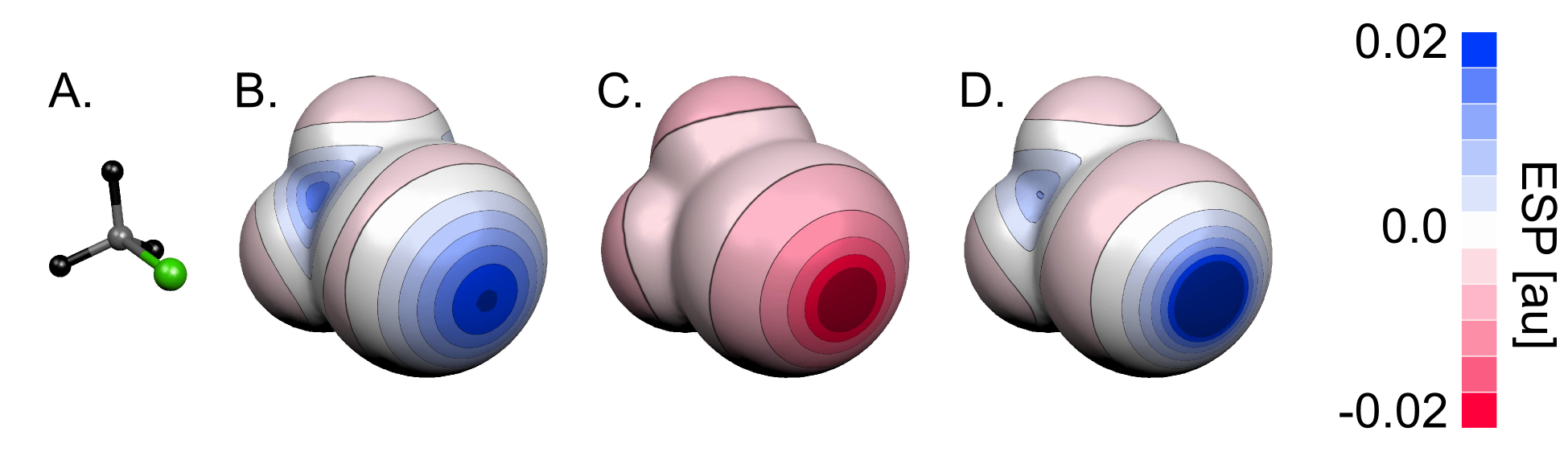}
\caption{A ball-and-stick model of trifluorobromomethane (CF$_3$Br) is 
provided (A). A comparison of its molecular electrostatic potential (ESP)
derived at QM (B), MM (C), and MM+ESH (D) levels is shown and projected on 
the surface of 0.001 au electron density. Note that the coloring goes from 
negative reds, through neutral white, to positive blues, and that it is 
identical for all the three surfaces.}
\label{fig:meps}
\end{figure}

The ESP anisotropy is important \cite{Kramer14}, for example, for description
of electron lone pairs or, more importantly with respect to this review, 
for halogen and other \sgh ~bonds. Figure \ref{fig:meps} shows MEPs of 
trifluorobromomethane calculated at the QM level (B) and using atomic partial charges
(C). The MEPs indicate that no \sgh ~exists on top of the molecular 
mechanical model of bromobenzene and that the ESP is progressively negative.

This shortcoming leads to a complete failure of MM calculations on XB 
complexes \cite{Dobes11}, and this fallacy has become paradigmatic in 
the field of XB \cite{Politzer08}. Below, we present a few routes 
to rectification.

\subsubsection{Methods with Off-Center Point Charges}
\label{sssec:esh}

\begin{figure}[tb]
\includegraphics{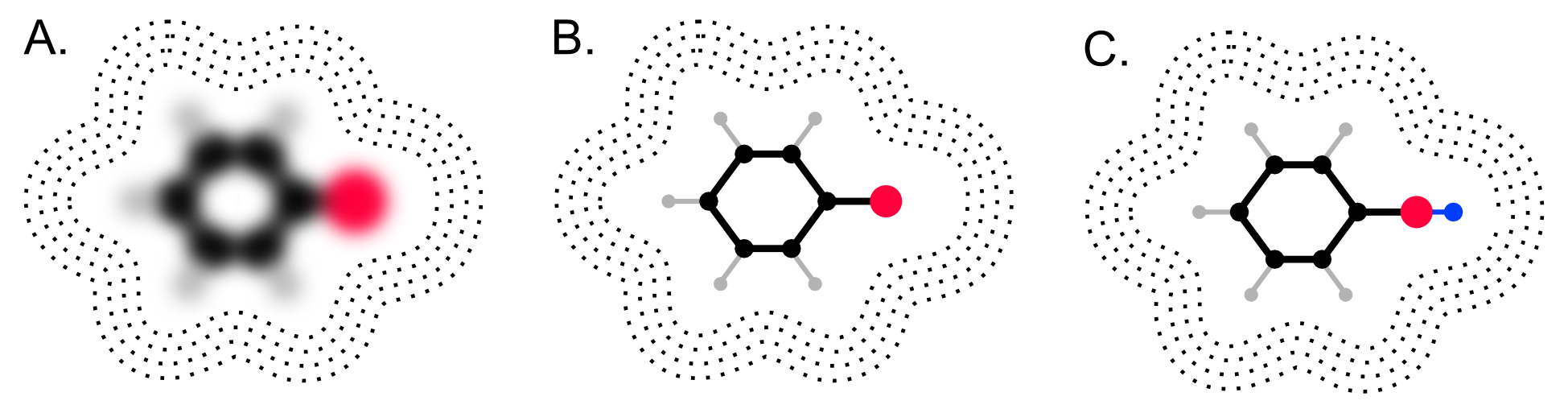}
\caption{A scheme of the off-center point-charge philosophy shown for 
bromobenzene. The quantum chemical character is represented by the blurred
atoms (A). The atom-centered point charges are calculated to represent 
the QM-based electrostatic potential (ESP) calculated at a large number 
of grid points (B). An additional site (in blue) is attached to the bromine
(in red) (C) to capture the intrinsic ESP anisotropy while still using 
the same grid of QM-based ESP.}
\label{fig:esh}
\end{figure}

The first attempts to correct for the ESP anisotropy appeared in a number 
of flavors most likely independently from several research groups within 
a single year. They collectively used an off-center positive point 
charge \cite{Vinter94} located near the halogen atom to mimic 
the \sgh ~\cite{Ibrahim11, Rendine11, Kolar12, Jorgensen12}. The extra
sites with a negative charge were previously used in description of oxygen 
and nitrogen electron lone pairs, which are negative \cite{Dixon97}, and 
it seemed natural to utilize their positive counterparts for halogen 
\sghs. The major concerns about these corrections focused on 
the properties of the extra charge, called the explicit \sgh ~(ESH).
The ESP of CF$_3$Br derived at the MM level enhanced with an ESH is shown 
in Figure \ref{fig:meps}. Visual inspection reveals that the MM+ESH model 
is somewhat \emph{``harder''} than the QM model, with smoother ESP 
variations. It is, however, better by far than the original MM model,
which tends toward an overall negative halogen ESP. The intellectual 
strategy of the ESH model is depicted in Figure \ref{fig:esh}. It highlights
the fact that one does not need to rely on atomic centers to represent 
a reference (QM-based) ESP.

From the MM point of view, the ESH particle may carry mass, charge, LJ 
potential, and other parameters. When the ESH has a non-zero mass, 
implementation into a MM code is straightforward. In contrast, mass-less 
particles are more difficult to treat, because the particle mass appears 
in the denominator in Newton's equations of motion. In these cases, the ESH 
is represented by a ghost- or dummy-atom, the force on which is redistributed 
to other atoms with non-zero mass in a precise way \cite{gmxManual, Berendsen81}.
A better known example of the use of mass-less dummy atoms is the TIP4P water 
model \cite{Jorgensen83}.
Such an approach is available in program packages such 
as Gromacs \cite{Pronk13, Pall14} and Amber \cite{Case14}. It must be noted 
that the mass of the ESH seems to have a negligible effect on the quality of 
halogen interactions.

The role of the ESH position and charge has been emphasized by numerous 
researchers. Several charge models differing in the complexity of the input 
data have been proposed \cite{Kolar12}. The ESH charge may be fitted by 
the RESP procedure, which requires a grid of electrostatic potential points.
The grid is usually derived from the electron density calculated at a QM level.
For the Amber family of force fields \cite{Cornell95}, HF/6-31G* is 
recommended \cite{Wang04}. All of the atoms are fitted at the same time, 
resulting in an all-fit model. The work of Ibrahim \cite{Ibrahim11} and 
Rendine \etal \cite{Rendine11} relied on this scheme. Compared with MM 
calculations, this QM-based charge derivation represents a bottleneck for 
high-throughput virtual screenings, and simplification is desired.

If the partial atomic charges of a molecule are known (e.g., from a database),
the charge of the halogen without ESH can be split into two 
contributions -- the charge of the ESH and the remainder of the halogen. 
Typically, a charge of approximately 0.2 e is used for ESH. Jorgensen and 
Schyman applied this model to enhance the OPLS-AA force 
field \cite{Jorgensen12}. They used a charge of 0.075 e for chlorine \sgh,
and 0.10 e for the bromine one. This approach effectively replaces the atomic
point charge with a dipole moment, the magnitude of which can be tuned by 
the ESH-halogen distance and charge splitting. Because no RESP fitting of
the charges is performed, the model is referred to as no-fit. In terms 
of reproducing the QM-based ESP, the all-fit model performs better 
than the no-fit model \cite{Kolar12}.

To be complete, yet another scheme has been presented. The ESH charge is 
taken as an adjustable parameter and the rest of atomic point charges is 
fitted by RESP. Hence the scheme is referred to as rest-fit \cite{Kolar12}.
The quality of the resulting ESP is somewhere between the all-fit 
and no-fit schemes. The advantage lies in more refined user control 
of the ESH charge, and consequently its Coulomb interaction, which 
can be arbitrarily enhanced if needed. 

The position of the ESH can be either fixed or flexible. The fixed ESH 
requires one additional parameter (the ESH-halogen distance), while the 
flexible ESH requires a few more parameters, including force constants 
for stretching the ESH-halogen bond and bending carbon-halogen-ESH angle.

In the all-fit model, the ESH charge and ESH-halogen distance are not 
independent, and no unique solution exists for their choice \cite{Kolar12}.
We suggested that a distance between 1.2 to 1.5 \AA ~would be reasonable. 
A recent study proposed 1.33 \AA \cite{Titov15}, which agrees with 
the previous findings. Slightly higher equilibrium distances were used 
in OPLS-AA, i.e., 1.6 \AA ~for the ESH of chlorine and bromine and 1.8 \AA
~for the ESH of iodine. Even higher distances were derived by Ibrahim, who
proposed that the ESH-halogen distance is equal to the van der Waals 
radius of the respective halogen \cite{Ibrahim11} (about 2.1 \AA ~for bromine).
This leads to an abnormal situation in which a typical halogen bond between 
a bromine and a carbonyl oxygen with XB length of about 3 \AA ~is modelled by
a ghost atom, which is less than 1 \AA ~from the carbonyl oxygen and deeply 
penetrates into its van der Waals sphere.

Some variants add a repulsive wall onto the ESH via LJ 
parameters \cite{Ibrahim11}. The higher the ESH-halogen distance, the more 
such repulsion is needed for the stability of computer simulations. Its role 
in the XB quality has not yet been systematically studied.

The off-center point charge models improve the structure and dynamics of 
protein-ligand complexes. On the other hand, there is still much room 
for improvement in terms of thermodynamics and liquid properties. For 
instance, the OPLS-AAx study by Jorgensen and Schyman \cite{Jorgensen12}
reported minor or no improvement in the density of liquid monohalogenated 
benzenes when compared to the uncorrected force field and experimental 
results.

\subsubsection{Electric Multipole Expansion}
\label{sssec:multipoles}

Another class of molecular mechanics of XB goes beyond the partial charge 
approximation and uses electric multipole expansion. The halogen ESP 
anisotropy can be described by a sum of multipoles, namely by a point charge
(monopole), dipole, quadrupole, etc. An isolated halogen atom has its electric
multipoles averaged to zero. However, a permanent quadrupole moment is 
an intrinsic property of isolated halogens with the valence electron 
configuration p$_x^2$p$_y^2$p$_z^1$. It is thus common to truncate the multipole
expansion after the quadrupoles \cite{Torii10, Jahromi13}. It was shown, however,
that the point electric quadrupole sufficiently represents the electrostatic 
situation on a covalently bonded halogen \cite{Torii03} and that excluding 
the dipole from the expansion increases the certainty of the quadrupole by 
about one order of magnitude \cite{Titov15}.

Torii and Yoshida \cite{Torii10} fitted the atomic point charges of an entire 
molecule with a quadrupole moment located only on the halogen. They found 
that the zz component of a traceless quadrupole (where the z-axis coincides 
with the R--X covalent bond) is insensitive to aromatic substituents, and 
the substitution effect is rather manifested by the variations of the atomic 
point charges. Titov \etal \cite{Titov15} compared the performance of off-center
charge model and the multipole expansion in reproducing QM-based molecular 
electrostatic potential. Their major finding was that both models performed 
equally well when the non-halogen part of the molecule was described in 
the same way. As opposed to off-center charges, the use of multipole expansion
in molecular dynamics software is limited.

\subsubsection{Aspherical Interatomic Potentials}
\label{sssec:asphericalLJ}

Shing Ho's research group adopted a completely different strategy to tackle
XB. Rather than searching for optimal parameters for current MM functional 
forms, they derived a new potential energy function to cover the anisotropy 
of the XB interaction \cite{Carter12}. Both the LJ and the electrostatic 
terms were modified. Altogether, seven parameters per halogen are needed,
which creates a model that is more complicated than the point-charge-based
and multipole-based models. Parameterization of bromine was performed 
employing correlated QM calculations and validated on experimental studies
of four-way DNA junctions with bromouracil \cite{Carter12}. Recently, 
the parameter set was completed with the addition of chlorine and 
iodine \cite{Scholfield14}. Some of the parameters are coupled, which 
agrees with findings from studies of extra-charge models \cite{Kolar12}. 
However, the potential energy function has been implemented only into 
the Amber \cite{Case14} program package so far, which is somewhat limiting.

\subsection{Methods for Virtual Screening}
\label{ssec:virtualScreening}

At early stages of drug discovery, the importance of computational models 
has been increasing rapidly. A common task of computer-aided drug design 
is to rank drug candidates according to their propensity to become 
a drug \cite{Klebe06, Schneider10}. Such virtual screenings typically 
work with huge databases of compounds, and the use of fast-scoring methods
is imperative. The scoring functions are also employed in determination 
of an unknown target-ligand binding pose, which is a prerequisite for 
many computational studies. A number of possible ligand orientations 
in the binding site is generated by 
an algorithm \cite{Meng92, Morris98, Lang09}, and the poses are subsequently
scored according to their presumed affinity \cite{Stahl01, Schneider02}.

Regarding XB, several methods have been developed that are even more 
approximate (and thus faster) than those used in molecular dynamics 
simulations. An empirical type of scoring function that recognizes 
the halogen bonding pattern on the basis of distance and angular parameters 
has been developed in F. Hoffmann-La Roche AG \cite{Kuhn11}. The covalent 
and noncovalent interactions are converted into a network that is used 
for scoring the target-ligand binding. This scoring function, named 
ScorpionScore, is a combination of pairwise and many-body terms. 
Interestingly, similar importance was assigned to halogen bonding 
and cation$\ldots \pi$ interactions.

Three variants of a knowledge-based scoring function (XBPMF) have been 
developed by Liu \etal \cite{Liu13}. XBPMF was trained on experimental 
X-ray geometries found in the Protein Data Bank. Such conversion 
of structural information into energy dependence represents a general 
strategy in scoring function design \cite{Muegge99, Golhke00}, and it became 
clear that one-dimensional (distance-dependent) potential might not 
be sufficiently accurate for XB. Liu \etal ~extended their previous work 
on two-dimensional potentials of hydrogen bonding \cite{Zheng11} to XB. They 
performed tests on a set of 162 protein-ligand complexes with known binding 
affinities (PDBbind, ver. 2012 \cite{Wang04b}). When compared to several 
popular scoring functions, XBPMF performed moderately well. It was rather 
weak in selecting the best binding pose, but strong in ranking ligands binding
to a particular protein.

\begin{figure}[tb]
\includegraphics{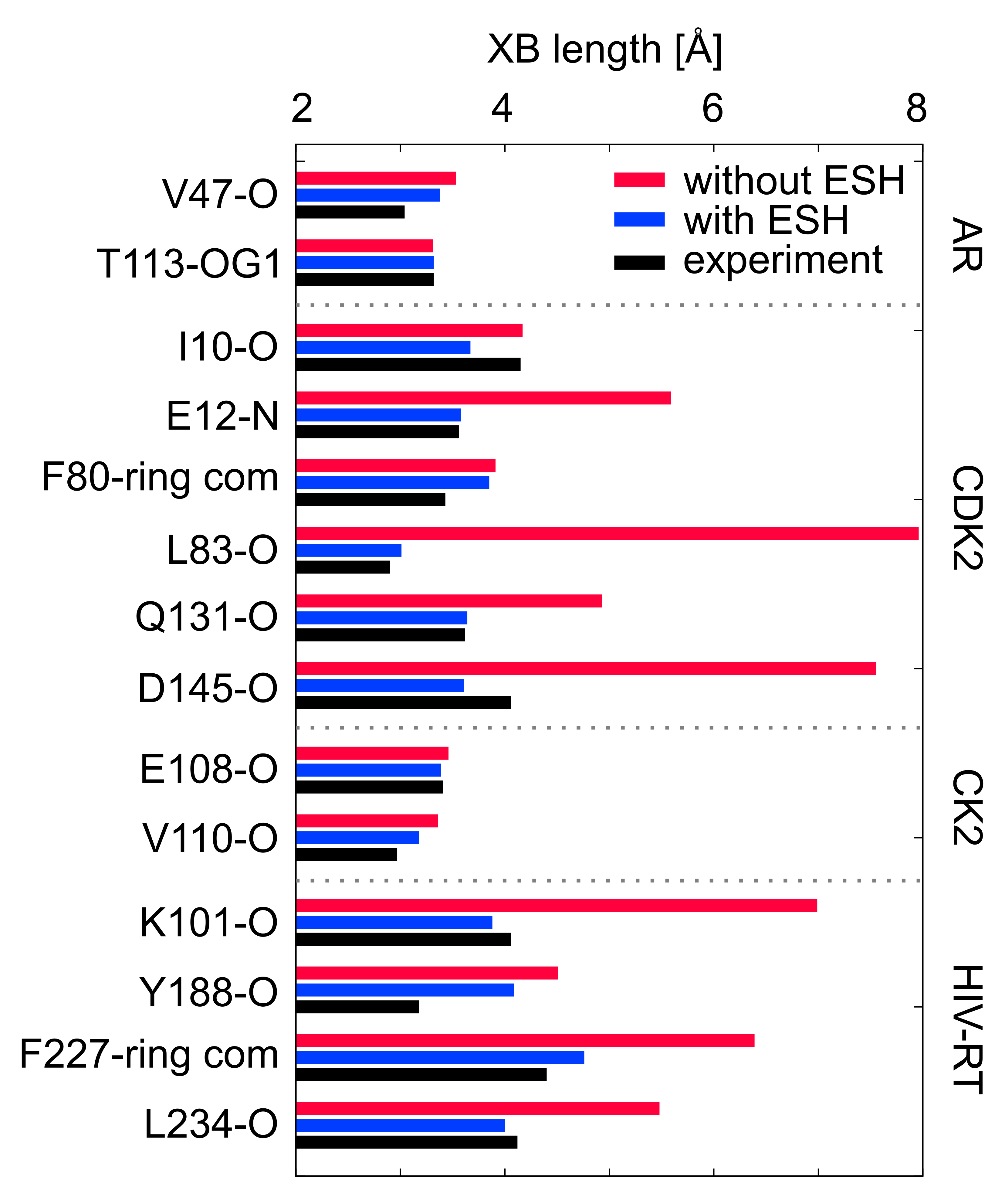}
\caption{Length of halogen-oxygen contacts with various amino acids of four 
proteins (AR, aldose reductase; CDK2, cyclin-dependent kinase 2; CK2, protein
kinase CK2; HIV-RT, human immunodeficiency virus reverse transcriptase) averaged
over all of the ligands. The results without the ESH model (red) and with the ESH
model (blue) are compared with the experimental X-ray crystallography (black).
\emph{com} stands for center of mass.}
\label{fig:docking}
\end{figure}

Of course, if a docking/scoring technique employs a molecular mechanical force 
field, one can enhance the performance by the off-center charges 
(see Section \ref{sssec:esh}). This was done \cite{Kolar13} for the first time 
for the DOCK6 program package \cite{Lang09}, using the no-fit scheme as described
by Kolář \etal \cite{Kolar12}. On a set of about 100 experimental protein-ligand
geometries, it was shown that the number and quality of halogen-oxygen contacts 
are improved. Figure \ref{fig:docking} shows the average length of halogen-oxygen
contacts between four proteins studied and their ligands. Docking with the ESH 
model performed better than docking without, as compared with the X-ray 
experimentally derived geometries.

F. Böckler's research group developed another physics-based scoring function. 
Using second-order Møller-Plesset perturbation theory (MP2), they designed 
a simplified model of halogen interaction with a protein-backbone carbonyl 
oxygen \cite{Zimmermann15}. The so-called XBScore is an extension of their 
previous systematic study of protein backbone interactions \cite{Wilcken12}. 
A polynomial dependence of the halogen bond geometry (XB length and angle) 
is introduced in XBScore. Training of the scoring function was performed on 
halobenzene$\ldots$N-methylacetamide complexes, and validation was performed 
on the full Protein Data Bank. XBScore was particularly successful 
in re-identification of existing protein-ligand halogen bonds \cite{Zimmermann15}. 

Although not fully documented, recent versions of some commercial docking/scoring 
packages like Glide \cite{Friesner06} and Gold \cite{Jones95, Verdonk03} offer
a limited description of halogen bonds. As admitted in the manual pages, 
the XB is expected to have only a minor effect on the results, unless 
a strong XB is established.

\subsection{Semiempirical QM Scoring Functions}
\label{ssec:sqmScoring}

It should be mentioned here that the binding affinity of a ligand to a protein 
represents a delicate balance between the free energy gained upon ligand binding
and free energy needed to desolvate both the ligand and the protein’s active site.
Furthermore, entropy should be considered. The problem is that free energy components
in a composite scoring function are larger than the experimental absolute 
binding free energy, and the difference between a tight-binding inhibitor 
(with nanomolar activity) and the weakest detectable inhibitor is only about 
5~kcal/mol. Therefore, a reliable scoring function must consist of different 
energy terms describing the particular interaction as accurately as possible.

We have focused on the binding free energy between protein and ligand. 
Because it is necessary to consider the whole protein having several thousands
of atoms, the use of QM methods is limited, and MM methods do not provide 
results that are accurate enough in all situations. SQM methods present 
a compromise between accuracy and computing economy. For a long time,
the use of SQM energies
was limited by the computing power. However, nowadays is has become more accessible
to run these kind of calculations for thousands of molecules \cite{Pecina15b},
which  makes this approach competitive with more established virtual 
screening techniques.

The use of SQM methods in drug design was pioneered by Merz's 
group \cite{Raha04, Peters06} who used AM1 and later PM3 methods. Because 
neither of these methods describe halogen bonding, the use of the respective
scoring function for ligands containing halogens is limited. In our laboratory,
we developed a novel scoring function based on the corrected PM6 SQM 
method \cite{Lepsik13}. The most recent modification of the method, PM6-D3H4X,
describes halogen bonding between ligand and protein (as well as other types 
of noncovalent interactions) well, and the method is thus suitable for 
assessing interactions between proteins and halogenated ligands.


\section{Computational Studies}
\label{sec:studies}

\subsection{Directionality of $\sigma$-Hole Interactions}
\label{ssec:directionality}

The ability of halogen bonds to adopt certain geometric patterns with 
a higher likeliness than others is one of their most characteristic properties.
Certainly, the broad applications of XB in biosciences and material sciences 
are partially thanks to it \cite{Politzer10}. Experimentally determined 
structures of XB complexes show that the R--X$\ldots$Y angle in the halogen
bond, where X is a halogen atom and Y is an electron donor, tends to be close
to 180\degree, i.e., a linear arrangement. Two 
surveys \cite{Lommerse96, Bauza13} of the Cambridge Structural 
Database \cite{Allen02} and one \cite{Auffinger04} of the Protein Data 
Bank \cite{Bernstein78} suggested that the tendency to linearity is stronger
for heavier halogens than lighter ones. However, due to the lower relative 
crystallographic occurrence of XBs involving iodine, the data for iodine 
are prone to larger statistical uncertainties than those of chlorine and
bromine, and thus there is some difficulty in obtaining quantitative trends.

\begin{figure}
\includegraphics{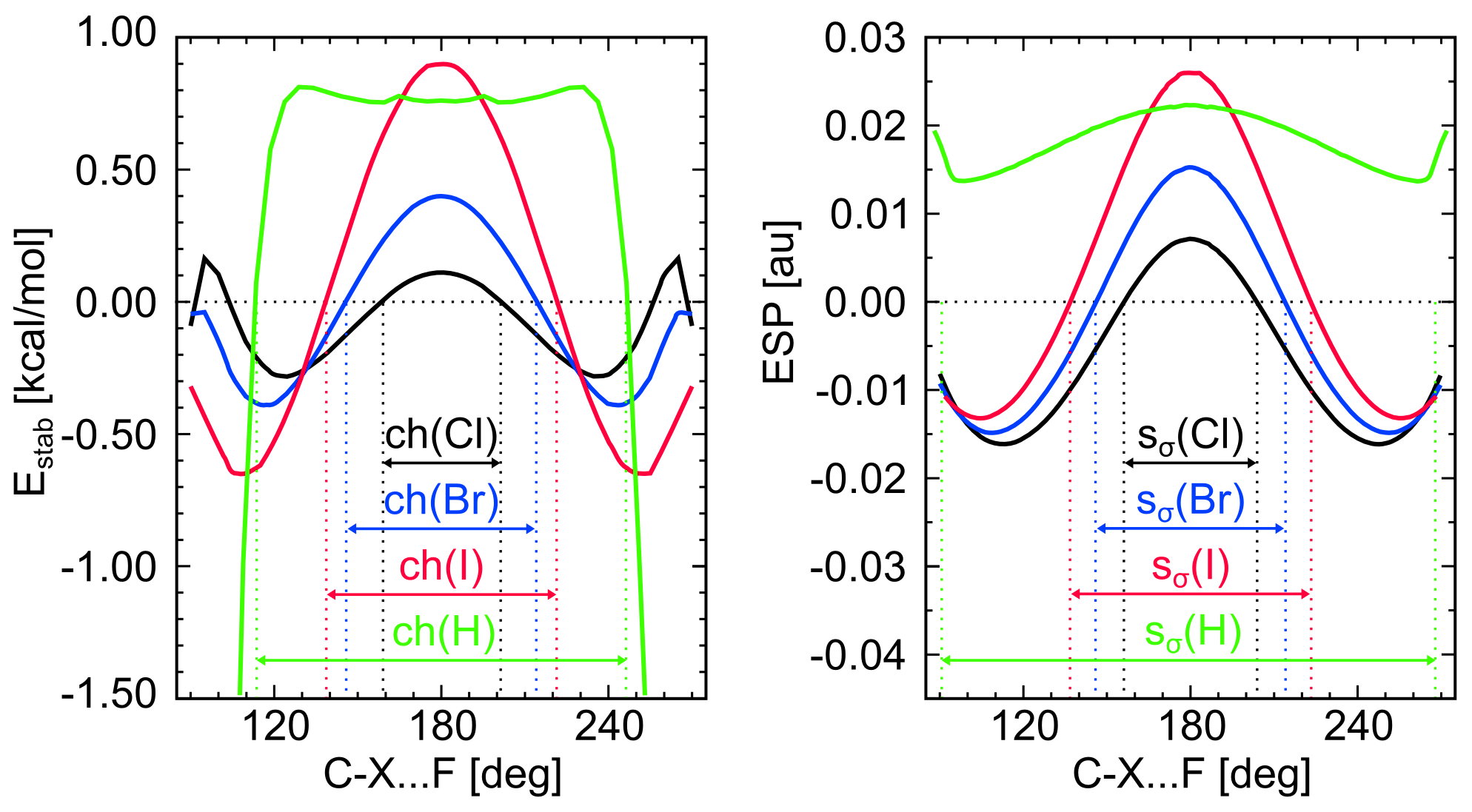}
\caption{Left panel: The stabilization energy profiles of halobenzene 
complexes with hydrogen fluoride calculated at the B3LYP-D3/def2-QZVP 
level. The angular channel, where the interaction is attractive, is 
labeled $ch(X)$, where X stands for the halogen. Right panel: The electrostatic 
potential (ESP) angular profiles calculated at 0.001 au electron density 
surface using the same computational method. The angular variants of the 
sizes of \sgh ~are labeled $s_{\sigma}(X)$, where X stands for the halogen.}
\label{fig:directionality}
\end{figure}

We have investigated the directionality of halogen bonding in a series of 
stand-alone halogenated aromatics and halogen benzene complexes with argon
and hydrogen fluoride. Figure \ref{fig:directionality} shows the angular 
dependences of the stabilization energy of chloro-, bromo-, and iodobenzene
with hydrogen fluoride calculated at the B3LYP-D3/def2-QZVP level with 
the effective core potentials on iodine \cite{Kolar14a}. These were obtained
by varying the C--X$\ldots$F angle between 90\degree and 270\degree, where 
180\degree corresponds to a linear XB. The X$\ldots$F--H angle was kept
180\degree.
  
The stabilization energy profiles depend on the atomic number of 
the halogen donor. The interaction with the hydrogen fluoride is attractive
in a linear conformation but repulsive in a perpendicular direction with 
respect to the C--X covalent bond. Such behavior resembles the ESP surface 
of halogens. As a consequence, by varying the C--X$\ldots$F angle, the complex 
maintains attractive interaction. The heavier the halogen involved in the
XB, the more variability is allowed to maintain the stability 
(i.e., positive stabilization energy). Moreover, the size of \sgh
~($s_\sigma$) reflects such changes (Figure \ref{fig:directionality}, right).
In the SAPT sense, the angular variations of the halogen bond are caused not
only by the electrostatic energy, but also by the exchange-repulsion energy;
the angular variation of the dispersion energy is considerably lower. This 
is consistent with another approach of dissecting dispersion interaction 
described by El Kerdawy \etal \cite{Kerdawy13}. Based on two-dimensional 
scans at the frozen-core CCSD(T)/aug-cc-pVQZ level using argon atom as a probe, they 
found that the sum of dispersion and repulsion is significantly angular
dependent.

In our previous work, we stated \cite{Kolar14a} that \emph{``$\ldots$the channel
allowing the approach of HF to halogenbenzenes (with the attractive interaction
energy) is the narrowest for Cl and the broadest for I. Alternatively these
values indicate that the directionality of halogen bonds is largest 
for chlorobenzene and becomes the smallest for iodobenzene,''} 
and by \emph{``the directionality,''} we meant the channel of attractive 
interaction energy. We admit that our use of the terminology was not 
particularly fitting, and we investigated the problem further \cite{Kolar14b}.
We have shown that one must distinguish between such directionality (in our 
work, represented by the channel) and the tendency to linearity. 
The above-mentioned experiments suggest, indeed, that the tendency to 
linearity should increase with increasing atomic number of the halogen.

The energetic penalty brought upon an angular distortion increases with 
increasing size of the \sgh. As shown in Figure \ref{fig:directionality},
the ESP profile is steepest in case of iodine, hence \emph{``the force 
acting on the electron donor is largest when iodine is involved, favoring
the linear arrangement of the XB more than in the cases of bromine and 
chlorine.''} \cite{Kolar14b}. In other words, the gradient of the ESP is 
higher for heavier halogens, which consequently determines the linear geometry
of the XB. Tsuzuki \etal ~drew similar conclusions \cite{Tsuzuki12} based 
on computational investigation of complexes of halogenbenzenes with pyridine.
Furthermore, they found that the magnitude of angular dependence is dependent
on the XB strength. For weaker XBs, the angular variations of stabilization
energy were weaker than for stronger XBs. The strength of XBs was modulated
by changing the XB length \cite{Tsuzuki12}.

\begin{figure}[tb]
\includegraphics{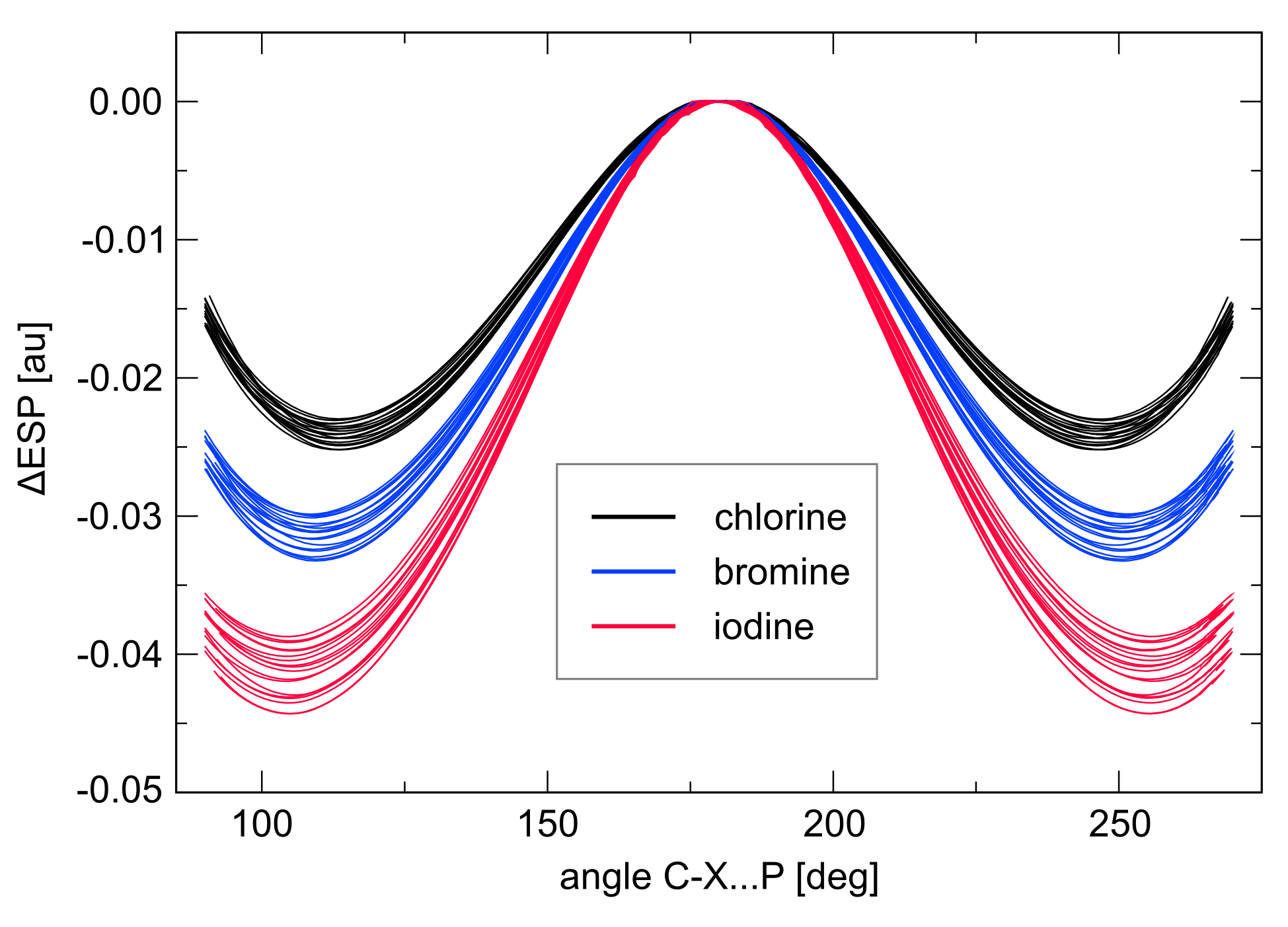}
\caption{An aggregation of the electrostatic potential (ESP) profiles from 
our previous work \cite{Kolar14a} normalized with respect to a 180\degree 
~angle. The ESP profiles were derived on the surface of electron density 
of 0.001 au. Chlorobenzene (black), bromobenzene (blue), and iodobenzene 
(red) derivatives are shown.}
\label{fig:allProfiles}
\end{figure}

It is intriguing to see the effect that an aromatic substitution has on 
the angular dependence of ESP. Figure \ref{fig:allProfiles} shows 
the relative ESP angular profiles calculated on the surface of electron 
density of 0.001 au for all benzene derivatives as previously 
described \cite{Kolar14a} (i.e., with --CH$_3$, --F, --NO$_2$ or --CN groups
in the meta- and/or para- positions with respect to the halogen). For sake 
of comparison, the profiles were normalized so that the magnitude of 
the \sgh ~was equal to 0 au (i.e., the ESP for angle 180\degree). 
The curves naturally cluster with respect to the halogen involved in a possible XB. 
While the exchange of the halogen has a rather strong effect on the ESP 
profile and its steepness, aromatic substitutions, on the other hand, vary 
the profiles considerably less. Consequently, the tendency of a halogenated 
aromatic to adopt a linear XB geometry is affected much more by the nature 
of the halogen involved in the XB than by the chemical substitutions on 
the aromatic ring, which is rather surprising.

Protein backbone carbonyl oxygens are electron donors with high natural 
abundance. Thus, an exploration of their propensity for halogen bonding 
is desirable. Three-dimensional spherical scans were performed 
to characterize the directionality in sub-optimal XB 
arrangements \cite{Wilcken12}. N-methylacetamide was chosen to model 
the protein backbone, and halogenbenzenes as the ligand. Unrelaxed scans 
were carried out at the MP2/TZVPP level using constrained monomer 
geometries. It was confirmed that, indeed, the interaction energy is 
angular dependent and decreases notably upon even slight XB deformation.
A decrease of about 50--60\% was reported for angular deviations of 
about 30\degree, but it must be kept in mind that secondary interactions 
may contribute to the stabilization, especially for large distortions. 
Relaxation upon distortion would increase the complex stability. These 
results were later utilized in design of a scoring function of halogen 
bonds \cite{Zimmermann15}.

Analogously to XB, other \sgh ~interactions seem to be driven by 
ESP anisotropy \cite{Iwaoka02}. Work by Adhikari and 
Scheiner \cite{Adhikari12} concluded that the propensity to linear 
arrangement is much higher for halogen, chalcogen, and pnicogen bonds 
than for hydrogen bonds. Based on SAPT decomposition, it was further 
revealed that, similarly to XB, the angular sensitivity of other 
\sgh ~interactions is strongly determined by the exchange-repulsion 
energy component \cite{Adhikari12}.

\subsection{Hydrogen vs. Halogen Bonding}
\label{ssec:hb}

Hydrogen and halogen bonds exhibit numerous similarities but also some 
important differences that can be important in bio- and material sciences.
As such, they are often the subject of direct comparisons. One of the most
important differences is their directionality. At the end of the last 
century, i.e., before the concept of halogen bonding had been widely 
spread, directionality of hydrogen bonds was considered an important property
of this noncovalent interaction \cite{Legon87, Bernstein95, Wood09}.

Based on the S66 \cite{Rezac11a} and X40 \cite{Rezac12a} benchmark data sets, 
the strength of the two interactions is comparable. The total stabilization
energies of HB and XB are generally similar, but this is not true for their
decomposition (e.g., by the SAPT scheme). This represents another very 
important difference between the two interactions. The electrostatic 
energy is, no doubt, the most attractive contribution to HB, while 
dispersion and induction energies are less important (although they cannot
be neglected). The situation is different for XB, and its characteristic
properties are due to the concerted action of dispersion, electrostatic,
and induction/charge-transfer energies.

HB has been considered a \sgh ~interaction by some 
researchers \cite{Politzer10, Shields10, Clark13, Politzer13b}. However, 
the nature of \sgh ~as a region of positive ESP on hydrogen differs 
from that on halogens. As detailed by Clark \cite{Clark13}, the hydrogen
\sgh ~is caused by a shift of electron density along the R--H covalent bond.
As a consequence, the hydrogen nucleus is exposed because there are no 
other electrons to shield it (the only one participates in the covalent bond).
Another difference mentioned by Clark \cite{Clark13} is that, in terms of ESP,
the hydrogen is generally positive everywhere.

Wolters \etal \cite{Wolters12} highlighted some differences between HB and XB
from the molecular orbital perspective. Based on relativistic DFT, they 
found that there is a significantly stronger HOMO-LUMO interaction in XB. 
They also noted that the heteronuclear dihalogens possibly involved in XB 
are less polar that hydrogen halides available for HB.

The literature on the directionality of hydrogen and halogen bonds is quite 
rich \cite{Platts96, Metrangolo05, Voth09, Politzer10, Murray10, Shields10, 
Tsuzuki12, Kolar14a}. As stated by Shields \etal \cite{Shields10}, 
the different sensitivity of the interaction strength of halogen and 
hydrogen bonds on the angular deviations can be attributed to the presence
or absence of lone electron pairs in the valence shell of the halogen 
or hydrogen. Figure 18 shows the stabilization energy profiles of halobenzene
and benzene complexes with hydrogen fluoride. The comparison arises from 
B3LYP-D3/def2-QZVP calculations, as previously described in 
detail \cite{Kolar14a}. The hydrogen analogue profile is notably shallower 
than the halogen one, which points to the low sensitivity of benzene complexes
with hydrogen fluoride on angular deviations, which is in full agreement with
the results for benzene analogues and pyridine \cite{Tsuzuki12}.

\begin{figure}[tb]
\includegraphics{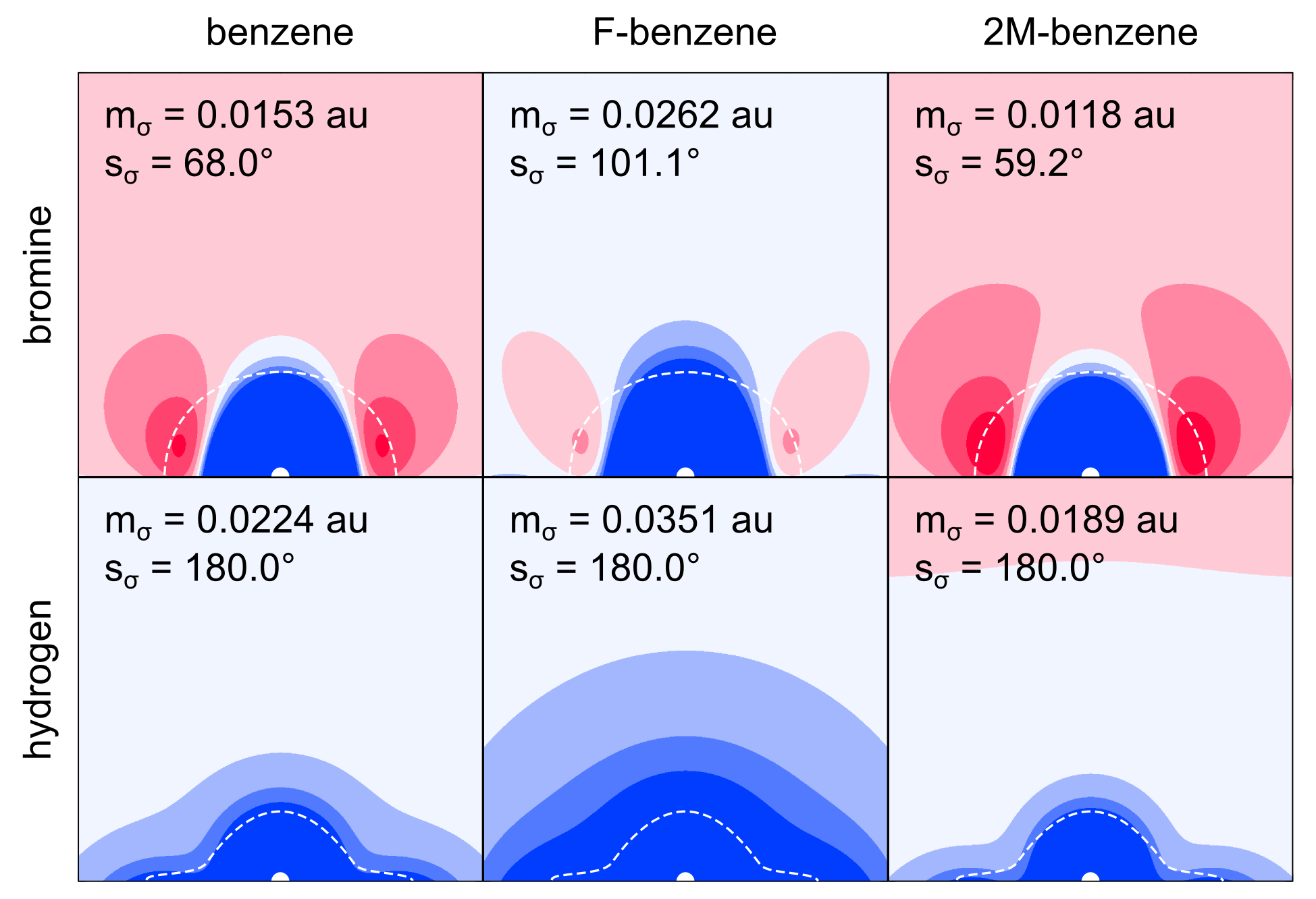}
\caption{Two-dimensional electrostatic potential maps of substituted 
bromobenzenes (upper panels) and benzenes (lower panels) calculated 
in the plane of the aromatic ring. Positive values are in blue and 
negative in red. All-hydrogen, para-fluoro-, and meta-dimethyl- analogues 
are shown in columns labeled benzene, F-benzene, and 2M-benzene, 
respectively. The magnitudes (m$_\sigma$) and sizes (s$_\sigma$) of the 
\sghs ~are provided. The dashed white line stands for the surface 
of 0.001 au electron density. The white hemicircle is either bromine 
(upper panels) or hydrogen (lower panels). Modified with permission from 
Ref. \cite{Kolar14a}. Copyright 2014 Royal Society of Chemistry.}
\label{fig:espHb}
\end{figure}

Furthermore, it is evident that the \sgh ~on halogens is typically 
a well-bounded positive region on the isodensity surface, while the ESP 
on hydrogens is positive everywhere (Figure \ref{fig:directionality}, 
right). This is also shown 
in two dimensions in Figure \ref{fig:espHb}. It is worth noting that, upon 
substitution of the aromatic ring, the hydrogen ESP changes differently 
than the halogen ESP. For example, when adding fluorine to the para-position
relative to bromine on bromobenzene, the \sgh ~magnitude increases from
0.015 to 0.026 au. Correspondingly, the \sgh ~magnitude on the benzene 
analogue changes from 0.022 to 0.035 au. The change of the \sgh ~size 
is, however, completely different: the halogen \sgh ~size increases, 
while it is virtually unchanged in the case of the hydrogen analogue 
(Figure \ref{fig:espHb}). 

Finally, a very important difference between hydrogen and halogen bonds 
concerns the solvation energy. HB is mostly more polar than XB, and 
consequently, the solvation/desolvation energy of HB complexes is higher
than that of XB complexes. This feature represents an important advantage 
of XB, particularly in protein-ligand association.

\subsection{Vibrational Spectra of Halogen-Bonded Complexes}
\label{ssec:spectra}

Formation of an R--X$\ldots$Y halogen bonded complex is accompanied by changes
in bond lengths and vibration frequencies of interacting subsystems. Larger 
changes are expected for the halogen donor R--X. For R--H$\ldots$Y hydrogen
bonding, which is similar to halogen bonding, it was believed for a long 
time that formation of HB is accompanied by lengthening of the R--H bond 
and lowering of the R--H vibration frequency (so-called red shift). These 
characteristics were considered  \emph{``fingerprints''} by which 
the formation of HB could be recognized. At the end of the last century, 
we showed \cite{Hobza00} that HB formation can be accompanied not only 
by red shift of R--H vibration frequency and elongation of the R--H bond,
but also by blue shift of this vibration and contraction of the bond. 
This finding initialized the introduction of a new definition 
of HB \cite{Arunan11}.

The first study on vibrational shifts in R--X$\ldots$Y halogen-bonded 
complexes suggested \cite{Politzer07b} the existence of only red shifts 
in R--X vibrations.  Subsequent studies \cite{Riley08b, Wang11}, however,
proved the existence of blue shifts in some halogen bonded complexes. 
The origin of blue and red shifts was investigated \cite{Murray08} 
in detail for 12 halogen bonded complexes in which the electron donor 
was either HCN or NH$_3$ and the halogen donors were XF, XCN, XNO$_2$, 
and XCF$_3$ (X=Cl, Br). The stabilization energies of all these complexes
were between 1 and 6.9~kcal/mol (at frozen-core MP2/6-311++G(3df,2p) level),
and all X$\ldots$Y distances were shorter
than the sum of the van der Waals radii of the respective atoms. Six of 
the complexes exhibited blue shifts (3 -- 17 cm$^{-1}$), which were always
accompanied by a decrease in R--X bond length. All of the NH$_3$ complexes 
displayed red shifts, even with those electron donors that exhibited 
a blue shift with HCN. Different shifts depend on the properties of 
the halogen donor and the electron donor. A necessary but not sufficient
condition for blue shifting is that the derivative of the \sgh
~donor's permanent dipole moment is opposite to the direction of 
the electric field induced by the acceptor \cite{Hermansson02, Qian02}. 
If this condition is satisfied, a further requirement concerning the nature
of the electron donor (contribution of the induced dipole moment derivatives
to the vibration shift should not be larger than that of the permanent
dipole) should be fulfilled. 

Our results on XB were fully consistent with these conditions developed 
earlier for blue- and red-shifting hydrogen bonds. The blue shifts of R--X 
stretching frequencies were also found for six complexes of anesthetics 
(chloroform, halothane, enflurane, and isoflurane) with 
formaldehyde \cite{Zierkiewicz11}. The binding energies of these complexes 
calculated at the CCSD(T)/CBS level were between 2.83 and 4.21~kcal/mol. 
In all complexes, the C--X bond length (where X = Cl, Br) was slightly 
shortened in comparison to the isolated XB donor, and an increase in 
the C--X stretching frequency was observed. The electrostatic interaction
was excluded as being responsible for the C--X bond contraction. On the other
hand, it was suggested that contraction of the C--X bond length could be 
explained in terms of the Pauli repulsion (the exchange overlap) between 
the electron pairs of the oxygen and halogen atoms in the complexes.

In conclusion, structural and vibration-frequency changes upon formation of 
a halogen bond are similar to those in hydrogen bonded complexes. The R--X
bond length can increase or decrease upon formation of a halogen bond, and
the respective vibration frequency can be shifted to lower (red shift) or 
higher (blue shift) energies. The changes in the bond length and vibration 
frequency are considerably lower for halogen bonded complexes than for 
hydrogen-bonded complexes, which is due to the different nature of 
the bonding (the presence of light hydrogen in HB).

\subsection{Structural Studies of Small Molecules}
\label{ssec:smallMolecules}

The utility of computational tools in understanding crystal packing and 
noncovalent-driven self-assembly is notably wide. Quantum chemistry is 
well-suited to study interactions of molecules with low spatial flexibility,
and the lack of solvation dynamics plays a favorable role.

Early surveys of the Cambridge Structural Database (CSD) \cite{Allen02} 
focused on halogen contacts were augmented by quantum chemical 
analyses \cite{Lommerse96, Allen97}. Ouvrard \etal ~inspected the interactions
of dihalogens \cite{Ouvrard03}. A comparative study of halogen and other 
\sgh ~bonds by Bauzá \etal \cite{Bauza13} combined the CSD survey and 
quantum chemistry. Based on BP86-D3/def2-TZVP calculations, Bauzá \etal ~
showed that for \sgh ~interactions with a nitrogen as the electron 
donor, halogens are energetically more favorable than chalcogens or pnicogens
as electron acceptors. In contrast, when the electron donor is an aromatic 
ring, pnicogens are preferred over halogens and chalcogens \cite{Bauza13}.

Iwaoka \etal ~used a combination of NMR spectroscopy, NBO
analysis, and the QTAIM approach to study Se$\ldots$O
intramolecular contacts \cite{Iwaoka04}. Because of a bond critical point (BCP)
identified between Se and O, the researchers concluded that the interaction 
is of covalent rather than electrostatic character, which is, however, 
limited to the QTAIM diction. It must be noted that in other schemes, 
the interaction might be classified differently.

We studied crystals of hexahalogenbenzenes (C$_6$F$_6$, C$_6$Cl$_6$, and 
C$_6$Br$_6$) and benzene (C$_6$H$_6$) using DFT and DFT-SAPT \cite{Trnka13}.
Halogenbenzene crystals are rich in halogen-halogen contacts, some of which
could be classified as a dihalogen bond. Such interaction is characterized 
by contact of the positive \sgh ~of one halogen with the negative belt
of a second halogen. To our surprise, we found that the dihalogen bond does
not contribute significantly to the stability of the crystals. The dispersion
energy, both in the SAPT sense and Grimme's empirical one, seems to be 
the leading source of stabilization. The differences in the sublimation 
energy of the crystals can be explained by the dispersion interactions.

Substantial pairwise stabilization energies were found for crystal structures
of diiodine complexes with diazabicyclo[2.2.2]octane (DABCO) and 
1,3-dithiole-2-thione-4-carboxylic acid (DTCA) \cite{Deepa14}. Furthermore,
similar complexes with other molecules (Cl$_2$, Br$_2$, IF, ICH$_3$, N$_2$)
were examined using the CCSD(T)/CBS and DFT-SAPT methods. Surprisingly, 
the stabilization energy correlated with the classical concept of electric
quadrupole moments of the electron acceptors. Comparing the quadrupoles of
X$_2$ molecules, we found that they have different signs for dihalogens 
and nitrogen. The quadrupoles of the halogens can be schematically written
as $+--+$, and those of nitrogen as $-++-$. Evidently, the notation 
reflects the concept of the \sgh. Furthermore, there exists 
a close correlation between $V_{max}$ and the quadrupole value. Finally,
in contrast to findings from many other DFT-SAPT studies of XB complexes,
the induction energy of the complexes of diatomics was the leading SAPT term.

Chalcogen and pnicogen bonding in heteroborane crystals has been studied 
computationally \cite{Pecina15}. $\sigma$-hole analyses and CCSD(T)/CBS calculations
were complemented by DFT-SAPT energy decompositions. In our opinion, this is 
a reliable way of studying noncovalent interactions without making compromises,
and the final picture seem to be quite complete. Heteroboranes are prone 
to halogen bonding in a comparable way as other, more typical halogen donors.
Positive \sghs ~were identified on halogens incorporated as exo-substituents,
as well as on chalcogens and pnicogens incorporated directly into the borane 
cage. Further experiments and calculations justified that these \sgh
~bonds are essential for the crystal packing of heteroboranes \cite{Fanfrlik14}.

\subsection{Halogen Bonds in Biomolecules and Medicinal Chemistry Applications}
\label{ssec:medicinalChemistry}

Halogen bonds have been proven to exist in numerous biomolecular 
complexes \cite{Auffinger04, Sirimulla13}, and their specificity was 
recognized quite early \cite{Cody84}. Moreover, XB has found a place 
in medicinal chemistry and, specifically, in protein-ligand interactions.
The reason is that the incorporation of a halogen onto the ligand scaffold 
frequently leads to enhanced drug activity. A review of the medicinal chemistry
applications of XB was recently published by F. Böckler's
group \cite{Wilcken13}. Many newly developed drugs contain a halogen atom,
because the presence of halogens is beneficial in several aspects. 
For example, it improves oral absorption. In addition, halogens fill 
hydrophobic cavities in the binding site, and they facilitate crossing 
the blood-brain barrier and prolong the drug lifetime. Recently, also 
the specific interaction of the halogens has appeared to contribute. 
As mentioned previously, the tunability of the XB (which is higher than 
that of HB) represents an important advantage of XB and can be applied 
to enhance protein -- ligand interactions.

Lu \etal ~studied halogen interactions in biological complexes \cite{Lu10}
both by analyzing crystallographic data and by calculations. In addition 
to halogen bonds, they investigated other kinds of halogen interactions.
They found that oxygen and aromatic $\pi$ systems are the preferred electron
donors in biomolecules. Their observations were rationalized by MP2 
calculations on small complexes and by composite quantum mechanics/molecular
mechanics (QM/MM) calculations of three halogen$\ldots \pi$ complexes (PDB
ID: 1WBG \cite{Hartshorn05}, 1ZOE \cite{Battistutta05}, and 
2PIT \cite{Estebanez07}).

The energetics and character of protein-ligand halogen bonds were 
investigated computationally on a small series of X-ray data \cite{Riley11b}
comprising the ligand binding domains of the thyroid hormone 
receptor \cite{Darimont98}, ADP-ribosyl transferase ART2 \cite{Mueller02},
and protein kinase CK2 \cite{Battistutta05, Battistutta07, Sarno11} 
(PDB IDs: 1BSX, 1GXZ, 1ZOH, 2OXD, 2OXY, and 3KXN). As noted in the study,
the \emph{``best case scenarios''} rather than all available motifs were 
selected intentionally, so that the halogen bonds studied had favorable 
geometric properties such as a C--X$\ldots$O angle larger than 165\degree.
A ratio of stabilization energies calculated at the Hartree-Fock and MP2 
levels were calibrated by DFT-SAPT calculations of model complexes. 
The ratio was used to estimate the role of dispersion energy. 
The stabilization energies were significant, and consequently 
the researchers claimed that \emph{``the halogen bonds are strong 
interactions that play an important role in the binding of small molecule 
ligands to proteins.''} However, according to recent studies, the favorable
energetics of XB may be compensated for by other contributions, such 
as solvation/desolvation changes.

\begin{figure}[tb]
\includegraphics{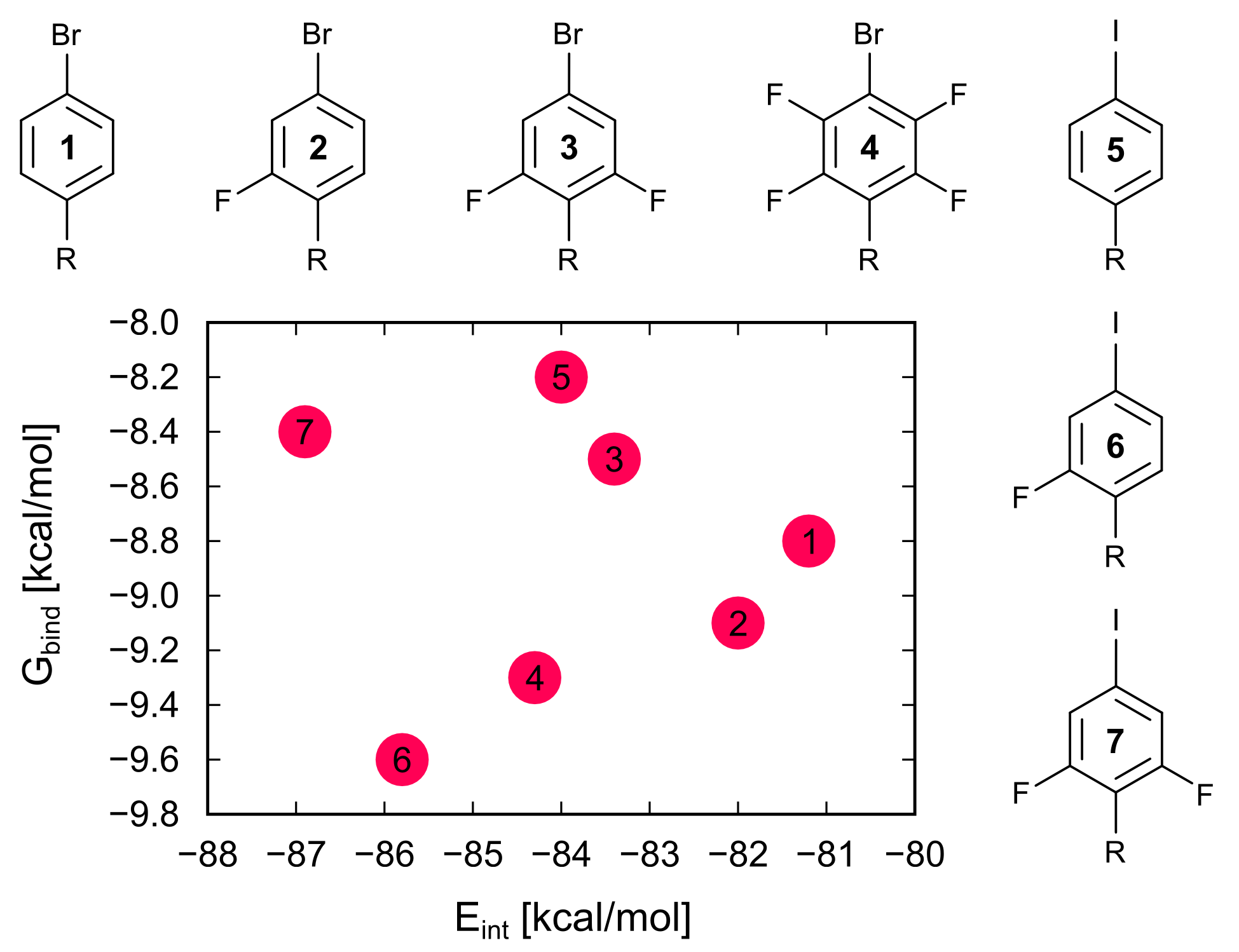}
\caption{Structural formulas of aldose reductase 
inhibitors \cite{Fanfrlik13} and the relationship between the experimental 
binding free energy ($G_{bind}$) and the vacuum interaction energy ($E_{int}$)
calculated at a composite DFT/PM6-D3H4X level.}
\label{fig:aldose}
\end{figure}

In our recent study, we modulated \cite{Fanfrlik13} aldose reductase (AR) 
inhibition by halogen bond tuning. A series of AR inhibitors derived from 
a known AR binder exhibiting a halogen bond between its bromine atom and 
the oxygen atom of the Thr113 side chain of AR was investigated. The strength
of the halogen bond was tuned by bromine–iodine substitution and 
by fluorination of the aromatic ring (Figure \ref{fig:aldose}). Our 
conclusions were supported by X-ray crystallography and IC50 measurements. 
Tuning the strength of the halogen bond with a monoatomic substitution 
decreased the IC50 value by about 1 order of magnitude (i.e., the inhibition 
activity increased). The composite DFT/PM6-D3H4X calculations revealed that 
the protein–ligand vacuum interaction energy increases upon substitution of 
iodine for bromine or upon the addition of electron-withdrawing fluorine 
atoms to the aromatic ring. However, there was no obvious relationship between 
the inhibition activity and the magnitude of the \sgh ~or the vacuum 
interaction energy (Figure \ref{fig:aldose}). Given the almost-identical 
binding poses of all of the inhibitors in crystal structures of AR-inhibitor 
complexes, we claim that this is because fluorination affects other properties 
of the ligand in addition to the halogen \sgh, and speculate that 
the inhibitor solvation free energy is decisive in our case study.

To pursue our hypotheses, we designed a halogen-to-hydrogen bond substitution 
(Br or I atoms to an --NH$_2$ group) in AR inhibitors \cite{Fanfrlik15}. 
The substitution of halogen with a polar group is expected to strengthen 
the gas-phase-like stabilization energy (providing a suitable HB complex can 
be formed), but it simultaneously increases the desolvation penalty. 
The combination of X-ray, advanced QM/SQM/MM scoring, molecular dynamics-based 
free energy simulations, and IC50 measurements showed that i) the geometry of 
the complex with XB and HB remained unchanged; ii) structurally, the XB was 
replaced by an HB; and iii) although the HB is significantly stronger than 
the original XB in a vacuum, the total binding affinity decreased due to 
the larger desolvation penalty of the --NH$_2$ group. Thus, the solvent (in 
this case, water) plays a very important role in XB of protein-ligand complexes
when competing with HB.

Carter and Shing Ho evaluated the free energy difference between XB and HB 
using differential scanning calorimetry of DNA junctions \cite{Carter11}. 
Bromouracil was incorporated into DNA, forming an intrastrand halogen bond, 
which consequently determined the ternary structure of the junctions in 
the crystalline environment \cite{Voth07}, consistent with the situation 
in solution at room temperature. The researchers claimed that the higher 
stabilization of the junction by halogen bonds compared to hydrogen bonds 
was enthalpic in origin, which was further supported by quantum chemical 
calculations at the MP2 level. To understand the role of solvation and steric
effects, bromine was replaced with a methyl group—a chemical group of similar
size. The researchers concluded that in the case of the DNA junctions, 
the enthalpic preference of XB over HB lies, to the first approximation, 
in the direct interaction of the electron donor and acceptor \cite{Carter11}. 
Enthalpy-entropy compensation plays a role \cite{Carter13}. 

The structure-activity relationship (SAR) of two class of ligands, namely 
GAGA-A/benzodiazepine binders and serine protease factor Xa inhibitors, was 
studied by employing AM1-based field properties \cite{Gussregen12}. An important
finding was that the field properties calculated at a SQM level revealed new 
trends that were inaccessible by conventional MM-based SAR models.

Several examples of how introduction of XB can lead to improvement in ligand 
activity have been reported to date: hepatitis C protease 
inhibitors \cite{Llinas10}; carboxamide inhibitors of 
integrin-$\alpha$4$\beta$1 \cite{Carpenter10}; and inhibitors of human 
cathepsin L \cite{Hardegger11, Hardegger11b}, isoforms 1 and 4 of CDC2-like 
kinase \cite{Fedorov11}, and phosphodiesterase type 5 \cite{Xu11}. 
Particularly compelling results have been obtained for lead-optimization 
of HIV reverse transcriptase inhibitors \cite{Himmel05, Bollini11}. 
The force-field-based free energy perturbation Monte Carlo method identified 
a compound with 55 pM anti-HIV activity (EC50 to the wild-type virus). 
This finding was somewhat serendipitous, because although no XB-related 
corrections were adopted for the OPLS-AA force field at that time, 
a directional chlorine-oxygen contact was predicted in the protein-ligand 
structure \cite{Bollini11}. Finally, the design of a halogen-enriched fragment
library by Wilcken \etal \cite{Wilcken12b} as a means for introduction of 
novel drug molecules utilizing halogen bonding should be mentioned.


\section{Concluding Remarks}
\label{sec:remarks}

We have attempted to review theoretical approaches to the study of XB and 
other \sgh ~interactions. The importance of both has been rising, 
especially in the fields of supramolecular chemistry and computer-aided 
drug design. We believe that in the near future, even more attention will 
focus on these kinds of noncovalent interactions. While research in past 
centuries has focused on covalent chemistry, future researchers will have 
to explore the world of noncovalent interactions, simply because such 
interactions are heavily involved in phenomena related to many recent 
groundbreaking discoveries. A broad arsenal of computational and experimental
tools is already available to aid in understanding such phenomena.

In our opinion, accurate \ai calculations combined with \sgh
~characteristics and energy decomposition schemes are the pathway that one
should follow to analyze \sgh ~complexes. From the perspective of 
reference \ai calculations, CCSD(T)/CBS remains 
the \emph{``gold standard.''} The density functional theory provides 
the M06-2X functional with its very promising results on both halogen 
bonded and \sgh ~bonded complexes, although its use for other kinds 
of noncovalent interactions is not so promising. There are still, however, 
a few challenges to be addressed in future studies. Regarding computational 
methods, the largest space for improvement seems to be in the fast and 
efficient DFT, SQM, and MM methods. The more approximate the approach, 
the more balanced its contributions have to be to cover a wide range of 
intermolecular interactions. Focusing on a single type of noncovalent 
interaction might not be the best choice.

Furthermore, the role of solvent in \sgh ~interactions has been 
studied mostly from the perspective of supramolecular 
chemistry \cite{Beale13}. For example, it was shown that XB vacuum 
interaction energies may correlate with thermodynamics in organic 
solvents \cite{Chudzinski12}. For medicinal chemistry applications, 
water is the exclusive solvent, and the behavior of XB in water is 
still poorly understood. It is well-known that the halogen atoms have 
electron donors in common with hydrogen bonds. Thus, the competition 
between halogen bonds and hydrogen bonds in biomolecule-ligand binding 
is largely affected by hydrogen bonding in water. We have recently shown
that the thermodynamics of the two interactions are of different 
natures \cite{Fanfrlik15}, but we do not think that the issue of solvation 
changes has been resolved. Clearly, benchmark calculations on small complexes
would be welcome. Also, MM force fields, which are traditionally used 
to determine solvation free energies, have not yet been examined thoroughly 
enough in description of the thermodynamics of \sgh ~complexes.

Another challenge lies in reliable predictions of the properties 
of \sgh ~complexes. Medicinal chemistry still suffers from a lack 
of quantitative predictions of ligand affinities or pharmacokinetic properties.
One reason might be that halogen bonding is not the decisive interaction between
the ligand and its target, but makes only a minor contribution to the total 
affinity. In this case, the computational models in use fail not only due 
to improper description of the XB (local effect), but also due to an overall 
deficiency in description of the global effects.

We are quite optimistic that these questions, as well as many others, will 
be addressed soon, given how XB and \sgh ~interactions have gained 
the spotlight during the last decade.


\section{Biographies}

\subsection{Michal H. Kolář}

Michal H. Kolář received his Ph.D. in 2013 from Charles University in Prague,
and from the Institute of Organic Chemistry and Biochemistry, Academy of Sciences
of the Czech Republic. With Pavel Hobza
he focused on theoretical and computational description of
noncovalent interactions. In 2012 his 
work on molecular modeling of halogen bonds was awarded by the French 
Embassy in the Czech Republic, and Sanofi-Aventis. He is a recipient of the 
Humboldt Research Fellowship for Postdoctoral Researchers. From 2014 he worked
with Paolo Carloni on RNA-ligand recognition in Forschungszentrum Jülich, Germany.
In 2016 he moved to Helmut Grubmüller's group at the Max-Planck Institute for
Biophysical Chemistry, Göttingen, Germany, pursuing his interest in 
biomolecular recognition and computer simulations.
Michal H. Kolář is active in science popularization; his efforts were 
recognized by the Czech Learned Society in 2012. In his spare time he 
likes traveling by train and dislikes tomatoes.

\subsection{Pavel Hobza}

Pavel Hobza is a holder of Distinguished Chair at the Institute of Organic 
Chemistry and Biochemistry, Academy of Sciences of the Czech Republic, 
Prague, and is a professor of chemistry at Charles University in Prague and 
Palacký University in Olomouc. He obtained his Ph.D. with Rudolf Zahradník
in 1974 at the Academy of Sciences in Prague. After postdoctoral studies 
in Montreal (with Camille Sandorfy), Erlangen (with Paul von Rague Schleyer)
and Munich (with Edward W. Schlag), he spent several periods as visiting 
professor in Montreal, Munich and Pohang (Republic of Korea). Dr. Hobza has 
authored or co-authored more than 500 papers and three books. These studies 
focus mainly on noncovalent interactions and their role in chemistry,
biodisciplines and nanosciences. According to Thomson Reuters in 2014 and 2015,
P.H. is ranked among the top 1\% of researchers with the most cited documents 
in the field of chemistry.


\section*{Acronyms and Abbreviations}

\begin{description}
\item[AM1] Austin model 1
\item[AM1-BCC] AM1 with bond charge correction
\item[au] atomic units
\item[B3LYP] three-parameter Becke-Lee-Yang-Parr functional
\item[BCP] bond critical point
\item[BSSE] basis set superposition error
\item[CBS] complete basis-set limit
\item[CC] coupled-cluster theory
\item[CCSD(T)] coupled-cluster theory with iterative single and double, and
perturbative triple excitations
\item[CSD] Cambridge Structural Database
\item[CT] charge transfer
\item[DABCO] diazabicyclo[2.2.2]octane
\item[DFT] density functional theory
\item[DFTB] density-functional tight binding
\item[DTCA] 1,3-dithiole-2-thione-4-carboxylic acid
\item[ESH] explicit \sgh
\item[ESP] electrostatic potential
\item[FCI] full configuration interaction
\item[HB] hydrogen bond
\item[HF] Hartree-Fock or hydrogen fluoride (depending on the context)
\item[HOMO] highest occupied molecular orbital 
\item[LJ] Lennard-Jones
\item[LUMO] lowest unoccupied molecular orbital
\item[MEP] molecular electrostatic potential
\item[MM] molecular mechanics
\item[MSE] mean signed error
\item[MP2] second-order Møller-Plesset perturbation theory
\item[MUE] mean unsigned error
\item[NBO] natural bond orbital
\item[NDDP] neglect of diatomic differential overlap
\item[PDB] Protein Data Bank
\item[PM3] parameterized model 3
\item[PM6] parameterized model 6
\item[PM6-D3H4X] PM6 with dispersion correction (3rd geenration), hydrogen-bonding
correction (4th generation) and halogen bonding correction
\item[QCISD] quadratic configuration interaction with single and double excitations 
\item[QM] quantum mechanics
\item[QTAIM] quantum theory of atom in molecules
\item[RESP] restricted electrostatic potential
\item[RMSD] root-mean-square deviation
\item[RMSE] root-mean-square error
\item[SAPT] symmetry-adapted perturbation theory
\item[SQM] semiempirical quantum mechanics
\item[XB] halogen bond
\end{description}


\begin{acknowledgement}
We thank Dr. Fanfrlík, Dr. Lo and, Dr. Řezáč for providing us data for some 
of the figures. This work was part of Research Project RVO: 61388963 of 
the Institute of Organic Chemistry and Biochemistry, Academy of Sciences 
of the Czech Republic. This work was also supported by the Czech Science 
Foundation [P208/12/G016] and the operational program Research and Development
for Innovations of the European Social Fund (CZ 1.05/2.1.00/03/0058). 
MHK is thankful for the support provided by the Alexander 
von Humboldt Foundation.
\end{acknowledgement}

\providecommand*\mcitethebibliography{\thebibliography}
\csname @ifundefined\endcsname{endmcitethebibliography}
  {\let\endmcitethebibliography\endthebibliography}{}

\begin{tocentry}
\includegraphics{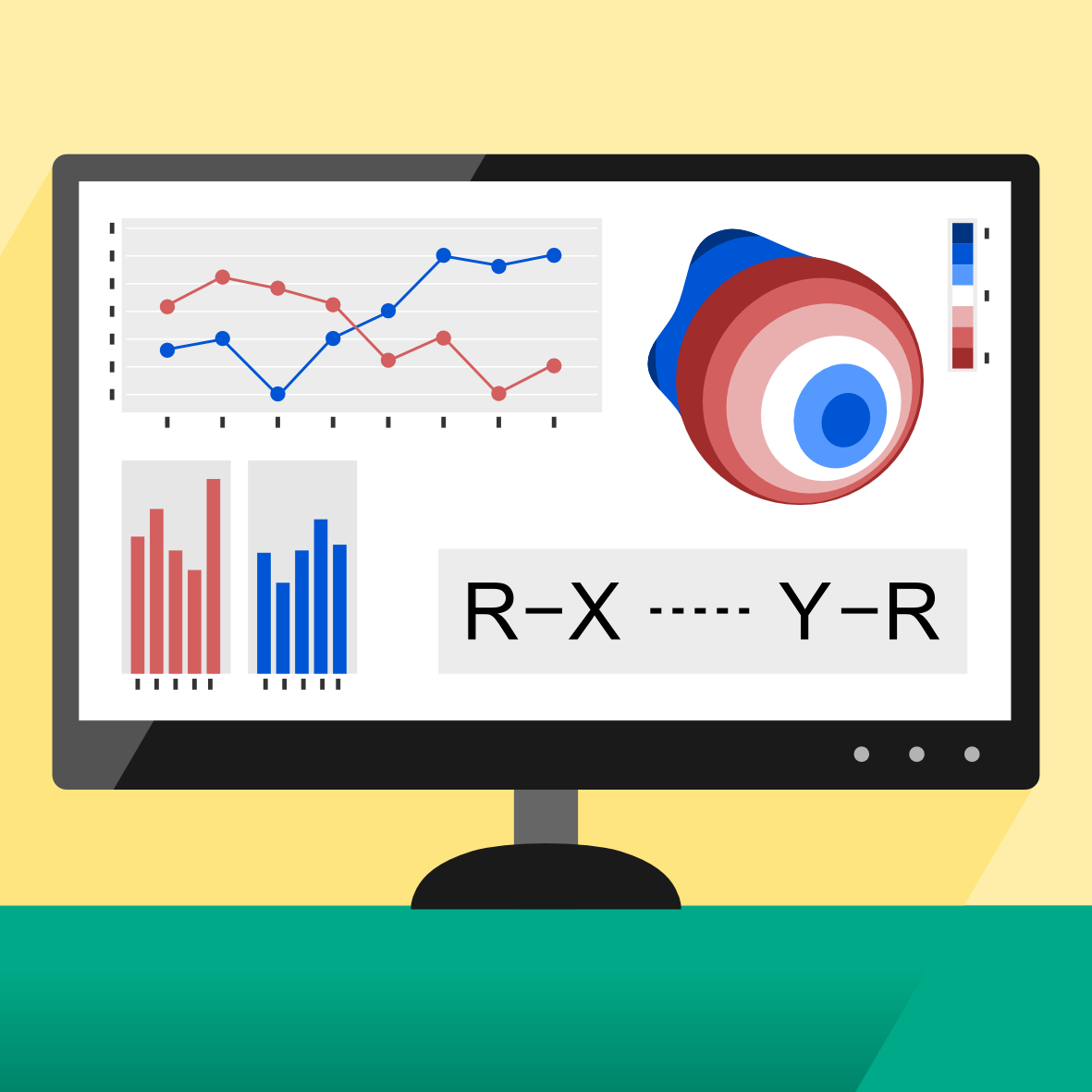}
\end{tocentry}

\end{document}